\documentclass[acmsmall,screen,nonacm,natbib=false]{acmart}

\setcopyright{none}

\RequirePackage[
  datamodel=acmdatamodel,
  style=acmnumeric,
  backref=true,
  backrefstyle=three,
  uniquename=false,
  uniquelist=false
  ]{biblatex}
\usepackage{software-biblatex}

\title{Classifying Capabilities (Extended Version)} %

\author{{Cao Nguyen} Pham}
\email{nguyen.pham@epfl.ch}
\orcid{https://orcid.org/0009-0005-2543-3309}
\author{Oliver Bračevac}
\email{oliver.bracevac@epfl.ch}
\orcid{https://orcid.org/0000-0003-3569-4869}
\author{Yichen Xu}
\email{yichen.xu@epfl.ch}
\orcid{https://orcid.org/0000-0003-2089-6767}
\author{Yaoyu Zhao}
\email{yaoyu.zhao@epfl.ch}
\orcid{https://orcid.org/0000-0003-2257-1413}
\author{Martin Odersky}
\email{martin.odersky@epfl.ch}
\orcid{https://orcid.org/0009-0005-3923-8993}
\affiliation{%
  \institution{EPFL}
  \city{Lausanne}
  \country{Switzerland}
}

\makeatletter
\def\@parfont{\sffamily\bfseries}
\makeatother

\usepackage{multicol}
\usepackage{enumitem}
\usepackage{bcprules}
\usepackage[only,llbracket,rrbracket]{stmaryrd}
\usepackage{commands}

\newcommand{\capless}{\textsf{Capless}}
\newcommand{\calculus}{\capless{}(\ensuremath{\mathcal{K}})}

\usepackage{mdframed}
\usepackage{wrapfig}
\usepackage{needspace}

\usepackage{mathtools}

\usepackage{listings}
\usepackage{lstfiracode} %

\newcommand{\btHL}[1]{\colorbox{yellow!20}{#1}}
\lstdefinelanguage{dotty}{
  keywords={erased, val, var, if, then, in, handle,
    return, def, match, case, new, type, trait,
     package, object, given, eff,
     pretype, class, extends, extension, infix, else,
     box, unbox, try, catch, import, throw, throws, using, enum,
     use, any, box, unbox, extension, this, abstract, final,
     sealed, override, private, protected, transparent, inline, opaque, open, tracked, lazy, uses, uses_init, update, consume,
     only, except},
  keywordstyle=\bfseries\color{magenta!80!black},
  sensitive=true,
  comment=[l]{//},
  morecomment=[s]{/*}{*/},
  commentstyle=\color{green!40!black},
  stringstyle=\color{green!60!black},
  morestring=[b]',
  morestring=[b]",
  moredelim=**[is][\btHL]{`}{`},
  columns=fullflexible,
  alsoletter={@},
}
\usepackage{tcolorbox}
\tcbuselibrary{listings,skins,breakable}

\newtcblisting{scalacode}{
  enhanced, breakable,
  colback=gray!4, colframe=gray!4, boxrule=0pt,
  borderline west={1.5pt}{0pt}{blue!20!gray!40},
  arc=0pt, left=2pt, right=1pt, top=1pt, bottom=1pt,
  before skip=3pt, after skip=3pt,
  listing only,
  listing options={language=dotty, style=FiraCodeStyle,
    basicstyle=\footnotesize\ttfamily,
    columns=fullflexible, keepspaces=true, showstringspaces=false,
    extendedchars=true, breaklines=false,
    aboveskip=0pt, belowskip=0pt},
}

\newtcblisting{scalacodesmall}{
  enhanced,
  colback=gray!4, colframe=gray!4, boxrule=0pt,
  borderline west={1.5pt}{0pt}{blue!20!gray!40},
  arc=0pt, left=2pt, right=1pt, top=1pt, bottom=1pt,
  before skip=3pt, after skip=3pt,
  listing only,
  listing options={language=dotty, style=FiraCodeStyle,
    basicstyle=\scriptsize\ttfamily,
    columns=fullflexible, keepspaces=true, showstringspaces=false,
    extendedchars=true, breaklines=false, mathescape=true,
    aboveskip=0pt, belowskip=0pt},
}

\usepackage{tikz}
\usepackage{cancel}
\usetikzlibrary{positioning,fit}

\newcommand{\TODO}[1]{{\color{red}\textbf{([TODO]: {#1})}}}
\renewcommand{\TODO}[1]{}

\makeatletter
\expandafter\gdef\csname ver@amsthm.sty\endcsname{2020/05/29 v2.20.6}
\makeatother
\usepackage[capitalize,nameinlink]{cleveref}
\RemoveFromHook{label}[firstaid/cleveref]
\crefname{appendix}{appendix}{appendices}
\Crefname{appendix}{Appendix}{Appendices}

\usepackage{yfonts,lettrine}

\AtBeginDocument{\setlength{\DefaultFindent}{0.5em}}
\makeatletter%
\begin{document}

\begin{abstract}
	Capture checking in Scala~3
	enables lightweight and practical effect and resource tracking
	by recording capabilities in types.
	However, the system offers no way to reason about \emph{kinds} of capabilities.
	Natural constraints such as
	``retaining only the control-flow capabilities of this closure'' or
	``excluding all thread-local capabilities from this argument''
	become inexpressible.
	Both arise in the Scala~3 standard library:
	\lstinline|Try| re-throws caught exceptions,
	so it retains only the control-flow capabilities of its body,
	and \lstinline|Future| must not capture thread-local resources.
	The inability to state these constraints has kept parts of the library outside capture checking.

	We introduce \emph{capability classifiers}:
	a tree-structured, user-extensible hierarchy of tags
	that classify capabilities by their semantic role.
	\emph{Projections} filter capture sets by classifier,
	supporting both inclusion (\lstinline|c.only[C]|) and exclusion (\lstinline|c.except[C]|).
	The tree structure enables decidable disjointness reasoning:
	classifiers on separate branches are guaranteed to be disjoint
	regardless of unknown extensions elsewhere in the hierarchy.
	We formalize classifiers as an extension of System Capless, a core calculus for capture checking,
	introducing a classifier kind algebra based on intersection, union, and subtraction
	of classifier subtrees.
	We extend the operational semantics to model exception interception and
	establish type safety, effect safety, and handler coverage via a big-step proof,
	fully mechanized in Lean~4.
	Classifiers are implemented in the Scala~3 capture checker,
	and we demonstrate their use on standard library types
	and real-world effect exclusion patterns.
\end{abstract}

\maketitle

\section{Introduction}\label{sec:intro}

Capture checking~\cite{boruchgruszecki2023capturing} provides a lightweight discipline for
tracking effects as capabilities.
A \emph{capability} is a value whose possession grants the authority to perform an effect
or access a resource, for example a file handle, a network socket, a mutable reference,
or a token that permits throwing an exception.
By recording which capabilities a term may use in its \emph{capture set},
the type system enforces lexically scoped bounds on side effects:
a function's type signature declares which capabilities it may use,
and the compiler statically checks that these bounds are respected.
This approach has been integrated into Scala~3,
and has been applied to large parts of the Scala standard library,
including the collections~\cite{xu2025capless}.

Two notable parts of the standard library have raised difficulties when adopting capture checking:
exception handling via \lstinline|Try[T]| and asynchronous computation via \lstinline|Future[T]|.
The difficulty lies not in tracking \emph{individual} capabilities,
but in reasoning about \emph{kinds} of capabilities.

Consider \lstinline|Try[T]| \cite{scalaTry}: its constructor executes a closure and catches any thrown exceptions,
packaging the result as a success-or-failure value.
\begin{scalacode}
val result: Try[String] = Try(file.read())
val content: String = result.get // may re-throw IOException
\end{scalacode}
The constructor takes its argument by name, so \lstinline|file.read()| is not evaluated at
the call site. Instead, it is implicitly wrapped in a closure that the constructor runs
internally.
In this example, \lstinline|Try| catches exceptions thrown by \lstinline|file.read()| and wraps
them as a value. The resulting \lstinline|Try| is pure with respect to most capabilities,
since e.g., the file handle is not retained.
However, it is \emph{not} pure with respect to control-flow capabilities such as
\lstinline|CanThrow| \cite{odersky21saferexceptions} labels: calling \lstinline|.get| may re-throw the caught exception,
which exercises that capability.
Without a way to express ``the control-flow subset of the closure's captures,'' the type of
\lstinline|Try|'s constructor must be either too permissive (unsoundly claiming purity) or too conservative
(retaining all captures).

A dual problem arises with \lstinline|Future[T]|.
Its constructor dispatches a closure to a thread pool,
where it may execute on an arbitrary carrier thread.
\begin{scalacode}
boundary: label =>
  Future:            // should reject: label is thread-local
    break(0, label)  // non-local return to another thread's stack
    file.read()
\end{scalacode}
As with \lstinline|Try|, the \lstinline|Future| constructor takes its body by name, and
Scala's colon syntax passes the indented block as that argument.
Here, \lstinline|boundary| introduces a \lstinline|Label| capability that is tied to the
originating thread's stack. Using it inside the \lstinline|Future| closure would attempt
a non-local return across threads, which must be rejected.
More generally, capabilities such as thread-local storage, stack-based control flow,
and native interop must not leak into the closure.
Yet existing capture checking provides no mechanism to \emph{exclude}
an entire category of capabilities from a capture set.
One can only enumerate specific capabilities as an upper bound.

Both problems share a common root: capture sets track capabilities by identity,
but lack the vocabulary to classify capabilities by \emph{kind}.
There is no way to filter a capture set to retain only capabilities of a certain kind,
nor to exclude all capabilities belonging to another.
These are precisely the operations needed to give accurate constraints to \lstinline|Try| and \lstinline|Future|,
and, as we demonstrate, to many patterns beyond the standard library.
The missing abstraction is not better tracking of individual capabilities, but operators that
project and exclude whole categories of capabilities.

We propose \emph{capability classifiers}:
a tree-structured, user-extensible system of tags that classify capabilities by their semantic role.
Each capability is associated with a classifier drawn from a tree-structured universe.
Capture sets are enhanced with \emph{projections} that filter by classifier \emph{kinds},
which consist of classifier subtree inclusions (\lstinline|c.only[C]|) and exclusions (\lstinline|c.except[C]|).
With classifiers, the previously problematic signatures become precise:
\begin{scalacode}
object Try:
  def apply[T](body: () => T): Try[T]^{body.only[Control]}
object Future:
  def apply[T](body: () -> {any.except[ThreadLocal]} T): Future[T]^{body}
\end{scalacode}
That is, \lstinline|Try| retains only control capabilities from its argument, while
\lstinline|Future| requires its argument to exclude thread-local ones.

The tree structure of classifiers is chosen to fit the setting that Scala-like languages
inhabit: an open ecosystem of separately compiled modules.
The tree is open, since any classifier can be extended with sub-classifiers in a separate
module, yet disjointness remains decidable and stable under such extensions:
classifiers on different branches stay disjoint no matter which sub-classifiers other
modules add later.
Hierarchical classification, filtering, and exclusion each have precedents in effect
systems, for instance as boolean algebras \cite{lutze23flix,gao25invalidation} or
lattices of qualifiers \cite{lee24qualifiers}.
Our contribution is their integration with capture checking, where the filtered sets
contain references to program values rather than abstract effect labels, under the
modularity constraints that separate compilation imposes on a language like Scala.
We explain this design point in Section~\ref{sub:why-tree}.

Beyond the standard library, we show that classifiers enable many practical patterns:
restricting parallel computations to read-only capabilities, modeling privilege qualifiers
on transaction objects, and encoding the effect exclusion patterns catalogued by
Lutze et al.~\cite{lutze23flix} in real-world codebases.
The result is a lightweight extension to capture checking that makes previously awkward
library interfaces precise without abandoning the existing capability-based account.

To summarize, our contributions are as follows.
\begin{itemize}[leftmargin=1.2em,labelsep=0.4em]
	\item We introduce capability classifiers for Scala~3 capture checking, motivated by
	      standard-library examples and real-world effect exclusion patterns that require
	      reasoning about capabilities by kind rather than by identity
	      (Section~\ref{sec:informal}).
	\item We design a tree-structured, user-extensible classifier hierarchy with a decidable
	      kind algebra supporting union, intersection, subtraction, and disjointness, chosen
	      to preserve modular extension under separate compilation
	      (Section~\ref{sub:classifiers-kinds}).
	\item We formalize capture checking with capability classifiers as System \calculus{}, an
	      extension of System Capless~\cite{xu2025capless} with kind-annotated projections and
	      capture kinding, and prove type safety with a checked big-step semantics, fully
	      mechanized in Lean~4. In contrast to prior capture-checking metatheories based on
	      syntactic progress and preservation, the big-step account lets us reason directly
	      about closure properties of evaluation, yielding not only type and effect safety,
	      but also used label prediction, capture prediction and handler coverage
	      (Sections~\ref{sec:calculus} and~\ref{sec:metatheory}).
	\item We implement classifiers in the Scala~3 capture checker, demonstrate them on
	      standard-library types and on real-world effect exclusion patterns, discuss the
	      resulting design decisions, and report on the impact of classifiers in the
	      capture-checked standard library (Sections~\ref{sec:impl}
	      and~\ref{sec:discussion}).
\end{itemize}
Finally, we review related work in Section~\ref{sec:related} and conclude in Section~\ref{sec:conclusion}.

\section{The Case for Classifiers: Capability Kinds in Scala}\label{sec:informal}

Capture checking \cite{boruchgruszecki2023capturing,xu2025capless,dottycc}
extends Scala's type system to track \textit{capabilities}, informally
values ``of interest'', such as file handles, access-permission tokens,
or mutable data structures.
In capture checking, types also record which capabilities are used or
held, or \textit{captured}, by values of that type.
These \textit{capturing types} have the form \lstinline|T^{x1, ..., xn}|,
which includes the shape type \lstinline|T| and the \textit{capture set} \lstinline|{x1, ..., xn}|.
Capture sets give an upper bound on capabilities that can be accessed by values of the type.
A pure type, written \lstinline|T^{}| or just \lstinline|T|, cannot capture any capabilities.

By making capability dependencies visible in types, capture checking serves as
a lightweight effect system: the capabilities a function closes over precisely
describe the effects it may perform, enabling fine-grained
control over what each piece of code is allowed to do.

The conventional function type \lstinline|A => B| is shorthand for
\lstinline|(A -> B)^{any}|\footnote{The universal capability \lstinline|any| was referred to as \textbf{cap} in older works.},
a function that may capture arbitrary capabilities.
In contrast, the pure function type \lstinline|A -> B| carries an empty capture set
and therefore cannot close over any capability.
For example, in the following code, \lstinline|run| declares that its callback \lstinline|task|
may only use the \lstinline|net| capability:

\begin{scalacode}
def main(fs: Filesystem^, net: Network^): Unit =
  def run(task : () ->{net} Unit) = task()
  run(() => net.send(fs.read("secret")))  // error: fs not in {net}
\end{scalacode}

\noindent
Here \lstinline|task| has type \lstinline|() ->{net} Unit|,
a shorthand for \lstinline|(() -> Unit)^{net}|.
The capture set \lstinline|{net}| makes explicit that this function may only use
the \lstinline|Network| capability held by \lstinline|net|.
Any attempt to capture \lstinline|fs|, which has type \lstinline|Filesystem^| (i.e., \lstinline|Filesystem^{any}|),
is rejected because \lstinline|fs| is not in the capture set.

Capabilities are extremely versatile: they can express concepts from
many different domains, such as exceptions, I/O, mutation, and more.
However, capture sets may only mention capability \textit{identities}.
This means it is not possible to:
\begin{enumerate}
	\item limit entire \textit{kinds} of capabilities based on system
	      specifications or user definitions (e.g., all control-flow-related
	      capabilities);
	\item refer to a \textit{subset} of capabilities captured by a given
	      reference, filtered by such kinds (e.g., only the read-only
	      subset of a mutable data structure); or
	\item exclude capabilities from a capture set.
\end{enumerate}
At the heart of this problem lies the lack of a classification mechanism
for capabilities and the inability to include or exclude them based on
that classification.
We introduce \textit{classifiers}, a hierarchical system and algebra for
classifying capabilities.
In this section, we give an informal introduction to classifiers in Scala
and showcase the need for them in the standard library.

\subsection{Capability Classifiers}

A \textit{classifier} is a tag drawn from a tree-structured hierarchy that
describes the semantic role of a capability.
Each classifier has exactly one parent but may have arbitrarily many children.
Because every path from a classifier to the root is unique, a capability
can be assigned to exactly one classifier while still being
recognized as belonging to every ancestor of that classifier.
The root of the tree is the built-in classifier \lstinline|Capability|,
provided by the standard library.
The standard library also provides two children of \lstinline|Capability|:
\lstinline|SharedCapability|, for capabilities safe to share across scopes,
and \lstinline|ExclusiveCapability|, for exclusively owned resources.

\begin{figure}[t]
	\begin{minipage}[t]{0.55\textwidth}
		\vspace{0pt}%
		\begin{scalacodesmall}
trait ThreadLocal extends Classifier, SharedCapability
trait Control     extends Classifier, ThreadLocal

class CanThrow[-T] extends Control

case class Environment(file: File, exc: CanThrow[Int])

def runOnNewThread(task: () ->{any.except[Control]} Unit)
\end{scalacodesmall}
	\end{minipage}\hfill
	\begin{minipage}[t]{0.42\textwidth}
		\vspace{0pt}%
		\begin{scalacodesmall}
def f(env: Environment^) =
  // ok: env.file is not Control
  runOnNewThread(() => env.file.read())

  // error: exc is Control
  runOnNewThread:
    given CanThrow[Int] = env.exc
    throw 1
\end{scalacodesmall}
	\end{minipage}
	\vspace{-8pt}
	\caption{Classifiers in Scala: declarations (left) and excluding the
		\lstinline|Control| category with \lstinline|.except| (right).}
	\label{fig:classifier-decls}
\end{figure}

\paragraph{Defining classifiers.}
In Scala, classifiers are represented as traits.
A new classifier is introduced by defining a trait that extends the marker
trait \lstinline|Classifier| and a parent classifier
(Figure~\ref{fig:classifier-decls}, top left).
This declares \lstinline|Control| as a child of \lstinline|ThreadLocal|.
Because classifiers form an open tree, new children can be added to any
classifier in a separate module without modifying existing definitions.

\paragraph{Classifying capabilities.}
A Scala class is associated with a classifier by extending the classifier
trait \textit{without} the \lstinline|Classifier| marker
(Figure~\ref{fig:classifier-decls}, left).
\lstinline|CanThrow| is a capability~\cite{scalaCanThrow} that tracks
possibly-thrown exceptions of type \lstinline|T|.
Every instance of it is thereby tagged with the \lstinline|Control| classifier,
which lets the type system distinguish it from unrelated capabilities such as
file handles.
The \lstinline|-T| syntax declares that \lstinline|CanThrow| is contravariant:
the more precise \lstinline|T| becomes, the smaller the set of values that can be thrown.

\paragraph{Projections: \lstinline|.only| and \lstinline|.except|.}
Given a capture reference \lstinline|a|,
\lstinline|a.only[C]| denotes the subset of capabilities reachable through
\lstinline|a| whose classifier is \lstinline|C| or any descendant of \lstinline|C|.
Dually, \lstinline|a.except[C]| denotes the subset whose classifier
is \textit{not} \lstinline|C| or any descendant of \lstinline|C|.
These two \textit{projections} also apply to the root capability
\lstinline|any| to constrain an arbitrary capture set by classifier.
Projections can be chained:
\lstinline|any.only[SharedCapability].except[Control]| restricts to shared
capabilities outside \lstinline|Control|.

With these tools in hand, consider the example of
Figure~\ref{fig:classifier-decls} (right).
An environment bundles a file handle and an exception-throwing capability.
A function that dispatches work to a new thread must forbid its closure from
using thread-local capabilities, such as exception throwing, and the \lstinline|.except| projection
expresses this constraint directly.
The capture set \lstinline|{any.except[Control]}| is polymorphic:
it accepts any capability \textit{except} those under \lstinline|Control|.
Accessing the \lstinline|File| is allowed because files are not
\lstinline|Control|, while throwing exceptions
using the \lstinline|CanThrow| capability \cite{odersky21saferexceptions}
is rejected at compile time.

\subsection{Case Study: \lstinline|Try[T]| Captures Control Capabilities}\label{sub:case-study-try}

\lstinline|Try[T]| \cite{scalaTry} is a sum type representing either a successful value or an
arbitrary runtime exception, enabling fluid conversion between exception-based
and monadic error handling.
Under previous capture checking, the constructor and re-throwing accessor
had the signatures in Figure~\ref{fig:try-sigs} (left).

\begin{figure}[t]
	\begin{minipage}[t]{0.44\textwidth}
		\vspace{0pt}%
		\begin{scalacodesmall}
object Try: /* before classifiers */
  def apply[T](body: () => T): Try[T]

class Try[+T]:
  def get(): T = /* re-throw if needed */
\end{scalacodesmall}
	\end{minipage}\hfill
	\begin{minipage}[t]{0.54\textwidth}
		\vspace{0pt}%
		\begin{scalacodesmall}
object Try:
  /* with classifiers */
  def apply[T](body: () => T): Try[T]^{body.only[Control]}
\end{scalacodesmall}
	\end{minipage}
	\vspace{-8pt}
	\caption{\lstinline|Try[T]| signatures before classifiers (left) and with
		classifiers (right).}
	\label{fig:try-sigs}
\end{figure}

\noindent
The constructor \lstinline|Try.apply| takes a computation with arbitrary captures, yet the
resulting \lstinline|Try[T]| is declared \textit{pure}.
At first glance this seems correct: \lstinline|Try.apply| evaluates \lstinline|body|,
catches any exception, and stores the outcome;
it does not retain a reference to the closure itself.

However, this signature overlooks a particular class of capabilities that
govern exception throwing.
\lstinline|CanThrow| capabilities \cite{odersky21saferexceptions,scalaCanThrow}
track possibly-thrown exceptions through types, and
\lstinline|boundary/break|'s \lstinline|Label| capability
\cite{scalaBoundary} tracks the ability to perform non-local returns.
Both are classified under \lstinline|Control|.
When \lstinline|Try.get| re-throws a stored exception, it is effectively
\textit{using} the \lstinline|Control| capabilities that were in scope when
the exception was originally thrown.
Therefore \lstinline|Try[T]| is not truly pure:
it retains the \lstinline|Control| capabilities of the closure.
With classifiers, this dependency can be stated precisely
(Figure~\ref{fig:try-sigs}, right).
The projection \lstinline|body.only[Control]| selects only the subset of
\lstinline|body|'s capture set whose classifier is \lstinline|Control| or a
descendant of \lstinline|Control|.
Non-control capabilities captured by the closure (e.g., file handles) are
excluded from the result type, reflecting the fact that \lstinline|Try|
does not propagate them.

\subsection{Case Study: \lstinline|Future[T]| Cannot Handle Thread-Locals}\label{sub:future-threadlocal}

\lstinline|Future[T]| \cite{scalaFuture} represents a computation that may execute on an
arbitrary worker thread.
Because the task scheduler is free to migrate the computation between carrier threads,
the closure submitted to a \lstinline|Future| must not rely on any thread-local state.
Thread-local resources include thread-local storage, local stack
manipulation (\lstinline|Control| capabilities, coroutines), and native C
interop.

With classifiers, this restriction is expressed directly in the
\lstinline|Future| constructor:

\begin{scalacode}
object Future:
  def apply[T](body: () ->{any.except[ThreadLocal]} T): Future[T]^{body}
\end{scalacode}

\noindent
The \lstinline|.except[ThreadLocal]| projection removes any
capability whose classifier falls under \lstinline|ThreadLocal|.
Crucially, because exception handling is tied to the thread-local call stack,
\lstinline|Control| is defined as a child of \lstinline|ThreadLocal|, so
\lstinline|.except[ThreadLocal]| excludes \lstinline|Control| capabilities
as well.
This illustrates how the tree structure pays off: a single exclusion of an
ancestor classifier automatically covers all of its descendants, without
requiring the programmer to enumerate them.

Beyond these standard-library examples, Lutze et al.\ \cite{lutze23flix} have
demonstrated the practical value of effect exclusions in existing code.
We revisit such patterns in Section~\ref{sub:fine-grained-control}.
 
\section{System \calculus{}: Capture Checking with Classifiers}\label{sec:calculus}

We formalize capability classifiers in System \calculus{}.
Our work is built upon System \capless{}~\cite{xu2025capless},
a core calculus for capture checking.
It models a lambda calculus that tracks capabilities in types.
The calculus follows monadic normal form (MNF)~\cite{hatcliff94generic}, where intermediate computations are
bound by \(\LET\) expressions and both functions and their arguments are restricted to variables.
This does not reduce expressiveness: any application \(t\ u\) can be rewritten as
\(\LET x_1 = t \IN \LET x_2 = u \IN x_1\ x_2\).
Types are stratified into \emph{shape types} $S$ (function types, type variables, $\top$)
and \emph{capturing types} $T$, which pair a shape type with a \emph{capture set}.
Existential types $\exists c. T$ allow existentially quantifying a capture set,
enabling abstraction over unknown capabilities.
The calculus supports three forms of polymorphism: term abstraction over values,
type abstraction over type variables bounded by shape types,
and capture abstraction over capture variables bounded by capture sets.
For example, we can type the original \lstinline|Try[T]| constructor using System \capless{} as follows:
\[\text{Try.apply}: \forall[T <: \top]\ \forall[\bs{c}: *]\ \forall(\text{body}: (\forall(u: \text{unit})\ T)\capt\bs{\set{c}})\ \text{Try}[T]\capt\bs{\set{}} \]
The constructor is polymorphic in the body's return type \(T\), as well as in its capture set \(\bs{c}\).
The \(: *\) bound indicates that the capture set is unbounded: the body can capture any variables in scope.
The return type \(\textsf{Try}[T]\capt\bs{\set{}}\) indicates that the resulting \lstinline|Try[T]| has no captures, which
is incorrect as previously mentioned. We will see how this is correctly typed in System \calculus{} in the next section.

\begin{figure}[tbp]\begin{flushleft}\footnotesize\noindent
		\begin{minipage}[t]{0.31\textwidth}
			\begin{flalign*}
				x,\,y,\,z         \tag*{\textbf{Variable}}                                                                   \\
				{c}                 \tag*{\textbf{Capture Variable}}                                                         \\
				{\new{\cls{\CLS, \CLSS}\in\mathcal{K}}}                 \tag*{\textbf{Classifier Trait}}                     \\
				{\new{\cls{\KIND{}}}\coloneqq}\  & \tag*{\textbf{Classifier Kind}}                                           \\
				                                 & \cls{\emptyset}                    \tag*{empty}                           \\
				                                 & \cls{\CLS - \overline{\CLS_i}}                    \tag*{subtree}          \\
				                                 & \cls{\KIND{}\vee\KIND{}}                    \tag*{union}                  \\
				{\new{\pi}}\coloneqq\            & \tag*{\textbf{Projected Capture}}                                         \\
				                                 & \sproj{\theta}{\cls{\KIND{}}}  \tag*{projection}                          \\
				\bs{C},\,\bs{D}\coloneqq\        & \bs{\set{{\pi_1,\cdots,\pi_n}}}     \tag*{\textbf{Capture Set}}           \\
				\G\coloneqq\                     & \tag*{\textbf{Context}}\label{syn:context}                                \\
				                                 & \emptyset                    \tag*{empty}                                 \\
				                                 & \G, x: T                    \tag*{term binding}\label{syn:term-binding}   \\
				                                 & \G, X<:S                    \tag*{type binding}                           \\
				                                 & {\G, c:B}                    \tag*{capture binding}                       \\
			\end{flalign*}
		\end{minipage}\hfill
		\begin{minipage}[t]{0.34\textwidth}
			\begin{flalign*}
				s,\,t,\,u\coloneqq\              & \tag*{\textbf{Term}}                                                      \\
				                                 & x                                      \tag*{variable}                    \\
				                                 & v                              \tag*{value}                               \\
				                                 & x\,y                              \tag*{application}                      \\
				                                 & x[S]                              \tag*{type application}                 \\
				                                 & {x[c]}                              \tag*{capture application}            \\
				                                 & \LET x = t \IN u                  \tag*{let}                              \\
				                                 & {\LET \langle c, x \rangle = t \IN u}    \tag*{existential let}           \\
				v\coloneqq\                      & \tag*{\textbf{Value}}                                                     \\
				                                 & \lambda^{\new{\bs{C}}}(x: T)t                      \tag*{function}        \\
				                                 & \lambda^{\new{\bs{C}}}[X<:S]t                      \tag*{type function}   \\
				                                 & {\lambda^{\new{\bs{C}}}[c:B]t}                    \tag*{capture function} \\
				                                 & \PACK\langle \bs{C}, x\rangle                    \tag*{pack}              \\
				\theta\coloneqq\                 & \tag*{\textbf{Capture}}                                                   \\
				                                 & x                      \tag*{variable}                                    \\
				                                 & {c}                      \tag*{capture variable}                          \\
			\end{flalign*}
		\end{minipage}\hfill
		\begin{minipage}[t]{0.31\textwidth}
			\begin{flalign*}
				B\coloneqq\                & \tag*{\textbf{Capability Bound}}                                        \\
				                           & {\new{\cls{\KIND{}}}}                    \tag*{kind}                    \\
				                           & \bs{C}                    \tag*{capture set}                            \\
				E,\,F\coloneqq\            & \tag*{\textbf{Existential Type}}                                        \\
				                           & {\exists c\new{: B}.\,} T                    \tag*{existential}         \\
				                           & T  \tag*{type}                                                          \\
				T,\,U\coloneqq\            & \tag*{\textbf{Type}}                                                    \\
				                           & S\capt \bs{C}                    \tag*{capturing}                       \\
				                           & S                    \tag*{pure}                                        \\
				R,\,S\coloneqq\            & \tag*{\textbf{Shape Type}}                                              \\
				                           & \top                    \tag*{top}                                      \\
				                           & X                    \tag*{type variable}                               \\
				                           & \forall(x: T)E             \tag*{function}                              \\
				                           & \forall[X<:S]E             \tag*{type function}                         \\
				                           & {\forall[c: B]E}             \tag*{capture function}                    \\
			\end{flalign*}
		\end{minipage}
		\vspace{-14pt}
		\caption{Abstract syntax of System \calculus{}. Changes compared to System \capless{}~\cite{xu2025capless} are \tnew{marked}.}\label{fig:syntax-paper}
	\end{flushleft}\end{figure}\normalsize

\begin{figure*}[htbp]
	\footnotesize
	\setlength{\afterruleskip}{3pt plus 1pt minus 1pt}

	\flushleft{\textbf{Capture Kinding} \quad \jbox{\new{\G \vdash \bs{C}: \cls{\KIND}}}}

	\begin{multicols}{3}
		\infrule[\ruledef{k-var}]
		{x : S\capt \bs{C} \in \G\andalso\G \vdash \bs{C}\PROJ[\cls{\KIND_1}]: \cls{\KIND_2}}
		{\G \vdash \bs{\set{\sproj{x}{\cls{\KIND_1}}}}: \cls{\KIND_2}}

		\infrule[\ruledef{k-sub}]
		{\G \vdash \bs{C}: \cls{\KIND}\andalso
			\cls{\subk{\KIND}{\KIND'}}}
		{\G \vdash \bs{C}: \cls{\KIND'}}

		\infrule[\ruledef{k-cvar}]
		{c : \bs{C}\in\G\andalso\G \vdash \bs{C}\PROJ[\cls{\KIND_1}] : \cls{\KIND_2}}
		{\G \vdash \bs{\set{\sproj{c}{\cls{\KIND_1}}}}: \cls{\KIND_2}}

		\infrule[\ruledef{k-union}]
		{\G \vdash \bs{C_1}: \cls{\KIND}\andalso
			\G \vdash \bs{C_2}: \cls{\KIND}}
		{\G \vdash \bs{C_1\cup C_2}: \cls{\KIND}}

		\infrule[\ruledef{k-cbound}]
		{c:\cls{\KIND_1}\in\G}
		{\G \vdash \bs{\set{\sproj{c}{\cls{\KIND_2}}}}: \cls{\KIND_1 \land \KIND_2}}

		\infrule[\ruledef{k-absurd}]
		{\cls{\KIND\EMPTY}}
		{\G \vdash \bs{\set{\sproj{\theta}{\cls\KIND}}}: \cls{\KIND'}}

		\infax[\ruledef{k-empty}]
		{\G \vdash \bs{\set{}}: \cls{\KIND}}

	\end{multicols}

	\flushleft{\textbf{Capture Subsetting} \quad \jbox{\new{\bs{C_1} \sqsubseteq \bs{C_2}}}}

	\begin{multicols}{2}

		\infrule[\ruledef{s-elem}]
		{\bs{C_1} \subseteq \bs{C_2}}
		{\bs{C_1} \sqsubseteq \bs{C_2}}

		\infrule[\ruledef{s-subkind}]
		{\cls{\subk{\KIND{}_1}{\KIND{}_2}}}
		{\bs{\{\sproj{\theta}{\cls{\KIND{}_1}}\}} \sqsubseteq \bs{\{\sproj{\theta}{\cls{\KIND{}_2}}\}}}

		\infrule[\ruledef{s-absurd}]
		{\cls{\KIND\EMPTY}}
		{\bs{\{\sproj{\theta}{\cls{\KIND{}}}\}} \sqsubseteq \bs{\{\}}}

		\infax[\ruledef{s-merge}]
		{\bs{\{\sproj{\theta}{\cls{(\KIND{}_1 \lor \KIND{}_2)}}\}} \sqsubseteq \bs{\{\sproj{\theta}{\cls{\KIND_1}}, \sproj{\theta}{\cls{\KIND{}_2}}\}}}
	\end{multicols}

	\flushleft{\textbf{Subcapturing} \quad \jbox{\subs{\G}{\bs{C_1}}{\bs{C_2}}}\quad (changed rules; full set in \Cref{fig:all-typing-full})}

	\begin{multicols}{2}
		\infrule[\ruledef{sc-var}]
		{x : S\capt \bs{C} \in \G}
		{\subs{\G}{\bs{\set{(x\new{\mid\cls{\KIND{}}})}}}{\bs{C\new{\PROJ[\cls{\KIND{}}]}}}}

		\infrule[\ruledef{sc-bound}]
		{c : \bs{C} \in \G}
		{\subs{\G}{\bs{\set{(c\new{\mid\cls{\KIND{}}})}}}{\bs{C\new{\PROJ[\cls{\KIND{}}]}}}}

		\infrule[\ruledef{sc-elem}]
		{{\bs{C_1}\new{\sqsubseteq} \bs{C_2}}}
		{\subs{\G}{\bs{C_1}}{\bs{C_2}}}

		\infrule[{\new{\ruledef{sc-proj}}}]
		{\G \vdash \bs{C}: \cls{\KIND{}}}
		{\subs{\G}{\bs{C}}{\bs{C\PROJ[\cls{\KIND{}}]}}}

	\end{multicols}

	\flushleft{\textbf{Bound Subtyping} \quad \jbox{\G \vdash B_1 <:_{\mathsf{B}} B_2}}
	\begin{multicols}{3}
		\infrule[\ruledef{b-set}]
		{\subs{\G}{\bs{C_1}}{\bs{C_2}}}
		{\G \vdash \bs{C_1} <:_{\mathsf{B}} \bs{C_2}}
		\infrule[\new{\ruledef{b-kind}}]
		{\cls{\subk{\KIND_1}{\KIND_2}}}
		{\G \vdash \cls{\KIND_1} <:_{\mathsf{B}} \cls{\KIND_2}}
		\infrule[\new{\ruledef{b-set-kind}}]
		{\G \vdash \bs{C}: \cls{\KIND}}
		{\G \vdash \bs{C} <:_{\mathsf{B}} \cls{\KIND}}

	\end{multicols}

	\flushleft{\textbf{Typing} \quad \jbox{\bs{C}; \G \vdash t : E}\quad (selected rules; full set in \Cref{fig:all-typing-full})}

	\begin{multicols}{3}
		\infrule[\ruledef{var}]
		{x: S\capt \bs{C} \in \G}
		{\bs{\set x}; \G \vdash x : S\capt\bs{\set x}}

		\infrule[\ruledef{abs}]
		{\bs{C\uplus \set{x}}; (\G, x: T) \vdash t : E\\ \G \vdash T\ \textbf{\textsc{wf}}}
		{\bs{\set{}}; \G \vdash \lambda^{\bs{C}}(x: T)t : (\forall(x: T)E)\capt \bs{C}}

		\infrule[\ruledef{app}]
		{\bs{C'}; \G \vdash x : (\forall (z: T)E)\capt \bs{C} \\
			\bs{C'}; \G \vdash y : T}
		{\bs{C'}; \G \vdash x\ y : \bs{[y/z]}E}

		\infrule[\ruledef{pack}]
		{\bs{C}; \G \vdash x : \bs{[D/c]}T \\
		\new{\G \vdash \bs D <:_{\mathsf{B}} {B}}}
		{\bs{\set{}}; \G \vdash \PACK\langle \bs D, x\rangle : \exists c\new{: B}.\,{T}}

		\infrule[\ruledef{let}]
		{\bs{C}; \G \vdash t : T\andalso \bs{C}; (\G, x: T) \vdash u : E \\
			\G \vdash \bs{C}, E\ \textbf{\textsc{wf}}}
		{\bs{C}; \G \vdash \LET x = t \IN u : E}

		\infrule[\ruledef{let-e}]
		{\bs{C}; \G \vdash t : \exists c\new{: B}.\, T \andalso \G \vdash \bs{C}, F\ \textbf{\textsc{wf}} \\
			\bs{C}; (\G, \new{c: B}, x: T) \vdash u : F}
		{\bs{C}; \G \vdash \LET \langle c, x \rangle = t \IN u : E}

	\end{multicols}

	\vspace{-10pt}
	\flushleft{\textbf{Subtyping} \quad \jbox{\G \vdash E_1 <: E_2}\quad (selected rules; full set in \Cref{fig:all-typing-full})}
	\begin{multicols}{2}
		\infrule[\ruledef{capt}]
		{\G \vdash S_1 <: S_2\\ \subs{\G}{\bs{C_1}}{\bs{C_2}}}
		{\G \vdash S_1\capt \bs{C_1} <: S_2\capt \bs{C_2}}

		\infrule[\ruledef{exist}]
		{(\G, c : \new{B_1}) \vdash T_1 <: T_2\andalso \new{\G \vdash B_1 <:_{\mathsf{B}} B_2}}
		{\G \vdash \exists c \new{: B_1}. T_1 <: \exists c \new{: B_2}.T_2}
	\end{multicols}

	\vspace{-15pt}
	\caption{Type system of System \calculus{} (condensed).
		Changes compared to System \capless{}~\cite{xu2025capless} are \tnew{marked}.
		The complete type system is given in \Cref{fig:all-typing-full}.}\label{fig:all-typing-paper}

\end{figure*}
 
\begin{figure*}[htbp]
	\footnotesize
	\setlength{\afterruleskip}{2pt plus 1pt minus 1pt}

	\flushleft{\textbf{Intersection} \quad \jbox{\cls{\KIND{}_1 \land \KIND{}_2 = \KIND}}\quad (selected rules; full set in \Cref{fig:kinds-full})}

	\begin{multicols}{3}

		\infrule[\ruledef{i-subtree-l}]
		{\cls{\CLS_1 \sqsubseteq \CLS_2}}
		{\cls{(\CLS_1 - \overline{\CLSS_1}) \land (\CLS_2 - \overline{\CLSS_2}) = \CLS_1 - (\overline{\CLSS_1}, \overline{\CLSS_2})}}

		\infrule[\ruledef{i-disjoint}]
		{\cls{\disj{\CLS_1}{\CLS_2}}}
		{\cls{(\CLS_1 - \overline{\CLSS_1}) \land (\CLS_2 - \overline{\CLSS_2}) = \emptyset}}

		\infrule[\ruledef{i-union-r}]
		{\cls{\KIND{} \land \KIND_1 = R_1}\andalso\cls{\KIND{}\land \KIND_2 = R_2}}
		{\cls{\KIND{} \land (\KIND_1 \lor \KIND_2) = R_1\lor R_2}}
	\end{multicols}

	\flushleft{\textbf{Subtraction} \quad \jbox{\cls{\ksub{\KIND_1}{\KIND_2} = \KIND}}\quad (selected rules; full set in \Cref{fig:kinds-full})}

	\begin{multicols}{2}

		\infax[\ruledef{st-tree}]
		{\cls{\ksub{(\CLS_1 - \overline{\CLSS_1})}{(\CLS_2 - \epsilon)} = \CLS_1 - (\CLS_2, \overline{\CLSS_1})}}

		\infrule[\ruledef{st-sub-r}]
		{\cls{\CLSS\sqsubseteq \CLS_2}\andalso \cls{\CLSS\sqsubseteq \CLS_1}\andalso \cls{\ksub{(\CLS_1 - \overline{\CLSS_1})}{(\CLS_2 - \overline{\CLSS_2})} = \KIND{}}}
		{\cls{\ksub{(\CLS_1 - \overline{\CLSS_1})}{(\CLS_2 - (\CLSS, \overline{\CLSS_2}))} = (\CLSS - \overline{\CLSS_1}) \lor \KIND{}}}

		\infrule[\ruledef{st-union-l}]
		{\cls{\ksub{\KIND_1}{(\CLS - \overline{\CLS_i})} = R_1}\andalso \cls{\ksub{\KIND_2}{(\CLS - \overline{\CLS_i})} = R_2}}
		{\cls{\ksub{(\KIND_1 \lor \KIND_2)}{(\CLS - \overline{\CLS_i})} = R_1 \lor R_2}}

		\infrule[\ruledef{st-union-r}]
		{\cls{\ksub\KIND{}\KIND_1 = R_1}\andalso \cls{\ksub{R_1}{\KIND_2} = R_2}}
		{\cls{\ksub\KIND{}{(\KIND_1 \lor \KIND_2)} = R_2}}

	\end{multicols}

	\medskip

	\flushleft{\textbf{Emptiness} \quad \jbox{\cls{\KIND\EMPTY}}}

	\begin{multicols}{3}

		\infax[\ruledef{e-empty}]
		{\cls{\emptyset\EMPTY}}

		\infrule[\ruledef{e-absurd}]
		{\cls{k \sqsubseteq \CLSS_i}}
		{\cls{k - (\CLSS_1, \dots, \CLSS_i, \dots, \CLSS_n)\EMPTY}}

		\infrule[\ruledef{e-union}]
		{\cls{\KIND_1\EMPTY}\andalso \cls{\KIND_2\EMPTY}}
		{\cls{(\KIND_1\lor\KIND_2)\EMPTY}}

	\end{multicols}

	\medskip

	\noindent
	\begin{minipage}[t]{0.48\textwidth}
		\flushleft{\textbf{Disjoint} \quad \jbox{\cls{\disj{\KIND_1}{\KIND_2}}}}
		\quad Defined as \(\cls{(\KIND_1\land \KIND_2)\EMPTY}\).
	\end{minipage}\hfill
	\begin{minipage}[t]{0.48\textwidth}
		\flushleft{\textbf{Subkinding} \quad \jbox{\cls{\subk{\KIND_1}{\KIND_2}}}}
		\quad Defined as \(\cls{(\ksub{\KIND_1}{\KIND_2})\EMPTY}\).
	\end{minipage}

	\medskip

	\flushleft{\textbf{Inclusion} \quad \jbox{\cls{\CLS \in \KIND}}}

	\begin{multicols}{3}
		\infrule[\ruledef{in-base}]
		{\cls{\CLS \sqsubseteq \CLS_0}}
		{\cls{\CLS \in (\CLS_0 - \epsilon)}}

		\infrule[\ruledef{in-nonexcl}]
		{\cls{\CLS \not\sqsubseteq \CLS_1}\andalso \cls{\CLS \in (\CLS_0 - (\CLS_2, \ldots, \CLS_n))}}
		{\cls{\CLS \in (\CLS_0 - (\CLS_1, \ldots, \CLS_n))}}

		\infrule[\ruledef{in-union}]
		{\cls{\CLS \in \KIND_i}}
		{\cls{\CLS \in (\KIND_1\lor \cdots \lor \KIND_i \lor \cdots \lor \KIND_n)}}
	\end{multicols}

	\flushleft{\textbf{Capture Set Projection}\quad $\bs{C}\PROJ[\cls{\KIND}] \coloneq \bs{\left\{\overline{\sproj{\theta_i}{\cls{\KIND_i\land \KIND}}}\right\}}$ where $\bs{C} = \bs{\left\{\overline{\sproj{\theta_i}{\cls{\KIND_i}}}\right\}}$}

	\flushleft{\textbf{Exclusive Union} \quad
		\(\bs{C \uplus \set{x_1, \ldots, x_n}} = \bs{C \cup \set{\sproj{x_1}{\top}, \ldots, \sproj{x_n}{\top}}}\)
		where \(\forall i, \phi. \bs{\sproj{x_i}\phi} \not\in \bs{C}\)}

	\vspace{-7pt}
	\caption{Kind algebra of System \calculus{} (condensed).
		The complete kind algebra is given in \Cref{fig:kinds-full}.}\label{fig:kinds4}

\end{figure*}
 
\begin{figure*}[htbp]
	\footnotesize

	\flushleft{\textbf{Semantic Domains}}
	\par\noindent
	\begin{minipage}[t]{0.21\textwidth}
		\begin{flalign*}
			s, t, u\coloneqq\  & \ldots \tag*{\textbf{Term}}  \\
			                   & \ls\ell \tag*{label}         \\
			v\coloneqq\        & \ldots \tag*{\textbf{Value}} \\
			                   & \ls\ell \tag*{label}
		\end{flalign*}
	\end{minipage}\hfill
	\begin{minipage}[t]{0.22\textwidth}
		\begin{flalign*}
			\theta\coloneqq\  & \ldots \tag*{\textbf{Capture}}               \\
			                  & \ls\ell \tag*{label}                         \\
			a \coloneqq\      & \langle\sta, w\rangle \tag*{\textbf{Answer}}
		\end{flalign*}
	\end{minipage}\hfill
	\begin{minipage}[t]{0.27\textwidth}
		\begin{flalign*}
			\sta\coloneqq\  & \tag*{\textbf{Label Context}}                    \\
			                & \emptyset                    \tag*{empty}        \\
			                & \sta,\ \ls\ell : S :: \cls{\CLS}  \tag*{binding}
		\end{flalign*}
	\end{minipage}\hfill
	\begin{minipage}[t]{0.25\textwidth}
		\begin{flalign*}
			w \coloneqq\      & \tag*{\textbf{Return Value}}                                \\
			                  & v              \tag*{return}                                \\
			                  & \BRK(\ls\ell, v)     \tag*{break}                           \\
			\ls F \coloneqq\  & \ls{\set{\ell_1, \ldots, \ell_n}} \tag*{\textbf{Label Set}}
		\end{flalign*}
	\end{minipage}

	\medskip
	\noindent
	\begin{minipage}[t]{0.47\textwidth}
		\flushleft{\tbs{\textbf{Potential Use Set}}\quad \jbox{\bs{\erase{v}}}}
		\begin{flalign*}
			\bs{\erase{\lambda^D(x: T)t}}   & = \bs{D} & \bs{\erase{\PACK\langle C, v\rangle}} & = \bs{\set{}}     \\
			\bs{\erase{\lambda^D[X <: S]t}} & = \bs{D} & \bs{\erase{\ell}}                     & = \bs{\set{\ell}} \\
			\bs{\erase{\lambda^D[c : B]t}}  & = \bs{D}
		\end{flalign*}
	\end{minipage}\hfill
	\begin{minipage}[t]{0.50\textwidth}
		\flushleft{\tls{\textbf{Runtime Labels}}\quad \jbox{\ls{C^*_\sta}} \quad \jbox{\ls{v^*_\sta}}}
		\begin{flalign*}
			\ls{\emptyset^*_\sta}                       & = \ls{\emptyset} \qquad\qquad
			\ls{(C_1 \cup C_2)^*_\sta}                    = \ls{(C_1)^*_\sta \cup (C_2)^*_\sta}                                                                                     \\
			\ls{\{\sproj{\theta}{\cls{\KIND}}\}^*_\sta} & = \begin{cases}
				                                                \ls{\set{\ell}} & \theta = \ls\ell,\ (\ls\ell : S :: \cls\CLS) \in \sta\text{ and }\cls{\CLS} \in \cls\KIND \\
				                                                \ls{\emptyset}  & \text{ otherwise}
			                                                \end{cases}
			                                            &                                                                                                                           \\
			\ls{v^*_\sta}                               & = \ls{\erase{v}^*_\sta}
		\end{flalign*}
	\end{minipage}

	\flushleft{\textbf{Big-step Evaluation (Selected)} \quad \jbox{\sta \vdash t \Downarrow_{\ls{F}} a}}

	\medskip
	\noindent
	\begin{minipage}[t]{0.44\textwidth}
		\infrule[\ruledef{e-val}]
		{}
		{\sta \vdash v \Downarrow_{\ls{F}} \langle \sta, v \rangle}
		\medskip
		\bigskip
		\infrule[\ruledef{e-app}]
		{\sta \vdash t_1 \Downarrow_{\ls{F}} \langle \sta', \lambda^{\bs D}(x: T)t \rangle \andalso \sta' \vdash t_2 \Downarrow_{\ls{F}} \langle \sta'', v \rangle\\[3pt]
			\ls{D^*_{\sta''}} \subseteq \ls F	\andalso \ls{v^*_{\sta''}} \subseteq \ls F \\[3pt]
			\sta'' \vdash \bs{[\erase{v}/x]}[v/x]t \Downarrow_{\ls{D^*_{\sta''} \cup\ v^*_{\sta''}}} a
		}
		{\sta \vdash t_1 t_2 \Downarrow_{\ls{F}} a}
		\medskip
		\bigskip
		\infrule[\ruledef{e-tapp}]
		{\sta \vdash t \Downarrow_{\ls{F}} \langle \sta', \lambda^{\bs D}[X <: S_0]t \rangle \andalso \ls{D^*_{\sta'}} \subseteq \ls F\\[3pt]
			\sta' \vdash [S/X]t \Downarrow_{\ls{D^*_{\sta'}}} a
		}
		{\sta \vdash t[S] \Downarrow_{\ls{F}} a}
		\medskip
		\bigskip
		\infrule[\ruledef{e-capp}]
		{\sta \vdash t \Downarrow_{\ls{F}} \langle \sta', \lambda^{\bs D}[c : B]t \rangle \andalso \ls{D^*_{\sta'}} \subseteq \ls F\\[3pt]
			\sta' \vdash \bs{[C/c]}t \Downarrow_{\ls{D^*_{\sta'}}} a
		}
		{\sta \vdash t\bs{[C]} \Downarrow_{\ls{F}} a}
	\end{minipage}%
	\hfill
	\begin{minipage}[t]{0.56\textwidth}
		\infrule[\ruledef{e-let-v}]
		{\sta \vdash t_1 \Downarrow_{\ls F} \langle \sta', v\rangle \\[3pt]
			\sta' \vdash \bs{[\erase{v}/x]}[v/x]t_2 \Downarrow_{\ls F} a}
		{\sta \vdash \LET x = t_1 \IN t_2 \Downarrow_{\ls F} a}
		\bigskip
		\medskip
		\infrule[\ruledef{e-ex-v}]
		{\sta \vdash t_1 \Downarrow_{\ls F} \langle \sta', \PACK\langle \bs D, v\rangle\rangle \\[3pt]
			\sta' \vdash \bs{[D/c]}\bs{[\erase{v}/x]}[v/x]t_2 \Downarrow_{\ls F} a}
		{\sta \vdash \LET \langle c, x \rangle = t_1 \IN t_2 \Downarrow_{\ls F} a}
		\bigskip
		\medskip
		\infrule[\ruledef{e-break}]
		{\sta \vdash t_1 \Downarrow_{\ls{F}} \langle \sta', \ls\ell \rangle \andalso \sta' \vdash t_2 \Downarrow_{\ls{F}} \langle \sta'', v \rangle\\[3pt]
			\ls{\ell} \in \ls F}
		{\sta \vdash t_1 t_2 \Downarrow_{\ls F} \langle \sta'', \BRK(\ls\ell, v) \rangle}
		\bigskip
		\medskip
		\infrule[\ruledef{e-app-b1}]
		{\sta \vdash t_1 \Downarrow_{\ls{F}} \langle \sta', \BRK(\ls\ell, v) \rangle}
		{\sta \vdash t_1 t_2 \Downarrow_{\ls F} \langle \sta', \BRK(\ls\ell, v) \rangle}
		\bigskip
		\medskip
		\infrule[\ruledef{e-app-b2}]
		{\sta \vdash t_1 \Downarrow_{\ls{F}} \langle \sta', v_1 \rangle \andalso \sta' \vdash t_2 \Downarrow_{\ls{F}} \langle \sta'', \BRK(\ls\ell, v) \rangle}
		{\sta \vdash t_1 t_2 \Downarrow_{\ls F} \langle \sta'', \BRK(\ls\ell, v) \rangle}

	\end{minipage}

	\vspace{-7pt}
	\caption{Checked big-step semantics of System \calculus{}: selected core rules. Remaining rules handle breaks in other constructs and can be found in Figure~\ref{fig:bigstep-full}.}\label{fig:bigstep}

\end{figure*}
 
The syntax and type system are presented in Figure~\ref{fig:syntax-paper} and Figure~\ref{fig:all-typing-paper}.
Changes compared to System \capless{} are marked \new{\text{accordingly}}.
System \calculus{} is parameterized by a universe of classifiers $\mathcal{K}$,
organized into a tree hierarchy.
Each base capability and scoped capability can be tagged with one classifier.
Captures are enhanced with \textit{projections}, which are filters on classifiers
based on a structure called a \textit{classifier kind}.

\subsection{Classifiers and Kinds}\label{sub:classifiers-kinds}

We embed the classifier tree $\mathcal{K}$ into an infinite tree with root $\cls{\top}$.
$\cls{k\sqsubseteq k'}$ denotes that $\cls{k'}$ is an ancestor of $\cls{k}$ by standard tree relations.
Naturally, $\cls{k \sqsubseteq \top}$ for all classifiers $\cls{k}$.
Furthermore, for any distinct classifiers $\cls{k_1}, \cls{k_2}$, we know that at most one of $\cls{k_1 \sqsubseteq k_2}$
and $\cls{k_2 \sqsubseteq k_1}$ holds. When neither is true, we say that $\cls{k_1}$ and $\cls{k_2}$ are \textit{disjoint}
(i.e., the subtrees rooted at $\cls{k_1}$ and $\cls{k_2}$ do not share any nodes).

Each node has a countably infinite number of children.
This is crucial for modularity, as a classifier cannot be ``singled out'' by excluding all its
known sub-classifiers: additional sub-classifiers can be added in another module, as in the
following example:

\noindent
\begin{minipage}[t]{0.55\textwidth}
	\vspace{0pt}%
	\begin{scalacodesmall}
// Module 1
trait A extends Classifier, SharedCapability
trait B extends Classifier, A // C likewise
def f(body: () ->{any.only[A].except[B].except[C]} Unit)
\end{scalacodesmall}
\end{minipage}\hfill
\begin{minipage}[t]{0.42\textwidth}
	\vspace{0pt}%
	\begin{scalacodesmall}
// Module 2, compiled separately
trait D extends Classifier, A
\end{scalacodesmall}
\end{minipage}
Function \lstinline{f} cannot assume that no sub-classifier of \lstinline{A} exists, since it can be
arbitrarily extended by another module. Therefore, classifiers naturally avoid the closed-world problem presented
by Lutze et al.~\cite{lutze23flix}. Flix overcomes this problem by treating the effect set top as a symbolic constant
during boolean unification, preventing negations from being translated into a finite positive set.
In System \calculus{}, each subtree is assumed to have infinitely many children, so this cannot happen.

Kinds $\cls{\KIND}$ represent filters for classifiers.
To model \lstinline|.only| and \lstinline|.except|, we work with subtrees of classifiers when constructing kinds.
They are represented as unions of ``holed classifier subtrees'':
a root subtree $\cls{k_0}$ and a list of excluded subtrees $\cls{k_1}, \ldots, \cls{k_n}$.
Each such subtree represents the set of all classifiers $\cls{k}$ that are sub-classifiers of $\cls{k_0}$
but are \textit{not} sub-classifiers of any remaining $\cls{k_i}$
(i.e., \(\cls{k \sqsubseteq k_0}\) and \(\forall i > 0, \cls{k \not\sqsubseteq k_i}\)).
The inclusion relation ($\cls{\CLS \in \KIND}$) determines whether a
classifier is included in a classifier kind. It is decidable, and is also used in the operational semantics.

We present in Figure~\ref{fig:kinds4} algorithmic rules for intersection ($\cls{\KIND_1 \land \KIND_2}$) and subtraction ($\cls{\ksub{\KIND_1}{\KIND_2}}$) of kinds.
The unambiguity of the rules depends on the previously mentioned sub-classifier-or-disjoint
property of the classifier tree.

The intersection of two subtrees can simply be computed as a new subtree whose exclusion list contains
both exclusion lists of the two inputs, and the new root is the deepest common classifier of
the two input roots. This is the smaller root if they have a sub-classifier relationship
(\ruleref{i-subtree-l} and \ruleref{i-subtree-r}), or none if the roots are disjoint \ruleref{i-disjoint}.
Compound cases are handled by computing the union of all intersections of pairs of subtrees.

Similarly, we handle subtraction of two subtrees individually, and
compound cases are handled by the distribution on the left-hand
side \(\cls{\ksub{(\KIND_1 \lor \KIND_2)}{\KIND} = (\ksub{\KIND_1}{\KIND}) \lor (\ksub{\KIND_2}{\KIND})}\)
\ruleref{st-union-l} and composition on the right-hand side
\(\cls{\ksub{\KIND}{(\KIND_1 \lor \KIND_2)} = \ksub{(\ksub{\KIND}{\KIND_1})}{\KIND_2}}\)
\ruleref{st-union-r}.
For subtrees, we compute recursively on the exclusion list of the
right-hand side. In the empty case \ruleref{st-tree}, we can simply append
the root to the left-hand side's exclusion list.
Otherwise, we follow the general equality of set subtractions
\(A\setminus (B\setminus C) = (A\setminus B) \cup (A\cap B\cap C)\).
The rules \ruleref{st-irrel-r}, \ruleref{st-absurd-r},
\ruleref{st-irrel-l}, \ruleref{st-sub-l}, and \ruleref{st-sub-r}
handle each case based on the sub-classifier/disjointness relationship
between the excluded node and the two roots.

Subkinding ($\cls{\subk{\KIND_1}{\KIND_2}}$) can be defined as empty subtraction ($\cls{(\ksub{\KIND_1}{\KIND_2}) \EMPTY}$),
and disjointness of kinds can be defined as empty intersection.
This is grounded in the fact that the subtraction operator satisfies the axioms of subtraction
algebras~\cite{schein1992difference,jun2004ideal}, which define an ordering relation.

There can be multiple representations of the empty set,
so $\cls{\KIND\EMPTY}$ is also defined algorithmically.
The interesting case is when a subtree contains an ancestor node in its exclusion list \ruleref{e-absurd}.

We use a single classifier name to represent a subtree with the given classifier as root
and no excluded roots, as well as a classifier kind containing that sole subtree.
Following this notation, $\cls{\top}$ is the classifier kind containing all classifiers.
The kind representing all classifiers except a subtree $\cls{k}$ can be written as $\cls{\top - k}$.

The definitions of intersection and subtraction are equivalent to the set semantics.
More details about the proofs can be found in Section~\ref{sub:set-semantics}.

\subsection{Capability Projections}

At the core of System \calculus{} lie \emph{projected captures} $\sproj{\theta}{\cls{\KIND}}$.
A projected capture consists of a capture \(\theta\), applied to a classifier kind filter
\(\cls{\KIND}\). It represents \textit{sub-parts} of \(\theta\) whose classifier \(\cls{k}\) matches the kind \(\cls{\KIND}\)
(i.e., \(\cls{k \in \KIND}\)).
For instance, we can correctly type the signature of the \lstinline|Try[T]| constructor (previously mentioned in Section~\ref{sub:case-study-try})
in System~\calculus{} as
\[\text{Try.apply}: \forall[T <: \top]\ \forall[\bs{c}: \cls{\top}]\ \forall(\text{body}: (\forall(u: \text{unit})\ T)\capt\bs{\set{c}})\ \text{Try}[T]\capt\bs{\set{\sproj{c}{\cls{\text{Control}}}}} \]
The return type captures a projection of the \lstinline|body|'s capture set \(\bs c\), filtered to only the classifier subtree rooted at \cls{\lstinline|Control|}.
The unbounded \(*\) is replaced with the \(\cls\top\) kind bound, allowing all capture sets.
Note that the capture set \(\bs{\set{\sproj{c}{\cls{\text{Control}}}}}\) is attached to the return type and
not the function: the constructor is \textit{pure} and does not capture any capabilities by itself.

For brevity we omit projections to $\cls{\top}$.
The notation $\bs{C}\PROJ[\cls{\KIND}]$ applies a further projection to each element in $\bs{C}$ by intersecting $\cls{\KIND}$
with each singleton's existing projection,
i.e.,
\[\bs{\set{\sproj{\theta_1}{\cls{\KIND_1}}, \ldots, \sproj{\theta_n}{\cls{\KIND_n}}}}\PROJ[\cls{{\color{blue}\KIND}}] = \bs{\set{\sproj{\theta_1}{\cls{\KIND_1\land{\color{blue}\KIND}}},\ldots,\sproj{\theta_n}{\cls{\KIND_n\land{\color{blue}\KIND}}}}}\]

Capture subsetting \(\bs{C_1} \sqsubseteq \bs{C_2}\) follows normal subsetting rules \(\bs{C_1} \subseteq \bs{C_2}\),
similar to System \capless{}~\cite{xu2025capless},
but is also extended to handle projections.
A projected capture is smaller than the same capture but with a super-kind
(i.e., a coarser filter), by \ruleref{s-subkind}.
Projection to an empty kind is equivalent to removing that capture from the set \ruleref{s-absurd}.
Finally, \ruleref{s-merge} allows us to treat one capture projected to a union of two kinds
as equivalent to two separate projected captures.

In System \capless{}, capture set variables can be either unbounded (as with the \lstinline|Try[T]| example above)
or bounded by other capture sets.
System \calculus{} replaces unbounded capture set variables with \emph{kind-bounded} variables: they can only be
instantiated with capture sets satisfying the given kind.
For example, the constructor for \lstinline|Future[T]| (Section~\ref{sub:future-threadlocal}),
which forbids capabilities under \lstinline|ThreadLocal|, can be typed as:
\[\text{Future.apply}: \forall[T <: \top]\ \forall[\bs{c} : \cls{\top - \text{ThreadLocal}}]\ \forall(\text{body}: (\forall(u: \text{unit})\ T)\capt\bs{\set{c}})\ \text{Future}[T]\capt\bs{\set{c}}\]
where the kind \(\cls{\top - \text{ThreadLocal}}\) represents the kind of all classifiers,
except the subtree rooted at \lstinline|ThreadLocal|.
The type system governs which capture sets can be instantiated under a kind bound through the \textit{capture kinding} process
described in the next section.

\subsection{Capture Kinding and Subcapturing}

Subcapturing rules \(\bs{C_1} <: \bs{C_2}\) largely mirror those of System \capless{}.
They follow the subsetting relation \(\bs{C_1} \sqsubseteq \bs{C_2}\), but with additional rules for
term \ruleref{sc-var} and capture variables \ruleref{sc-bound}, which refine captures based on their bounds.
Projections on the variables are applied to the corresponding bound capture sets.
The only new rule is \ruleref{sc-proj}, which allows us to equate any capture set \(\bs{C}\) to its
projected version \(\bs{C}\PROJ[\cls{\KIND}]\), if we can deduce that \(\bs{C}\) contains only captures under
kind \(\cls{\KIND}\), through a judgment called \textit{capture kinding}, which we shall describe below.

The capture kinding judgment \(\typs{\G}{\bs{C}}{\cls{\KIND}}\) is derived by looking up each projected capture
in the context.
For capture-set-bounded term and capture variables, we recursively check the definitions
(in \ruleref{k-var} and \ruleref{k-cvar}). Similar to subcapturing, projections of the capture
are applied to the upper bound.
For kind-bounded capture set variables in \ruleref{k-cbound},
the intersection of the projection and the kind bound is the most precise kind we can derive.
Subkinding is handled by \ruleref{k-sub}, with a special case for empty kinds in \ruleref{k-absurd}
to allow kinding judgments even on invalid references with empty projections.

Note that the simple addition of \ruleref{sc-proj} allows us to derive useful, non-trivial facts.
That projection preserves subcapturing
(\(\subs{\G}{\bs{C}}{\bs{D}}\) implies \(\subs{\G}{\bs{C}\PROJ[\cls{\KIND}]}{\bs{D}\PROJ[\cls{\KIND}]}\))
can be proven by induction on the original derivation and using composition of projections
(\(\bs{C}\PROJ[\cls{\KIND_1}]\PROJ[\cls{\KIND_2}] = \bs{C}\PROJ[\cls{\KIND_1 \land \KIND_2}]\)).
Empty-kinded sets can be proven to be empty
(\(\typs{\G}{\bs{C}}{\cls{\KIND}}\) where \(\cls{\KIND\EMPTY}\) implies \(\subs{\G}{\bs{C}}{\bs{\set{}}}\))
by first projecting \(\bs{C}\) to the empty kind \(\cls{\KIND}\) using \ruleref{sc-proj},
then using \ruleref{sc-elem} to conclude \(\subs{\G}{\bs{C}\PROJ[\cls{\KIND}]}{\bs{\set{}}}\),
where the underlying subset relation is obtained by repeatedly applying \ruleref{s-absurd} to each
projected capture.

Together, kinding and subcapturing allow refinement and filtering of variables in capture sets.
For example, with two disjoint child classifiers \(\cls{k_1}, \cls{k_2} \sqsubseteq \cls{k}\) and a kind-bounded
capture set variable \(c : \cls{k_1}\) in scope, we can derive that \(\bs{\set{c}}\PROJ[\cls{k_2}] <: \bs{\set{}}\),
i.e., a function capturing values of kind \(\cls{k_2}\) will not capture any capabilities of kind \(\cls{k_1}\).

\subsection{Typing and Subtyping}
Subtyping remains unchanged from System \capless{}~\cite{xu2025capless}: subtyping of capturing types \(S_1\capt \bs{C_1} <: S_2\capt \bs{C_2}\) requires both
subtyping on the shape types and subcapturing on the capture sets \ruleref{capt}, but otherwise
follows kernel System~\(\text{F}_\le\), unsurprisingly.
Capture set bound subtyping either follows subcapturing (when both bounds are capture sets), subkinding
(when both bounds are kinds), or capture kinding when we refine a kind bound to a capture-set bound \ruleref{b-set-kind}.

Similar to System \capless{}, the typing judgment \(\bs C; \G \vdash t : E\) states that \(t\) has the
existential type \(E\) under context \(\G\) with the \emph{use set} \(\bs C\).
The use set represents capabilities that \emph{may} be used during the evaluation of \(t\).
We follow System \capless{} for use set judgments: values can be typed with an empty use set as they are already
evaluated, variables are typed with themselves in the use set \ruleref{var}, and abstractions \ruleref{abs}, \ruleref{tabs}, \ruleref{cabs}
reify the body's use set into the function's \emph{capture set}, leaving an empty use set.
Note that the use set is recorded in the lambda term, which is used in the checked big-step semantics (Section~\ref{sub:big-step}).
Otherwise, typing is standard with regard to System \(\text{F}_\le\), and remains unchanged from System \capless{}.
Note that let typing \ruleref{let} and \ruleref{let-e} require the final use set $\bs C$ and return types to
be well-formed in the absence of the let-bound parameter, which we can achieve by widening its appearances in
capture sets (e.g., by applying \ruleref{sc-var}).

\subsection{Translation from Surface Language}\label{sub:translation}

Scala provides \lstinline|any| as syntactic sugar for the \emph{universal capability}:
a reference standing for ``some unspecified capability in scope''.
It does not appear in the formal calculus.
Instead, \lstinline|any| is \emph{desugared} into explicit quantification over a fresh capture
variable.
In a \emph{contravariant} input position, \lstinline|any| introduces a fresh \emph{capture
	abstraction}: the enclosing function becomes capture-polymorphic.
In a \emph{covariant} (output) position, \lstinline|any| becomes an \emph{existential type}.
Consider the \lstinline|Try.apply| constructor from Section~\ref{sub:case-study-try}:
\begin{scalacode}
def apply[T](body: () ->{any} T): Try[T]^{body.only[Control]}
\end{scalacode}
The contravariant \lstinline|any| on \lstinline|body|'s type introduces a fresh
capture variable $\bs{c} : \cls{\top}$.
It is translated to:
\[\text{Try.apply}: \forall[T <: \top]\ \forall[\bs{c} : \cls{\top}]\ \forall(\text{body}: (\forall(u: \text{unit})\ T)\capt\bs{\set{c}})\ \text{Try}[T]\capt\bs{\set{\sproj{c}{\cls{\text{Control}}}}}\]

The translation between
the surface language and System \capless{}
has been formalized in System Reacap~\cite{xu2025capless}, a surface calculus for capture checking
with \lstinline|any|, and applies directly
to System \calculus{}.
Section~\ref{sec:impl} describes the implementation.

\subsection{Big-Step Semantics}\label{sub:big-step}

We define \emph{checked} big-step semantics in Figures~\ref{fig:bigstep} and~\ref{fig:bigstep-ext}.
The judgment \(\sta \vdash t \Downarrow_{\ls{F}} a\) includes access checks (in \ls{green}) under label allowance \(\ls{F}\).
The big-step evaluation involves a label context $\sta$ that records the classifier and value type
associated with created labels throughout the evaluation. Therefore, evaluation always returns a
possibly-appended label context as part of its answer: either a return value $\langle \sta', v\rangle$
or a propagating break $\langle \sta', \BRK(\ls\ell, v)\rangle$.

Ignoring all access checks,
the judgments (\ruleref{e-val}, \ruleref{e-app}, \ruleref{e-tapp}, \ruleref{e-capp},
\ruleref{e-let-v}, \ruleref{e-ex-v}) describe a standard top-level (i.e., no store bindings),
substitution-based, call-by-value big-step reduction over the \emph{relaxed},
direct-style syntax used in the proofs (Figure~\ref{fig:relaxed-full}),
since substitution takes evaluation out of MNF.
Note that substituting a term variable $x$ involves both a capture-set substitution $\bs{[C/x]}$
and a regular term substitution $[v/x]$, as $x$ can appear in both contexts.

The access checks make the scoping discipline of labels explicit.
The allowance \(\ls F\) directly governs breaks: invoking a label \ruleref{e-break} is stuck
if the label is not included in the allowance.
At the outer typing-to-runtime boundary, the allowance is the runtime label set computed from the
static typing capture set: it records only which labels may be accessed during evaluation,
by comparing each label's projection kind to the current label context.
Within the rules, allowances are otherwise ordinary runtime label sets.
Intuitively, the runtime labels $\ls{v_\sta^*}$ (Figure~\ref{fig:bigstep})
of a $\lambda$-abstraction are the set of labels that it can reach when its body is evaluated.
Accordingly, $\lambda$-abstractions are entered under the label allowance dictated by their annotated capture sets.\footnote{We assume that $\lambda$-abstractions carry their annotated capture sets (cf.\ Figure~\ref{fig:all-typing-paper}).}
Application rules (\ruleref{e-app}, \ruleref{e-tapp}, \ruleref{e-capp}) check that
the abstractions' and their arguments' runtime labels are bounded by the current allowance.
This is what allows propagated breaks to stay well-scoped.
All other rules support the \textit{extended} System \calculus{}, which we describe in the next section.

\subsection{Boundary, Break, and Intercept}

To prove safety properties about the system (Section~\ref{sec:metatheory}), we extend System \calculus{} with additional
control-flow constructs (Figure~\ref{fig:extension-paper}): \textit{boundary}, \textit{break} (following \cite{xu2025capless,boruchgruszecki2023capturing}),
and a new construct \textit{intercept}.

\paragraph{Boundary and Break.}
Following \cite{xu2025capless,boruchgruszecki2023capturing},
$\BOUNDARY[S,\cls{\CLS}]\AS\langle c, x\rangle\IN t$ introduces a scope
with a fresh label $\ell$, which the body can use to transfer control (\textit{break})
disruptively out of the scope.
We extend this with a classifier $\cls{\CLS}$: the boundary rules
(\ruleref{e-bnd-v}, \ruleref{e-bnd-c}, \ruleref{e-bnd-p})
create a fresh label
with an arbitrary sub-classifier $\cls{\CLS_0} \sqsubseteq \cls{\CLS}$.
This non-determinism models the runtime class of a caught exception, which may be a concrete
subclass of the declared one.

Breaks can be invoked by applying a label value $l$ (of type $\BREAK[S]$) to the appropriate value $v$
of type $S$ (\ruleref{e-break} and statically \ruleref{t-break}), returning the disruptive answer $\BRK(\ls\ell, v)$.
This is propagated upwards, in a similar way to throwing exceptions (for example, as described by
\ruleref{e-app-b1} and \ruleref{e-app-b2} for applications; other propagation rules can be found in \ref{fig:bigstep-full}), until it is either caught by the corresponding
$\BOUNDARY$ \ruleref{e-bnd-c} or an $\INTERCEPT$ (described below), or reaches the top level.
Boundaries push the fresh label onto the persistent label context, but the label allowance
includes it only during the evaluation of the body: the label may escape
(e.g., inside returned abstractions), but invoking an escaped label is stuck.
Corollary \ref{cor:effect-safety} proves that well-typed closed terms
do not evaluate to $\BRK$.

\paragraph{Intercept.}
The construct $\INTERCEPT[S,\bs{C_t},\cls{\KIND}]\WITH h \IN t$ intercepts any
break that escapes $t$ whose label has classifier $\cls{\CLS} \in \cls{\KIND}$,
dispatching the caught label and its payload to the handler $h$.
Evaluation is deterministic: a matching break is handled, a non-matching one propagates,
and a normal return passes through unchanged (\ruleref{e-icp-m}, \ruleref{e-icp-p}, \ruleref{e-icp-v}).
The inner term runs under the allowance of its annotated use set and may thus invoke more labels
than the context allows, as long as they are caught: handlers are evaluated with the original
allowance \ruleref{e-icp-m}, while propagated labels are re-checked to be in the same allowance
\ruleref{e-icp-p}.
Interestingly, we can refine the use set of the \(\INTERCEPT\) expression:
after handling labels of kind $\cls{\KIND}$, only captures outside \(\cls\KIND\) may escape
from the body, together with the handler's own captures.
Specifically, for handlers that \emph{do not} re-throw (i.e., those that do not use the captured label),
we can drop matching labels from the use set of \(t\), so that only \(\bs{C_t}\PROJ[\cls{\ksub{\top}{\KIND}}]\)
is included \ruleref{t-icp-pass}.

The \(\INTERCEPT\) construct acts as a modular extension to the extended System \capless{} \cite{xu2025capless}.
It is a straightforward model for \emph{intercepting} exception handlers: handlers that intercept an existing
lexical exception handler,
as in the case of Scala's interactions between \lstinline|Try[T]| \cite{scalaTry} and \lstinline|boundary| \cite{scalaBoundary}.
The separation is meaningful: it precisely defines the requirements that intercepting handlers must uphold to
maintain \emph{scope-safety}, without modifying the existing assumptions of lexical handlers.

\paragraph{Encoding \lstinline|Try[T]|.}
Scala's \lstinline|Try[T]| can be encoded into the extended System \calculus{}.
Using Church encoding, we can represent the two cases of the container (a value of type \lstinline|T|,
or a $\BRK$ of unknown type) as follows, where the extra capture-set parameter $\bs C$ represents
the capture set of the $\mathsf{Try}$ value itself, to appear within parameter positions of the handlers:
{\small
\begin{align*}
	\mathsf{Try}(T, \bs{C}) \;=\;
	 & \forall[X{<:}\top].\;\forall[c{:}\cls{\top}].             \\
	 & \quad\Bigl(
	\forall\!\bigl(s{:}(\forall(x{:}T)\,X)\capt\bs{\{c\}}\bigr). \\
	 & \qquad
	\forall\!\bigl(f{:}\bigl(\forall[Y{<:}\top].\;
	\forall(b{:}(\BREAK[Y])\capt\bs{C})\,(\forall(r{:}Y)\,X)\capt\bs{\{c\}}
	\bigr)\capt\bs{\{c\}}\bigr).\;X
	\Bigr)\capt\bs{\{c : \top\}\cup\bs{C}}
\end{align*}}%
The constructor can then be encoded as taking a thunk $b$, running it under an $\INTERCEPT$ where
$\cls\KIND = \cls{\text{Control}}$, with the handler and final results wrapped accordingly into different
cases of the $\mathsf{Try}(T, \bs{\set{\sproj{b}{\cls{\text{Control}}}}})$.
The full development in Lean 4 is further explained in Appendix \ref{app:lean-try}.

\section{Metatheory}\label{sec:metatheory}

We establish soundness of System \calculus{} in three steps. First, the classifier kind algebra
admits a set-theoretic semantics that justifies extensional reasoning about projections and
handler dispatch. Second, unlike prior work on capture
calculi~\cite{boruchgruszecki2023capturing,xu2025capless}, which establishes soundness by small-step
preservation and progress, we use a checked big-step semantics, a first in that line of work.
This is useful for two reasons: (1) it makes the scoping discipline of labels explicit,
and (2) it lets us reason directly about what values reach and how this relates to
capture sets and classifiers. Third, we employ a standard safety proof for big-step semantics,
with relaxed typing rules to handle direct-style, non-MNF terms,
and an extension to answer typing that links returned breaks to label scoping, which allows us
to show the relation between static typing capture/use sets and the usage of labels at runtime.
The closest precedent is the second-class values
literature~\cite{osvald16secondclass,xhebraj22secondclass,thiemann2025secondclass},
which also reasons in big-step style about what function values may capture,
and logical-relations work on reachability types~\cite{bao25reachability},
which connects annotations to the values reachable at runtime.
Neither qualifies types by classifier-indexed capture sets governed by subcapturing,
nor tracks delimiter-scoped control.
We sketch the most important aspects here and refer to the Lean mechanization (Appendix \ref{app:mechanization}) for the full development.

\subsection{Set Semantics of Classifier Kinds}\label{sub:set-semantics}

The kind algebra (Figure~\ref{fig:kinds4}) admits a set-theoretic reading.
Define $\sem{\cls{\KIND}} = \set{\cls{\CLS} \mid \cls{\CLS \in \KIND}}$,
where membership is the decidable inclusion judgment of Figure~\ref{fig:kinds4}.
Then intersection, subtraction, emptiness, sub-kinding, and disjointness
correspond exactly to set intersection, difference, emptiness, inclusion, and disjointness.
This lets us carry out later arguments about classifier projection, exclusion,
and handler dispatch by ordinary set reasoning on kind denotations.

\begin{figure*}[tp]
	\footnotesize
	\setlength{\afterruleskip}{9pt plus 2pt minus 1pt}

	\noindent
	\begin{minipage}[t]{0.52\textwidth}
		\flushleft{\textbf{Syntax}}
		\begin{flalign*}
			s,\,t,\,u,\,h\coloneqq\  & \ldots \tag*{\textbf{Term}}                                                 \\
			                         & {\BOUNDARY[S, \cls{\CLS}] \AS \langle c, x\rangle \IN t}    \tag*{boundary} \\
			                         & {\INTERCEPT[R, \bs{C_t}, \cls{\KIND}] \WITH h \IN t}    \tag*{intercept}    \\
			R, S \coloneqq\          & \ldots \tag*{\textbf{Shape Type}}                                           \\
			                         & \BREAK[S] \tag*{break}
		\end{flalign*}
	\end{minipage}%
	\hfill
	\begin{minipage}[t]{0.44\textwidth}
		\flushleft{\textbf{Subtyping} \quad \jbox{\G \vdash E_1 <: E_2}}
		\infrule[\ruledef{break}]
		{\G \vdash S_2 <: S_1}
		{\G \vdash \BREAK[S_1] <: \BREAK[S_2]}
	\end{minipage}

	\flushleft{\textbf{Exception Handler Type Abbreviations}}
	\begin{align*}
		\HANDLER_\mathsf{pass}[E, \cls{\KIND}, \bs{C_h}, \bs{C_t}] \; & :=\; \forall [X {<:} \top]\, \forall [c : \bs{C_t\PROJ[\cls{\KIND}]}]\, \forall (b : \BREAK[X]\capt \bs{\set{c}})\, (\forall (y : X)\, E) \capt \bs{C_h}              \\
		\HANDLER_\mathsf{gen}[E, \cls{\KIND}, \bs{C_h}, \bs{C_t}] \;  & :=\; \forall [X {<:} \top]\, \forall [c : \bs{C_t\PROJ[\cls{\KIND}]}]\, \forall (b : \BREAK[X]\capt \bs{\set{c}})\, (\forall (y : X)\, E) \capt \bs{C_h \cup \set{c}}
	\end{align*}

	\flushleft{\textbf{Typing} \quad \jbox{\bs{C}; \G \vdash t : E}}

	\begin{multicols}{2}
		\infrule[\ruledef{t-break}]
		{\bs{C}; \G \vdash x : \BREAK[S]\capt \bs{C_x}\\[3pt] \bs{C}; \G \vdash y : S \andalso \G \vdash R\ \textbf{\textsc{wf}}}
		{\bs{C}; \G \vdash x\,y : R}

		\infrule[\ruledef{t-bnd}]
		{\bs{C \uplus \set{c, x}}; (\G, c : \cls{\CLS}, x : \BREAK[S]\capt \bs{\set{c}}) \vdash t : S \\[3pt] \G \vdash S\ \textbf{\textsc{wf}}}
		{\bs{C}; \G \vdash \BOUNDARY[S, \cls{\CLS}] \AS \langle c, x\rangle \IN t : S}

		\infrule[\ruledef{t-icp-pass}]
		{\bs{C_t}; \G \vdash t : E\\[3pt] \bs{C_h}; \G \vdash h : \HANDLER_\mathsf{pass}[E, \cls{\KIND}, \bs{C_h}, \bs{C_t}]}
		{\bs{C_t\PROJ[\cls{\ksub{\top}{\KIND}}] \cup C_h}; \G \vdash \INTERCEPT[E, \bs{C_t}, \cls{\KIND}]\WITH h \IN t : E}

		\infrule[\ruledef{t-icp-gen}]
		{\bs{C_t}; \G \vdash t : E\\[3pt] \bs{C_h}; \G \vdash h : \HANDLER_\mathsf{gen}[E, \cls{\KIND}, \bs{C_h}, \bs{C_t}]}
		{\bs{C_t \cup C_h}; \G \vdash \INTERCEPT[E, \bs{C_t}, \cls{\KIND}]\WITH h \IN t : E}
	\end{multicols}
	\vspace{-14pt}
	\caption{Extended System \calculus{} with boundaries and interceptions.}\label{fig:extension-paper}

\end{figure*}
\begin{figure*}[tp]
	\footnotesize

	\flushleft{\textbf{Big-step Evaluation (continued)} \quad \jbox{\sta \vdash t \Downarrow_{\ls{F}} a}}

	\medskip
	\noindent
	\begin{minipage}[t]{0.50\textwidth}

		\infrule[\ruledef{e-bnd-v}]
		{ \ls\ell \text{ fresh} \andalso \cls{\CLS_0} \sqsubseteq \cls\CLS \andalso t' = \bs{[\set{\ell}/c]}\bs{[\set{\ell}/x]}[\ls\ell/x]t \\[3pt]
			\sta, \ls\ell : S :: \cls{\CLS_0} \vdash t' \Downarrow_{\ls{F \cup \set{\ell}}} \langle \sta', v \rangle}
		{\sta \vdash \BOUNDARY[S, \cls{\CLS}] \AS \langle c, x \rangle \IN t \Downarrow_{\ls F} \langle \sta', v \rangle}

		\bigskip
		\infrule[\ruledef{e-bnd-c}]
		{ \ls\ell \text{ fresh} \andalso \cls{\CLS_0} \sqsubseteq \cls\CLS \andalso t' = \bs{[\set{\ell}/c]}\bs{[\set{\ell}/x]}[\ls\ell/x]t \\[3pt]
			\sta, \ls\ell : S :: \cls{\CLS_0} \vdash t' \Downarrow_{\ls{F \cup \set{\ell}}} \langle \sta', \BRK(\ls\ell, v) \rangle
		}
		{\sta \vdash \BOUNDARY[S, \cls{\CLS}] \AS \langle c, x \rangle \IN t \Downarrow_{\ls F} \langle \sta', v \rangle}

		\bigskip
		\infrule[\ruledef{e-bnd-p}]
		{ \ls\ell \text{ fresh} \andalso \cls{\CLS_0} \sqsubseteq \cls\CLS \andalso t' = \bs{[\set{\ell}/c]}\bs{[\set{\ell}/x]}[\ls\ell/x]t \\[3pt]
			\sta, \ls\ell : S :: \cls{\CLS_0} \vdash t' \Downarrow_{\ls{F \cup \set{\ell}}} \langle \sta', \BRK(\ls{\ell'}, v) \rangle
			\andalso \ls\ell \neq \ls{\ell'}}
		{\sta \vdash \BOUNDARY[S, \cls{\CLS}] \AS \langle c, x \rangle \IN t \Downarrow_{\ls F} \langle \sta', \BRK(\ls{\ell'}, v) \rangle}
	\end{minipage}%
	\hfill
	\begin{minipage}[t]{0.50\textwidth}
		\infrule[\ruledef{e-icp-v}]
		{\sta \vdash t \Downarrow_{\ls{(C_t)^*_\sta}} \langle \sta', v \rangle}
		{\sta \vdash \INTERCEPT[R, \bs{C_t}, \cls{\KIND}] \WITH h \IN t \Downarrow_{\ls F} \langle \sta', v \rangle}

		\bigskip
		\infrule[\ruledef{e-icp-m}]
		{\sta \vdash t \Downarrow_{\ls{(C_t)^*_\sta}} \langle \sta', \BRK(\ell, v) \rangle \andalso (\ell : S :: \cls\CLS) \in \sta' \\[3pt]
			\cls\CLS \in \cls{\KIND} \andalso \sta' \vdash h[S][\bs{\set{\sproj{\ls\ell}{\cls\KIND}}}]\ \ls\ell\ v \Downarrow_{\ls F} a}
		{\sta \vdash \INTERCEPT[R, \bs{C_t}, \cls{\KIND}] \WITH h \IN t \Downarrow_{\ls F} a}

		\bigskip
		\infrule[\ruledef{e-icp-p}]
		{\sta \vdash t \Downarrow_{\ls{(C_t)^*_\sta}} \langle \sta', \BRK(\ell, v) \rangle \andalso  (\ell : S :: \cls\CLS) \in \sta' \\[3pt]
			\cls\CLS \not\in \cls{\KIND} \andalso \ls\ell \in \ls F}
		{\sta \vdash \INTERCEPT[R, \bs{C_t}, \cls{\KIND}] \WITH h \IN t \Downarrow_{\ls F} \langle \sta', \BRK(\ell, v)\rangle}

	\end{minipage}

	\vspace{-9pt}
	\caption{Checked big-step semantics of System \calculus{}: boundary and intercept rules.}\label{fig:bigstep-ext}

\end{figure*}
 
\subsection{Runtime Relaxed Typing}\label{sub:relaxed-typing}

\begin{figure*}[tp]
	\footnotesize

	\flushleft{\textbf{Capture Kinding (for labels)} \quad \jbox{\sta; \G \vdash \bs{C} : \cls{\KIND}}}
	\begin{multicols}{2}
		\infrule[\ruledef{k-label}]
		{(\ls \ell : S :: \cls\CLS) \in \sta \andalso \cls\CLS \in \cls{\KIND_l} \andalso \cls\CLS \in \cls{\KIND}}
		{\sta; \G \vdash \bs{\set{\sproj{\ls \ell}{\cls{\KIND_l}}}} : \cls{\KIND}}

		\infrule[\ruledef{k-label-absurd}]
		{(\ls \ell : S :: \cls\CLS) \in \sta \andalso \cls\CLS \not\in \cls{\KIND_l}}
		{\sta; \G \vdash \bs{\set{\sproj{\ls \ell}{\cls{\KIND_l}}}} : \cls{\emptyset}}
	\end{multicols}

	\flushleft{\textbf{Relaxed Typing (Selected Rules)}\quad \jbox{\bs C; \sta; \G \vdash_\star t : E}}

	\begin{multicols}{2}
		\infrule[\ruledef{rt-label}]
		{(\ls\ell : S :: \cls{\CLS}) \in \sta \andalso \cls{\CLS \in \KIND}}
		{\bs\emptyset; \sta; \G \vdash_\star \ell : \BREAK[S]\capt \bs{\set{\sproj{\ell}{\cls{\KIND}}}}}

		\infrule[\ruledef{rt-app}]
		{\bs{C_1}; \sta; \G \vdash_\star t_1 : (\forall(z: S\capt \bs{C_0}) E)\capt \bs{C_1} \\[3pt]
			\bs{C_2}; \sta; \G \vdash_\star t_2 : S\capt \bs{C_2} \andalso \sta; \G \vdash \bs{C_2} <: \bs{C_0}}
		{\bs{C_1 \cup C_2}; \sta; \G \vdash_\star t_1\ t_2 : \bs{[C_2/z]}E}

		\infrule[\ruledef{rt-pack-var}]
		{\bs{C'}; \sta; \G \vdash_\star x : \bs{[D/c]}T \andalso \sta; \G \vdash \bs{D} <:_{\mathsf{B}} B}
		{\bs{\emptyset}; \sta; \G \vdash_\star \PACK\langle \bs D, x\rangle : \exists c: B. T}
		\infrule[\ruledef{rt-pack-val}]
		{\bs{\emptyset}; \sta; \G \vdash_\star v : \bs{[D/c]}T \andalso \sta; \G \vdash \bs{D} <:_{\mathsf{B}} B}
		{\bs{\emptyset}; \sta; \G \vdash_\star \PACK\langle \bs D, v\rangle : \exists c: B. T}
	\end{multicols}

	\flushleft{\textbf{Answer Typing}\quad \jbox{\ls{F}; \sta; \G \vDash_\mathsf{a}w : E}}

	\begin{multicols}{2}
		\infrule[\ruledef{sa-ret}]
		{\bs{\emptyset}; \sta; \G \vdash_\star v : E}
		{\ls{F}; \sta; \G \vDash_\mathsf{a}v : E}

		\infrule[\ruledef{sa-brk}]
		{(\ls\ell : S :: \cls{\CLS}) \in \sta \andalso
			\ls\ell \in \ls F \andalso
			\bs{\emptyset}; \sta; \G \vdash_\star v : S
		}
		{\ls{F}; \sta; \G \vDash_\mathsf{a} \BRK(\ls\ell, v) : E}
	\end{multicols}
	\vspace{-10pt}
	\caption{Capture Kinding for Labels, Relaxed typing definitions (selected rules) and Answer Typing rules used in the metatheory statements. The full relaxed typing judgments are given in Appendix \ref{app:relaxed-typing}.}
	\label{fig:sem-defs4}
\end{figure*}
 
Evaluation introduces labels and direct-style, non-MNF terms, which are not included in the original
System \calculus{} static typing.
To reason about runtime terms, we introduce \textit{relaxed typing} judgments
\(\bs C; \sta; \G \vdash_\star t : E\), which are additionally parameterized by the label context
and relax the MNF restrictions.
Relaxed typing rules largely follow regular typing rules.
We present the most interesting differences. The full set of rules can be found in Figure \ref{fig:relaxed-full}.

The two judgments are connected by type-preserving translations to and from MNF:
any typed MNF term (\(\bs C; \G \vdash t : E\)) is relax-typed as is
(\(\bs C; \sta; \G \vdash_\star t : E\)),
and MNF normalization (Definition \ref{def:mnf-normalization}) maps any relax-typed term
back to the original typing with label rules added (\(\bs C; \sta; \G \vdash \widehat{t} : E\)).
Therefore, relaxed typing works as a proof device: our safety theorem (Theorem~\ref{thm:safety})
maintains relaxed typing as an invariant during evaluation, before re-normalizing the final result
back into standard MNF typing.

\paragraph{Typing Labels}
The rule \ruleref{rt-label} allows typing a label value present in the label context \(\sta\)
at the corresponding \(\BREAK[S]\) shape type. The interesting part lies in the capture set of
the value: a label has a single classifier \(\cls\CLS\),
so it can be projected to any kind \(\cls\KIND\), so long as it contains the classifier (i.e., $\cls\CLS \in \cls\KIND$).
This is similarly presented in the additional capture kinding rules for labels, where
\ruleref{k-label} allows kinding a (valid projection of a) label at any kind containing its classifier,
while \ruleref{k-label-absurd} sees a projection outside of the label's classifier as the empty set.
This allows us to freely change projections of a label within a capture set, as long as they remain
valid, while retaining the accessibility of the label itself during computation of runtime labels.

\paragraph{Non-MNF Terms}
Normal typing rules easily transfer to non-MNF terms: note that judgments such as
$\bs D; \G \vdash x : S\capt\bs C$  for variables can always be stated more precisely as
\(\bs{\set{x}}; \G \vdash x : S\capt\bs C\). We can always relax such premises into
\(\bs C; \G \vdash t : S\capt\bs C\), knowing that when \(x\) shows up in the use set,
its value is being \textit{used}, and so its \textit{potential use set}
(i.e., its annotated capture set during static typing) will be charged.
There are, however, two exceptions requiring careful consideration:
term application \ruleref{app} and packing \ruleref{pack}.
For application, we are more flexible about which capture set gets substituted into the dependent
result type: while the original typing implicitly widens the argument variable's type to \(T\) and
substitutes \(\bs{\set{y}}\) for $z$, the relaxed rule makes the widening step explicit \ruleref{rt-app}.
For packing, generalizing to terms risks making the calculus no longer call-by-value, so
distinct rules for variable \ruleref{rt-pack-var} and value packing \ruleref{rt-pack-val} are introduced.

\subsection{Main Results}

The central result is safety of checked evaluation.
Given a well-typed term under the top-level (i.e., empty) context
and a label allowance of at least the runtime labels of the use set,
evaluation never gets stuck and produces a
well-typed answer in the result label context.

\begin{theorem}[Safety of Checked Evaluation]\label{thm:safety}
	If\;
	$\bs{C}; \sta; \emptyset \vdash t : E$\;
	and\; $\ls{C^*_\sta} \subseteq \ls F$\;
	and\; $\sta \vdash t \Downarrow_{\ls F}^?\ a$,
	then evaluation does not get stuck,
	and $\exists w, \sta' \supseteq \sta$ such that $a = \langle \sta', w \rangle$,
	and $\ls F; \sta'; \emptyset \vDash_{\mathsf{a}} w : E$.
\end{theorem}
where \(\sta \vdash t \Downarrow_{\ls F}^?\ a\) means that evaluation terminates with
answer \(a\) or gets stuck (i.e., it does not diverge).
The proof proceeds by induction on the big-step derivation.
The full development can be found in Appendix \ref{app:safety}
and is mechanized in Lean 4 (Appendix \ref{app:lean-safety}).
Note that in the case of returning a value, we immediately obtain type safety:
\begin{corollary}[Type Safety]\label{cor:type-safety}
	If\;
	$\bs{C}; \sta; \emptyset \vdash t : E$\;
	and\; $\ls{C^*_\sta} \subseteq \ls F$\;
	and\; $\sta \vdash t \Downarrow_{\ls F}\ \langle \sta', v \rangle$,
	then $\bs{\emptyset}; \sta'; \emptyset \vdash_\star v : E$.
	Equivalently, $\bs{\emptyset}; \sta'; \emptyset \vdash \widehat{v} : E$.
\end{corollary}

Furthermore, from the links between checked evaluation's label allowance and answer typing, we can derive
additional properties about labels, breaks, and capture sets.

\begin{corollary}[Effect Safety]\label{cor:effect-safety}
	If $\bs{\emptyset}; \sta; \emptyset \vdash t : E$
	and $\sta \vdash t \Downarrow_{\ls{\emptyset}} \langle \sta', w \rangle$,
	then $\exists$ $v$ such that $w = v$.
\end{corollary}

\begin{corollary}[Used Label Prediction]\label{cor:use-pred}
	If\;
	$\bs{C}; \sta; \emptyset \vdash t : E$\;
	and\; $\sta \vdash t \Downarrow_{\ls{C^*_\sta}}^?\ a$,
	then evaluation does not get stuck.
\end{corollary}
That is, evaluation of $t$ relies only on access to runtime labels of $\bs C$.
In other words, the typing judgment's use set correctly predicts the set of labels used during evaluation of $t$.

\begin{corollary}[Capture Prediction]\label{cor:capture-pred}
	If $\bs{D}; \sta; \emptyset \vdash t : S\capt\bs C$
	and $\sta \vdash t \Downarrow_{\ls{D^*_\sta}} \langle \sta', v \rangle$, then
	\[
		\sta'; \emptyset \vdash \bs{\lfloor v \rfloor} <: \bs{C}
		\qquad\text{and}\qquad
		\ls{\lfloor v \rfloor^*_{\sta'}} \subseteq \ls{C^*_{\sta'}}
	\]
\end{corollary}

Recall that a potential use set \(\bs{\lfloor v \rfloor}\) bounds the captures that a value reaches upon
usage. We can predict the potential use set of the answer returned by
the evaluation of a term from its static typing judgment.
This is similar to the Capture Prediction for Terms lemma in $\text{CC}_{<:\square}$ \cite{boruchgruszecki2023capturing}.

\begin{corollary}[Handler Coverage]\label{cor:classifier-sound}
	If $\bs{D}; \sta; \G \vdash t : E$ and $\sta \vdash t \Downarrow_{\ls{D^*_\sta}} \langle \sta', \BRK(\ls\ell, v) \rangle$, then:
	\begin{enumerate}
		\item \emph{Breaks.} \(\ls\ell \in \ls{D^*_{\sta'}}\).
		\item \emph{Intercepts.}
		      If \(t = \INTERCEPT[R, \bs{C_t}, \cls\KIND] \WITH h \IN t'\), and \((\ls\ell : S :: \cls\CLS) \in \sta\),
		      then either \(h\) was evaluated, or \(\cls\CLS \in \cls{\ksub{\top}\KIND}\).
	\end{enumerate}
\end{corollary}

Type-level kind annotations thus predict both which labels may be reached by returned values
and how \lstinline{intercept} partitions propagating breaks into handled and escaping ones.

Two further consequences, found in Appendix \ref{app:corollaries}, clarify how the control operators behave.
Fresh labels introduced by a boundary are local and never appear in the final answer.
And for the pass rule of \lstinline{intercept}, if the handler's own captures exclude labels of
the intercepted kind, then no break whose classifier belongs to~$\cls{\KIND}$ can escape.
 
\section{Implementation of Capability Classifiers in Scala}\label{sec:impl}

This section describes how the Scala~3 capture checker implements the classifiers of
System \calculus{} (Section~\ref{sec:calculus}), and how they fare in practice.

\paragraph{Projections in Source Code}
The \lstinline|.only| and \lstinline|.except| projections of Section~\ref{sec:informal} map
directly to the projected captures of Section~\ref{sec:calculus}.
A chain of projections denotes a single one: \lstinline|a.only[k1].except[l1]| stands for
\(\sproj{a}{(k_1 - l_1)}\).
Source code has no union syntax for kinds.
To project to a union of subtrees, the programmer writes a union of projected references:
\begin{lstlisting}[mathescape=true]
{a.only[k1].except[l1], a.only[k2].except[l2]} $\;\mathrel{\hat{=}}\; \set{\sproj{a}{(k_1 - l_1)\cup (k_2 - l_2)}}$
\end{lstlisting}
Every reference then projects to a single subtree, which keeps subkinding simple.
Capture subsetting compares a kind against the union of the kinds of all projections of
the same reference, by \ruleref{s-merge}.

\paragraph{Declaring Classifiers}
Since classifiers are the plain traits of Section~\ref{sec:informal}, the classifier tree of
Section~\ref{sub:classifiers-kinds} is a slice of Scala's nominal subtype hierarchy, rooted at
\lstinline|Capability|, and the compiler answers sub-classifier queries by subtype tests.
Scala traits allow multiple inheritance, so the compiler rejects any class that inherits two
unrelated classifier traits, preserving the sub-classifier-or-disjoint property that
Section~\ref{sub:classifiers-kinds} relies on.

\paragraph{Representation in the Capture Checker}
The compiler retrofits classifiers onto the existing capture checker~\cite{xu2025capless}, whose
data structures and constraint solver revolve around capture sets of references.
Classifiers add one new form of reference, the counterpart of the projected
captures \(\pi\) of Figure~\ref{fig:syntax-paper}:
\begin{lstlisting}[basicstyle=\fontsize{6.5}{7.8}\selectfont\ttfamily]
// x.only[o].except[e1]...except[en], the exclusions strictly below o
case class Classified(underlying: Capability, only: ClassSymbol, except: List[ClassSymbol]) extends DerivedCapability
\end{lstlisting}
A \lstinline|Classified| reference represents the projected capture \(\sproj{x}{(o - \overline{e})}\).
Its \lstinline|only| and \lstinline|except| fields form one holed subtree of
Section~\ref{sub:classifiers-kinds}.
Kinds never appear as objects of their own, and solver and set operations carry over unchanged.
The cost surfaces in coverage checking: a projection of \lstinline|b| may be covered by no
single reference and still be covered by several projections of \lstinline|b| together.
The checker decides such cases with the remainder computation of Figure~\ref{fig:remainder},
which subtracts each covering subtree from the covered one and accepts when nothing remains,
deciding subkinding as emptiness of subtraction.
Each branch implements one subtraction rule of Figure~\ref{fig:kinds-full}, named in its comment.

\paragraph{Normal Forms and Subcapturing}
A smart constructor maintains the invariants noted in the comments of \lstinline|Classified|:
chained \lstinline|.only| projections collapse to their least classifier, exclusions outside
\lstinline|only| drop, an exclusion covering \lstinline|only| empties the whole reference
(rules \ruleref{e-absurd} and \ruleref{s-absurd}), and the exclusion list stays pruned and sorted.
The checker never meets a degenerate kind.
Subcapturing \(\subs{\G}{\bs{C_1}}{\bs{C_2}}\) then asks, for each reference in
\(\bs{C_1}\), whether \(\bs{C_2}\) accounts for it in one of three ways.
A single reference subsumes it, following paths and variable bounds, rules \ruleref{sc-var} and
\ruleref{sc-bound}, and weakening the classifier by \ruleref{s-subkind}.
Several projections of the same reference cover it together, decided by the remainder
computation of Figure~\ref{fig:remainder}, by rule \ruleref{s-merge}.
Finally, the capture set of its type accounts for it recursively.

\paragraph{Cost of the Kind Algebra}
\Needspace{15\baselineskip}
\begin{wrapfigure}[12]{r}{0.56\textwidth}
	\vspace{-10pt}
	\begin{lstlisting}[basicstyle=\scriptsize\ttfamily,mathescape=true]
// (o1 - e1) \ (o2 - e2), as a union of subtrees
type Subtree = (Cls, List[Cls])                $\mkern-10mu$// $\color{green!40!black}o - \overline{e}$
def subtract(o1: Cls, e1: List[Cls],
             o2: Cls, e2: List[Cls]) =
  def go(ex: List[Cls]): List[Subtree] = ex match
    case Nil => (o1, o2 :: e1) :: Nil          $\mkern-10mu$// st-tree
    case l :: ls =>
      if !l.isSub(o2) then go(ls)              $\mkern-10mu$// st-irrel-r
      else if l.isSub(o1) then (l,e1)::go(ls)  $\mkern-10mu$// st-sub-r
      else if o1.isSub(l) then (o1,e1)::Nil    $\mkern-10mu$// st-sub-l
      else go(ls)                              $\mkern-10mu$// st-irrel-l
  go(e2).filterNot(isEmpty)
\end{lstlisting}
	\vspace{-8pt}
	\caption{The kind remainder computation, abridged.
		Comments name the subtraction rules of Figure~\ref{fig:kinds-full}.}
	\label{fig:remainder}
	\vspace{-10pt}
\end{wrapfigure}
Classifier comparisons are single subtype tests, and most kind operations are polynomial.
Membership and emptiness reduce to linear scans, intersection and disjointness to the pairwise
intersections of the two kinds' subtrees.
The exception is subtraction, and with it subkinding.
One call to the remainder computation of Figure~\ref{fig:remainder} costs \(O(e)\) subtype
tests for a subtrahend with \(e\) exclusions and returns at most \(e + 1\) subtrees, since
the \lstinline|st-sub-r| branch adds one subtree per exclusion it meets.
Subtracting a whole kind iterates the computation over its subtrees, so a kind with \(s\)
subtrees can grow to \(O(s\,(e+1)^{n})\) subtrees under a subtrahend with \(n\).
In practice the exponent stays small: annotation kinds hold one or two subtrees, exclusion
lists rarely go beyond a handful of classifiers, and the final \lstinline|filterNot| of
Figure~\ref{fig:remainder} normalizes away empty subtrees.
Accordingly, we observe no noticeable compile-time regression against the capture checker of
System \capless{}, whose checking phase takes under 15\% of total compilation
time~\cite[Table~1]{xu2025capless}.
A dedicated performance study across compiler and ecosystem is future work.

\paragraph{Capability Classes}
Section~\ref{sec:informal} tags a class with a classifier through its \lstinline|extends|
clause.
We call such classes \emph{capability classes}:
\begin{scalacode}
class P(val q: Q^) extends Control
\end{scalacode}
Every instance of \lstinline|P| is a capability classified under \lstinline|Control| and enters
capture sets as the projection \lstinline|any.only[Control]|.
The instance's captures must fit within that projection, so a classified class can hold only
capabilities under its classifier.
The field \lstinline|q| above must therefore either have a \lstinline|Control|-classified type
\lstinline|Q|, or constrain its capture set to a \lstinline|Control| subset, as in
\lstinline|q: Q^{any.only[Control]}|.
In exchange, clients may assume the tighter default: the shorthand \lstinline|P^| in parameter
position stands for \lstinline|P^{any.only[Control]}| rather than \lstinline|P^{any}|.
The declared classifier bounds the capabilities an instance holds rather than tagging them all
exactly.
Note that the classifier \lstinline|Control| does not bound a class at \textit{exactly} \lstinline|Control|,
but at the \textit{subtree rooted at \lstinline|Control|} (as the projection on \lstinline|any| indicates).
This looseness enables encapsulation: a class may keep private members at finer sub-classifiers, and
clients can track those subsets (Section~\ref{sub:this-subset}).

\paragraph{Capture Inference}
Capture checking runs as its own phase after type checking and local type inference,
solving subset constraints over capture set variables on the typed
program~\cite{xu2023boxinference, xu2025capless}.
An inferred set contains the exact references a term uses, such as the path
\lstinline|file.read|, and never a projection.
A projection such as \lstinline|file.only[Read]| covers a whole subset of the capabilities
reachable through \lstinline|file|, a coarser statement.
Writing one is a deliberate widening, which belongs in annotations at API boundaries,
in particular where the precise path is hidden (Section~\ref{sub:this-subset}).
Projections from annotations flow through the solver's subset constraints like any
other reference.

\paragraph{Standard-Library Coverage}
By \texttt{tokei}'s count of non-comment, non-blank lines in sources compiled under capture
checking, the standard library is 81.7\% capture-checked (45{,}146 of 54{,}992 lines) once classifiers and the accompanying library
changes are in place, up from 66.7\% (36{,}605 of 54{,}841 lines) on the classifier-free
Scala~3.8.0 baseline, which already incorporates System \capless{}.
Classifiers do not account for the whole difference, but without them types such as
\lstinline|Try| and \lstinline|Future| and their packages \lstinline|scala.util| and
\lstinline|scala.concurrent| cannot be typed precisely at all.
The annotation burden is small: typing these signatures precisely takes
14 \lstinline|.only| and 151 \lstinline|.except| projections across the whole library.

\paragraph{Impact on Existing Code}
Adopting classifiers changes almost no existing capture-checked code.
Fully polymorphic capture sets keep their meaning under the top classifier \lstinline|Capability|,
so the capture-checked collections of System \capless{} need no classifier-specific changes.
The one source-level incompatibility concerns capability declarations.
Scala seals \lstinline|Capability|, so a user-defined capability must extend one of its two
standard children of Section~\ref{sec:informal}, \lstinline|SharedCapability| or
\lstinline|ExclusiveCapability|.
The split serves Scala's tracking of exclusive capabilities~\cite{xu2026capybara}, but not classifiers.
Sealing the root does not conflict with the open classifier tree of
Section~\ref{sub:classifiers-kinds}, which gives every classifier unboundedly many children.
The checker never treats the two children as exhausting \lstinline|Capability|: excluding
both still leaves a nonempty kind.
 
\section{Discussion}\label{sec:discussion}

We discuss design decisions surrounding classifiers and their implementation.

\subsection{Why a Tree of Classifiers?}\label{sub:why-tree}

\paragraph{A separate kind}
Set-based~\cite{marino2009generic} and boolean-algebraic~\cite{lutze23flix,gao25invalidation} effect systems can express
a finite subset of effects, or the effect universe except a finite subset, but not an open
specification: for example, \lstinline|Control| capabilities include not only those defined
in the standard library, but also an unbounded number of capabilities in other modules.
By separating classifiers from capabilities, we provide a simple solution to classification: capabilities
may tag themselves with new or existing classifiers, and the type system treats
the set of capabilities in a particular classifier as unbounded.

We deliberately decouple capability classifiers from shape types for simplicity: classifiers
have a much more rigid structure than types.
This is reflected in System \calculus{}, where classifier and kind operations are algorithmic and total.
Furthermore, this separation allows the core capture calculus to evolve mostly independently, since the integration
surface is small (minor changes in subset calculation and one additional subcapturing rule).
The Scala implementation also benefits from this restriction: a single-inheritance trait system for
classifiers is simple to work with.

\paragraph{The need for hierarchies}
A system with flat classifiers (i.e., no sub-classifiers) was also considered.
However, classifiers tend to organize into hierarchies, especially permission classifiers such as
\lstinline|Read| and \lstinline|ReadWrite| (Section~\ref{sub:this-subset}).
Flat classifiers would instead require multiple tags on a single capability, as in a tagging system where
a read-write reference must carry separate read and write tags,
placing a heavy burden on users to tag their capability classes.
An open tree structure (or more flexible structures, e.g., lattices) models such hierarchies
more naturally.

\paragraph{Disjoint decompositions}
However, an open lattice is too flexible: we lose the ability to reason about disjointness.
Two sub-lattices upper-bounded by \(A\) and \(B\) in an open lattice are never disjoint, since a future
module can always create a new classifier \(C\) directly under \(A\) and \(B\).
Disjointness is useful in practice: capability classes that split their aggregated capabilities
into sub-groups need a way to guarantee that these groups are disjoint.
For example, a \lstinline|Database| capability may partition its usage by table,
modeled as sub-classifiers of \lstinline|Database|. Without disjointness, it is not possible to
assert that \lstinline|db.readUsers| uses \lstinline|db.only[UserTable]|
and not \lstinline|db.only[TransactionTable]|.
\noindent
\begin{minipage}[t]{0.53\textwidth}
	\vspace{0pt}%
	\begin{scalacodesmall}
trait Database extends Classifier, ExclusiveCapability
trait UserTable extends Classifier, Database
/* Similarly, TransactionTable, ProductTable ... */

class DBImpl extends Database:
  val readUsers: () ->{this.only[UserTable]} List[User]
\end{scalacodesmall}
\end{minipage}\hfill
\begin{minipage}[t]{0.44\textwidth}
	\vspace{0pt}%
	\begin{scalacodesmall}
// does not type check with a lattice!
val doesntTouchTransactions :
    () ->{db.except[TransactionTable]}
    List[User]
  = db.readUsers
\end{scalacodesmall}
\end{minipage}

Flix \cite{lutze23flix, lutze24associated} describes disjoint subsets with
effect variables or associated effects, but disjointness must be requested
explicitly. The Flix analogue of \lstinline|readUsers| subtracts every sibling
table from its effect, as in
\lstinline!List[User] \ {UserTable - TransactionTable - ProductTable}!.
Boolean-algebraic subtyping faces the same problem: the effect parameter \lstinline|UserTable|
must be a subtype of the negation of all preceding parameters.
InvalML \cite{gao25invalidation} provides the outer scope variable \(\omega\) to refer to union of
visible effect parameters in scope, but are limited to region variables specifically.
A classifier tree approach enjoys disjointness by definition.

\subsection{Fine-Grained Capability Classification on References} \label{sub:fine-grained-control}

Lutze et al.~\cite{lutze23flix} compile a set of real-world effect exclusion patterns.
We succesfully replicated all 59 case studies in Scala with classifiers. In this section,
we show how classifiers enable
fine-grained control over individual references that Flix's global exclusions cannot express.

\paragraph{Simple effect exclusion}
Many patterns simply restrict a closure parameter from performing certain effects.
We model this by creating a new classifier for the kind of effects to be disallowed.
For instance, from the \lstinline|amule| example\footnotemark[3]{},
we can model a function that must not invoke any \lstinline|CSafeIOException|:
\begin{scalacode}
trait CSafeIOException extends Classifier, Control
case class CFileDataIO[T](doRead : (T, Long) -> {any.except[CSafeIOException]} Long)
\end{scalacode}
In addition to preserving semantics, we can also mark \lstinline|CSafeIOException| as part of
\lstinline|Control| capabilities, such that external code may properly reject usage of C exception handlers.

A more involved example comes from \lstinline|dune|, where
the \lstinline|runWithErrHandler| function requires the body parameter \lstinline|f|
to not nest another \lstinline|runWithErrHandler| call,
and the handler function \lstinline|handleErrNoRaise| may not raise exceptions.
We capture the requirements by creating a classifier \lstinline|ErrHandler| and giving it to the
global capability \lstinline|RunWithErrHandler| (Figure~\ref{fig:exclusion}, left).

\begin{figure}[t]
	\begin{minipage}[t]{0.46\textwidth}
		\vspace{0pt}%
		\begin{scalacodesmall}
trait ErrHandler extends Classifier, Control

object RunWithErrHandler extends ErrHandler:
  type Fiber[_] // threaded computation
  type Raise    // exception-raising capability
  def apply[A](
    /* f must not create ErrHandlers */
    f: this.Raise ->{any.except[ErrHandler]} A,
    /* cannot use exceptions */
    handleErrNoRaise: Exception ->
      {any.except[Control]} Fiber[Unit],
  ): Fiber[A]
\end{scalacodesmall}
	\end{minipage}\hfill
	\begin{minipage}[t]{0.52\textwidth}
		\vspace{0pt}%
		\begin{scalacodesmall}
trait Commit extends Classifier, SharedCapability
trait Rollback extends Classifier, SharedCapability

trait Transaction:
  val commit: () ->{this.only[Commit]} Unit
  val rollback: () ->{this.only[Rollback]} Unit
  /* other operations... */

// runs body with the transaction, commit on end
def withTransaction[T, E^](
  body: (tx: Transaction^) ->
    (() ->{E, tx.except[Commit].except[Rollback]} T)): T
\end{scalacodesmall}
	\end{minipage}
	\vspace{-8pt}
	\caption{Effect exclusion in real-world APIs: \lstinline|RunWithErrHandler| from
		\lstinline|dune| (left) and \lstinline|withTransaction| from
		\lstinline|google-cloud-go| (right).}
	\label{fig:exclusion}
\end{figure}

Here, \lstinline|f| can throw any exceptions but cannot mention \lstinline|RunWithErrHandler|, since
it is under the \lstinline|ErrHandler| classifier.
Note, however, that the specification is ambiguous on whether \lstinline|handleErrNoRaise| can \textit{use}
another error handler in its body.
The declaration in Figure~\ref{fig:exclusion} disallows it (as \lstinline|Control| is a parent of \lstinline|ErrHandler|), but we can
allow this by adding \lstinline|any.only[ErrHandler]|
(to allow all error handlers) into \lstinline|handleErrNoRaise|'s capture set.
Going further, we can allow \emph{only} the same \lstinline|RunWithErrHandler| mechanism
to be used, by setting the capture set to \lstinline|{this, any.except[Control]}|.

\paragraph{Per-reference privilege restriction}
Where classifiers go beyond Flix is in restricting specific \textit{references} rather than
globally banning effect classes.
Another example from Lutze et al.~\cite{lutze23flix} highlights this:
the \lstinline|withTransaction| function from the \lstinline|google-cloud-go| example\footnote{The original definitions can be found in
	\href{https://github.com/amule-project/amule/blob/31aff1499f1e491930e2e6fa2ca8d08d538a5493/src/SafeFile.h\#L58}{\texttt{amule-project/amule}}
	and
	\href{https://github.com/googleapis/google-cloud-go/blob/a27528eb510ea3be29f5268cc6cf6343ece58881/firestore/doc.go\#L221}{\texttt{google/google-cloud-go}}
	GitHub repositories.}
takes a closure with a \lstinline|Transaction| as its input.
This closure can use the \lstinline|Transaction|, but \textit{must not} call \lstinline|commit| or
\lstinline|rollback| on it.
The Flix implementation solves this by completely banning usage of \textit{all}
\lstinline|Commit| and \lstinline|Rollback| effects, which is too conservative. By precisely
projecting the transaction's identity in the capture set, we can do better
(Figure~\ref{fig:exclusion}, right).
The \lstinline|withTransaction| function accepts a closure that takes a \lstinline|Transaction|,
and then \textit{captures} a limited variant of this transaction in its output closure.
Note that the parameter closure itself is \textit{pure}: it has no capture set, meaning it does
not perform any operations on its own other than pure computation.
The \emph{returned} function value however
may use \lstinline|tx|, but not its \lstinline|Commit| nor \lstinline|Rollback| subset.
Interestingly, by having a capture variable \lstinline|E| in its captures,
it can use other capabilities,
including \lstinline|Commit|/\lstinline|Rollback| capabilities of \textit{other} transactions.
Note that \lstinline|E| is defined outside of \lstinline|tx|'s scope, so it cannot contain \lstinline|tx|.

\subsection{Constraining Between Levels of Classifiers}\label{sub:this-subset}
Beyond exclusions on different classifiers, the flexibility of classifiers allows us to express complex restrictions
between levels of classifiers.
We present an example from Osvald et al.~\cite{osvald16secondclass}, which was also posed as a challenge
for capture-checking systems in~\cite{xhebraj22secondclass}.
In this example, we have a parallel map operation on a \lstinline|List| data structure that
should not perform any write operations, as non-deterministic writes are one of the most common
sources of data races. On the other hand, non-deterministic reads are safe in the absence of writes.

\begin{figure}[t]
	\begin{minipage}[t]{0.487\textwidth}
		\vspace{0pt}%
		\begin{scalacodesmall}
/* User-defined capability classifiers */
trait ReadWrite extends Classifier, SharedCapability
trait Read extends Classifier, ReadWrite


trait File(val name: String):
  // capability members, ensured to be erased
  erased val r: Read^
  erased val rw: ReadWrite^

  // operations guarded by capability members
  val read: () ->{r} Int
  val write: Int ->{rw} Unit

// runs f with an open file, closes it after use
def withFile[T](f: File^ => T): T
\end{scalacodesmall}
		\centerline{\footnotesize (a)}
	\end{minipage}\hfill
	\begin{minipage}[t]{0.483\textwidth}
		\vspace{0pt}%
		\begin{scalacodesmall}
class List[+T]:
  def parMap[U](f:$\!$T ->{any.only[Read]} U): List[U]

withFile: file =>
  val list = List(99,97,112,121,98,97,114,97)
  list.parMap: item =>
    file.read()    // ok
    file.write(42) // error: capturing ReadWrite
\end{scalacodesmall}
		\centerline{\footnotesize (b)}
		\vspace{2pt}
		\begin{scalacodesmall}
trait File(val name: String):
  erased private val r: Read^
  erased private val rw: ReadWrite^
  val read: () ->{this.only[Read]} Int
  val write: Int ->{this.only[ReadWrite]} Unit
\end{scalacodesmall}
		\centerline{\footnotesize (c)}
	\end{minipage}
	\vspace{-8pt}
	\caption{Read-only parallel map: (a) capability declarations, (b) their use,
		and (c) encapsulating the capability members.}
	\label{fig:parmap}
\end{figure}

We introduce two new classifiers (Figure~\ref{fig:parmap}) \lstinline|Read| and \lstinline|ReadWrite|, with the former a sub-classifier
of the latter.
Each \lstinline|File| now holds two compile-time-only capabilities corresponding to these classifiers, and
operations on the file now capture these capabilities.
Note that calling these operations will ``use'' the capability that was captured.

When type-checking the closure passed into \lstinline|parMap|, the presence of the \lstinline|file.rw| capability
causes the compiler to reject the call, due to the capture set \(\set{\text{\lstinline|file.r|}, \text{\lstinline|file.rw|}}\) not
satisfying the capture kind \lstinline|Read|. On the other hand, without the \lstinline|file.write| call,
the example would compile.

We can go further and encapsulate the compile-time capabilities
(Figure~\ref{fig:parmap}c), using
\lstinline|this.only[Read]| to refer to the subset of the File itself that contains
only \lstinline|Read| capabilities. The compiler guarantees that the \lstinline|File| implementation will
not capture \lstinline|rw| in \lstinline|read|. We however gain a clearer public API,
free of the ``administrative'' capabilities of the implementation.
On the other hand, \lstinline|parMap| should also be more permissive with capabilities that are
\textit{neither} reading or writing to references. Depending on API considerations,
they can be allowed by including \lstinline|any.except[ReadWrite]|
in \lstinline|f|'s capture set.

\subsection{Non-Lexical Exceptions with Lexical Effect Handlers} \label{sub:safer-exceptions}

Capture checking calculi~\cite{boruchgruszecki2023capturing, xu2025capless}
promote lexical effect handling, where scoped effects like exceptions~\cite{odersky21saferexceptions}
are assumed to be handled by tunneling~\cite{zhang19tunneling}.
This is, however, not the case for exception handling on the JVM~\cite{zhang16tunneledexceptions} or JavaScript,
two main compilation targets for Scala.
To maintain compatibility with these exception semantics, we can extend the \lstinline|CanThrow| model to
allow \textit{intercepting} handlers by including the \lstinline|CanThrow| capability as part of the thrown exception.
Any \lstinline|catch| clause can then rethrow the same exception type as the caught exception, while
effect safety is retained.

\begin{figure}[t]
	\begin{minipage}[t]{0.46\textwidth}
		\vspace{0pt}%
		\begin{scalacodesmall}
def logErrors(f: () => Unit): Unit =
  try
    f()
  catch case (exc: Exception)
             (using ct: CanThrow[exc.type]^) =>
    println(exc.toString())
    throw exc // intercept & rethrow
\end{scalacodesmall}
	\end{minipage}\hfill
	\begin{minipage}[t]{0.51\textwidth}
		\vspace{0pt}%
		\begin{scalacodesmall}
case class Failure(e: Exception, ct: CanThrow[e.type]^)
def captureException(f: () => Unit)
    : Option[Failure^{f.only[Control]}] =
  try
    f(); None
  catch case (exc: Exception)
             (using ct: CanThrow[exc.type]^) =>
    Some(Failure(exc, ct)) // store exc and its CanThrow!
\end{scalacodesmall}
	\end{minipage}
	\vspace{-8pt}
	\caption{Intercepting handlers: \lstinline|logErrors| rethrows the caught exception (left),
		\lstinline|captureException| stores it together with its \lstinline|CanThrow| capability (right).}
	\label{fig:intercepting}
\end{figure}

In Figure~\ref{fig:intercepting} (left), \lstinline|logErrors| prints exceptions thrown by \lstinline|f| and then rethrows them.
Note that apart from those captured by \lstinline|f|,
\lstinline|logErrors| does not require any additional capabilities:
exception safety is retained even when we rethrow the exception, since a handler
for it is already known to exist.
The \lstinline|CanThrow| capability has the singleton type of \lstinline|exc|, because the tested type \lstinline|Exception| is too coarse:
the original exception handler may not handle a different subtype of \lstinline|Exception|.

One question remains: what if we were to \textit{store} this capability?
In Figure~\ref{fig:intercepting} (right), \lstinline|captureException| \textit{stores} the \lstinline|CanThrow| capability
and returns it, allowing the caller to unpack the capability and rethrow the exception.
What does the resulting \lstinline|Failure| capture?
Classifiers provide a solution:
since \lstinline|CanThrow| capabilities have the \lstinline|Control| classifier,
we can filter the capture set down to \lstinline|f.only[Control]|, which contains
all possible \lstinline|CanThrow| capabilities.%
\footnote{An astute reader may recognize that this is the same situation as \lstinline|Try[T]| in Section~\ref{sec:informal}:
	this is exactly the reasoning behind the capture sets for its constructor.}

\subsection{Classifiers as Traits}\label{sub:classifier-traits}

The classifier universe presented in this paper is distinct from types for pragmatic reasons.
However, in the Scala implementation, classifiers are simple traits. While they are restricted, we can still
leverage this fact to impose additional constraints on implementing capabilities.

Scala's \lstinline|Unscoped| classifier is one such example: it represents capabilities that are self-contained and can be
returned from a scope, such as owned, mutable data structures.
The class restriction requires all \lstinline|Unscoped| values to capture only other \lstinline|Unscoped| values. This allows
the compiler, with uniqueness tracking~\cite{xu2026capybara}, to safely treat returned \lstinline|Unscoped| values as completely fresh,
enabling capabilities to be tracked in a non-scoped way.
Formalizing this extension is future work.

As another example, serializable closures require all their captures to also be serializable. Spores~\cite{miller14spores} track
this in Scala by strictly controlling captured variables through a DSL implemented as Scala macros, forcing the user to declare
all captures in advance.
With classifiers, we can fuse the serializability requirement and functionality into a single classifier trait:
\begin{scalacode}
trait Serializable extends Classifier, SharedCapability:
  def serialize(): Array[Byte]
  def deserialize(from: Array[Byte])
class Spore(inner: () ->{any.only[Serializable]} Unit)
\end{scalacode}

The \lstinline|Spore| class accepts closures that capture only serializable references. The compiler can then use this
information to safely derive serialization methods for such closures, while the usage of spores remains simple and
expressive: they are still ordinary function types.

\subsection{Limitations}\label{sub:limitations}

Classifiers rely on subkinding alone, with no kind variables, so the system cannot constrain
two unknown capture sets to be disjoint, e.g.,
a parallel combinator quantifies over two capture variables,
\begin{scalacode}
def fork[A, B, E1^, E2^](f: () ->{E1} A, g: () ->{E2} B): (A, B)
\end{scalacode}
but no bound makes \lstinline|E1| and \lstinline|E2| disjoint, so \lstinline|fork| cannot
promise non-interference.
Flix~\cite{lutze23flix} states this disjointness with effect differences
\lstinline|p1: T \ {e1 - e2}, p2: U \ {e2 - e1}|.
InvalML~\cite{gao25invalidation} states it as a negation bound \(\beta \leq \neg\alpha\)
between quantified effect variables, which transposed to our setting would be a capture-set
bound \(\bs{E_2} \leq \neg\bs{E_1}\), negating sets of references rather than kinds.
Capture checking addresses non-interference through an orthogonal separation mechanism
instead, formalized in System Capybara~\cite{xu2026capybara} and shipped as Scala's
separation checking~\cite{scalaSepChecking}.

The kind algebra checks given kinds and never solves for an unknown one.
In Flix, passing an \lstinline|IO| function to a parameter of effect
\lstinline|ef - Throw| makes boolean unification derive \lstinline|ef| from
\lstinline|IO = ef - Throw|.
Our checking never needs such a step, since a call site requires some instantiation below
its bound rather than a most general solution, and capture inference finds the least one.
Every exclusion pattern of Section~\ref{sub:fine-grained-control} type-checks this way.
The difference surfaces only at API boundaries, where projections must be written by hand
because inference never produces them (Section~\ref{sec:impl}).

Being subtree-based, classifier kinds don't have non-trivial meets: two sibling subtrees always meet
at an empty kind. Therefore, a capability cannot be classified at two sibling classifiers:
The compiler rejects \lstinline|class Log extends Read, Serializable| outright (Section~\ref{sec:impl}),
and requiring \lstinline|.only[Read]| and \lstinline|.only[Serializable]| on a capture set collapses it
to empty.
Qualifier lattices have exact meets: in System \(F_{<:}^{Q}\)~\cite{lee24qualifiers},
variables can be quantified as \(l_1 \wedge l_2\) where \(l_1, l_2\) are both base lattice elements
(corresponding to a single classifier), and can admit both declarations.
Neither algebra subsumes the other: the tree lacks meets, but enjoys complements and disjoint decompositions.

\section{Related Work}\label{sec:related}

\paragraph{Capability-based Effects}
Type-and-effect systems date back to Lucassen and Gifford's polymorphic
effects~\cite{lucassen88polymorphiceffects}.
Craig et al.~\cite{craig18wyvern} show that capabilities can be tracked in a type system to ensure
capability safety and reason about effects. Brachthäuser et al.~\cite{brachthauser20effekt} view
effect types as capabilities required from the calling context, enabling lightweight effect
polymorphism without effect variables, while coeffect systems~\cite{petricek14coeffects} study
contextual requirements more generally. This perspective is close to capture tracking, since a
capture set records which capabilities a term requires from its environment. Capture
calculus~\cite{boruchgruszecki2023capturing,xu2025capless,xu24separation} tracks captured
capabilities in Scala types, and our calculus builds directly on~\textcite{xu2025capless}.
\textcite{tang26rows} give a uniform encoding of row-based and capability-based systems into modal
effect types.

Capture checking is more than a lightweight effect system.
A capture set names program values, often paths such as \lstinline|body|, \lstinline|tx|, or
\lstinline|file.r|, so capture checking itself reasons about capabilities by the identity of
the values that carry them, keeping two file handles \lstinline|file1| and \lstinline|file2|
distinct.
Classifiers inherit this precision: a projection like \lstinline|tx.except[Commit]| restricts
one particular transaction rather than a whole effect class
(Sections~\ref{sub:fine-grained-control} and~\ref{sub:this-subset}).
The same capturing-types discipline reaches further, to sharing, lifetimes, aliasing, and
separation of resources~\cite{xu24separation,xu2026capybara}, which traditional effect systems
do not track directly.
TACIT builds agent security on the purity of empty capture
sets~\cite{DBLP:conf/cais/OderskyZXBP26}.

\paragraph{Effect Classification and Exclusions}
Flix~\cite{lutze23flix} introduces effect exclusion on a boolean algebra of effect sets, typing
functions that accept any effects \emph{except} a given set, with full Hindley-Milner inference
by boolean unification.
Checking there is equality-based, ours
subsumption-based~\cite{tang95effectsubtyping,wansbrough99polymorphictype}.
Boolean unification keeps the top of the algebra symbolic to stay sound in an open world,
and the classifier tree keeps the top of every subtree symbolic (Section~\ref{sub:why-tree}).
We replicate Flix's case studies in Section~\ref{sub:fine-grained-control}, and
Section~\ref{sub:limitations} states what Flix can express that classifiers cannot.

Boolean-algebraic type-and-effect systems in general incorporate
intersection and negation into effect typing~\cite{madsen20boolean, gao25invalidation}.
They are more expressive than our system because they allow effect set variables
in algebraic expressions, but also considerably more complex.
Classifiers can also represent exclusion of open sets of classifiers, which does not appear to be
representable with intersection and negation of finite sets.

Row typing offers another route to classifying and filtering effects.
R\'emy's record calculus attaches presence and absence flags to row
labels~\cite{remy89records}, and the Rose language of Morris and
McKinna~\cite{morris19rows} derives absence and disjointness from qualified
predicates over rows, so both can state that a named label is absent even from
an otherwise open row.
Neither calculus, as defined, can state absence of an open category: a row is
a finite list of named labels over one uniform tail, so it can mark finitely many
labels absent or close the remainder entirely, but cannot exclude a whole
open-ended category while leaving the rest open.
That is what \lstinline|any.except[Control]| expresses, excluding the
\lstinline|Control| subtree together with every sub-classifier that later
modules add.
Rows could recover this by kinding the label universe into categories, which
rebuilds the classifier tree on the label side.
The modal encoding of \textcite{tang26rows} discussed above suggests a formal
bridge between rows and capabilities.

Effect masking~\cite{biernacki18handlecare,lindley17frank,leijen2016koka,wu14handlersscope,tang25modal} provides a
way to skip a certain effect handler at runtime when multiple handlers are
installed. Biernacki et al.~\cite{biernacki18handlecare} formalize related selective
forwarding via a lift operator in a logical-relations semantics of handlers,
and Eff~\cite{bauer15eff}, Koka~\cite{leijen2016koka,leijen17roweffects},
Frank~\cite{lindley17frank}, and scoped handlers~\cite{wu14handlersscope} study handlers
over open effect rows.
Masking and lifting re-route the search for the nearest handler, whereas classifier
projections subtract categories from what a reference may reach. Our \lstinline|intercept|
construct is where the two perspectives meet: it lets lexically scoped
\lstinline|boundary|/\lstinline|break| control temporarily take on
nearest-handler semantics, intercepting matching breaks, while the
intercepted label's capture set still governs where that label may be
retained and used.

First-class names~\cite{xie22firstclass} distinguish scoped effect instances by explicit names, a
complement to classifiers, which group capabilities into extensible semantic categories with
subtree-based filtering.

Xu et al.~\cite{xu23making} discuss a subtraction operator and disjointness on types. Their
operations are algebraically similar to ours, but serve a type system with records and universal
quantification for reasoning about interfaces and overloading rather than capabilities or effects.

\paragraph{Classifying Specific Capability Classes}

Spores~\cite{miller14spores} constrain what may appear in a closure's captured environment,
enforcing serialization and concurrency safety. This is close to our
\lstinline|only| and \lstinline|except| operators, which also restrict which capability values a
closure may capture. Spores attach these restrictions to each closure
type, whereas our system factors them through a shared, open classifier hierarchy. We further
discuss spores and classifier traits in Section~\ref{sub:classifier-traits}.

Osvald et al.~\cite{osvald16secondclass} classify values by lifetime scope, and Xhebraj et
al.~\cite{xhebraj22secondclass} extend this with delayed stack reclamation. Our approach can provide
equivalent static guarantees (Section~\ref{sub:this-subset}), though similar stack-allocation
optimizations remain future work.

Fractional capabilities and related qualifiers can track ownership and mutability. Haller and
Odersky~\cite{haller10uniqueness} classify references as unique or borrowed. Fractional
permissions~\cite{boyland03fractional} split read and write access, and Gordon et
al.~\cite{gordon12immutability} use a mutability lattice. Marshall and
Orchard~\cite{marshall24ownership} encode fractional ownership with graded modal types. Qualified
types~\cite{lee24qualifiers} likewise attach lattice-based restrictions to values. We do not
directly tackle ownership, but classifiers can express closely related distinctions, as discussed in
Section~\ref{sub:this-subset}.

\paragraph{Tracking Exceptions in Types}

While exceptions have been present in programming languages since at least
Goodenough~\cite{goodenough75exceptions}, many languages omit them from static types. Java's checked
exceptions~\cite{gosling96java} face exception-polymorphism problems, motivating Scala effect
systems~\cite{rytz2012polymorphiceffects, rytz2014scalaeffects} and Swift~2's \lstinline|rethrows|
clause~\cite{swiftErrorHandling}. Wu et al.~\cite{wu14handlersscope} model exceptions with scoped
effect handlers. Related accounts appear in Eff~\cite{bauer15eff}, Koka~\cite{leijen2016koka}, and
Frank~\cite{lindley17frank}. Zhang et al.~\cite{zhang16tunneledexceptions,zhang19tunneling} argue
that intercepted exceptions are hard to reason about with higher-order functions and propose
``tunneling'' exceptions to statically known handlers. Lexical effect handlers in
Effekt~\cite{brachthauser20effekt} and Lexa~\cite{ma24lexa} generalize this idea. Odersky et
al.'s safer exceptions~\cite{odersky21saferexceptions} track exception handlers with capabilities. Lexical handlers
still do not match the non-lexical exception semantics of current Scala backends, and
Section~\ref{sub:safer-exceptions} discusses how our classifier-based account builds on safer
exceptions.

Delimited continuations can be encoded on top of existing exception
mechanisms~\cite{koppel2018capturing}, and tunneling can itself be implemented over Java
exceptions~\cite{zhang16tunneledexceptions}. Together with recent Scala continuation
work~\cite{bartrina25lexical,pham24stackcopying}, this suggests a route to capture-safe lexical
effect handlers across all backends. In Scala, this matters because any practical account has to
span the JVM, Native, and JS backends.

\paragraph{Dynamic Effect Handlers}
Plotkin and Pretnar~\cite{plotkin2013algebraiceffects} formalize algebraic effect handlers,
Kammar et al.~\cite{kammar13handlers} develop their programming idioms, and Yoshioka et
al.~\cite{yoshioka24abstracting} abstract over the choice of effect representation.
A general handler dispatches on the operations of one effect, whereas our \lstinline|intercept|
discharges the open family of labels under one classifier.
The route to general handlers in Scala runs through the continuation mechanisms discussed
above.
 
\section{Conclusion}\label{sec:conclusion}

Capture checking tracks capabilities by identity, but standard library types
such as \lstinline|Try| and \lstinline|Future| require reasoning about
capabilities by \emph{kind}: retaining only control-flow capabilities, or
excluding thread-local ones.
We introduced capability classifiers, a tree-structured hierarchy of tags with
\lstinline|.only|/\lstinline|.except| projections on capture sets, giving
precise types to constructs that were previously out of reach.
The tree structure ensures disjointness between branches while
remaining open to new classifiers in any module.
We formalized classifiers as an extension of System Capless, with a decidable
kind algebra and kind-annotated projections, and proved type safety via a
checked big-step semantics, establishing effect safety, capture prediction,
and handler coverage, all mechanized in Lean~4.
The system is implemented in the Scala~3 compiler.
Overall, classifiers make capture checking more expressive by letting the type
system capture the effect distinctions that practical code requires, further
strengthening the case for programming with effects and capabilities in a
mainstream language.

\section*{Acknowledgement}

This work is  supported by the SNSF Advanced Grant TMAG-2\_209506/1, ``Capabilities for Typing Resources and Effects''.

\printbibliography{}

\clearpage
\appendix
\counterwithin{figure}{section}
\crefalias{section}{appendix}
\crefalias{subsection}{appendix}
\crefalias{subsubsection}{appendix}

\section{The Formal System in Full}\label{app:formal-system}

This appendix completes the formal development of Section~\ref{sec:calculus}
and Section~\ref{sec:metatheory}.
The main text presents System \calculus{} in monadic normal form (MNF),
where the operands of applications and the bodies of packs are variables
(Figures~\ref{fig:syntax-paper}, \ref{fig:all-typing-paper} and
\ref{fig:extension-paper}).
For the big-step safety proof we work instead with the \emph{relaxed} type
system, a proof device that drops the variable restriction on operands and
pack bodies so that terms are in direct style.
Direct style is the more standard setting for a substitution-based big-step
proof.
The evaluation rules substitute evaluated values into term bodies, and
admitting arbitrary operands and pack bodies lets a single typing judgment
classify every intermediate configuration without renormalizing to MNF after
each substitution.
The relaxed judgment $\bs{C}; \sta; \G \vdash_\star t : E$ appears in
Figure~\ref{fig:sem-defs4}.
We prove safety for the relaxed system and transport it back to MNF.

The development is structured as a round trip.
\begin{enumerate}
	\item Every MNF-typed term is relaxed-typed
	      (Lemma~\ref{lem:mnf-to-relaxed}).
	\item The checked big-step semantics evaluates relaxed-typed terms, and
	      safety holds for the relaxed system
	      (Theorem~\ref{thm:safety-relaxed}).
	\item Any relaxed-typed answer can be normalized back into MNF without
	      changing its type (Lemma~\ref{lem:mnf-normalization}), and on MNF
	      terms the two systems agree (Lemma~\ref{lem:relaxed-to-mnf}).
\end{enumerate}
Chaining the three steps yields safety for the MNF system of the main text
(Theorem~\ref{thm:safety-mnf}).
Below we give the relaxed syntax and the full relaxed typing rules
(\Cref{app:relaxed-typing}), the correspondence between the two systems
(\Cref{app:mnf-relaxed}), the complete checked big-step evaluation rules
(\Cref{app:full-bigstep}), and the safety theorems with their corollaries
(\Cref{app:safety,app:corollaries}).
All results are mechanized in Lean~4, and \Cref{app:mechanization} relates
the pencil and paper formulation to the mechanization.

For reference, \Cref{fig:all-typing-full} reproduces the complete MNF type
system of System \calculus{}, of which \Cref{fig:all-typing-paper} in the main
text shows only a representative selection.
Likewise, \Cref{fig:kinds-full} reproduces the complete kind algebra
condensed in \Cref{fig:kinds4}.

\begin{figure*}[htbp]
\resizebox{\textwidth}{!}{\begin{minipage}{1.15\textwidth}
	\scriptsize

	\flushleft{\textbf{Capture Kinding} \quad \jbox{\new{\G \vdash \bs{C}: \cls{\KIND}}}}

	\begin{multicols}{2}
		\infrule[\textsc{k-var}]
		{x : S\capt \bs{C} \in \G\andalso\G \vdash \bs{C}\PROJ[\cls{\KIND_1}]: \cls{\KIND_2}}
		{\G \vdash \bs{\set{\sproj{x}{\cls{\KIND_1}}}}: \cls{\KIND_2}}

		\infrule[\textsc{k-cvar}]
		{c <: \bs{C}\in\G\andalso\G \vdash \bs{C}\PROJ[\cls{\KIND_1}]: \cls{\KIND_2}}
		{\G \vdash \bs{\set{\sproj{c}{\cls{\KIND_1}}}}: \cls{\KIND_2}}

		\infrule[\textsc{k-cbound}]
		{c:\cls{\KIND_1}\in\G}
		{\G \vdash \bs{\set{\sproj{c}{\cls{\KIND_2}}}}: \cls{\KIND_1 \land \KIND_2}}

		\infrule[\textsc{k-sub}]
		{\G \vdash \bs{C}: \cls{\KIND}\andalso
			\cls{\subk{\KIND}{\KIND'}}}
		{\G \vdash \bs{C}: \cls{\KIND'}}

		\infrule[\textsc{k-absurd}]
		{\G \vdash \bs{C}: \cls{\KIND}\andalso
			\cls{\KIND\EMPTY}}
		{\G \vdash \bs{C}: \cls{\KIND'}}

		\infrule[\textsc{k-union}]
		{\G \vdash \bs{C_1}: \cls{\KIND}\andalso
			\G \vdash \bs{C_2}: \cls{\KIND}}
		{\G \vdash \bs{C_1\cup C_2}: \cls{\KIND}}

		\infax[\textsc{k-empty}]
		{\G \vdash \bs{\set{}}: \cls{\KIND}}

	\end{multicols}

	\flushleft{\textbf{Capture Subsetting} \quad \jbox{\new{\bs{C_1} \sqsubseteq \bs{C_2}}}}

	\begin{multicols}{2}

		\infrule[\textsc{s-elem}]
		{\bs{C_1} \subseteq \bs{C_2}}
		{\bs{C_1} \sqsubseteq \bs{C_2}}

		\infrule[\textsc{s-subkind}]
		{\cls{\subk{\KIND{}_1}{\KIND{}_2}}}
		{\bs{\{\sproj{\theta}{\cls{\KIND{}_1}}\}} \sqsubseteq \bs{\{\sproj{\theta}{\cls{\KIND{}_2}}\}}}

		\infrule[\textsc{s-absurd}]
		{\cls{\KIND\EMPTY}}
		{\bs{\{\sproj{\theta}{\cls{\KIND{}}}\}} \sqsubseteq \bs{\{\}}}

		\infax[\textsc{s-merge}]
		{\bs{\{\sproj{\theta}{\cls{(\KIND{}_1 \lor \KIND{}_2)}}\}} \sqsubseteq \bs{\{\sproj{\theta}{\cls{\KIND_1}}, \sproj{\theta}{\cls{\KIND{}_2}}\}}}
	\end{multicols}

	\flushleft{\textbf{Subcapturing} \quad \jbox{\subs{\G}{\bs{C_1}}{\bs{C_2}}}}

	\vspace{0.3em}

	\begin{multicols}{2}
		\infrule[\ruledef{sc-trans}]
		{\subs{\G}{\bs{C_1}}{\bs{C_2}} \\ \subs{\G}{\bs{C_2}}{\bs{C_3}}}
		{\subs{\G}{\bs{C_1}}{\bs{C_3}}}

		\infrule[\textsc{sc-var}]
		{x : S\capt \bs{C} \in \G}
		{\subs{\G}{\bs{\set{(x\new{\mid\cls{\KIND{}}})}}}{\bs{C\new{\PROJ[\cls{\KIND{}}]}}}}

		\infrule[\textsc{sc-bound}]
		{c <: \bs{C} \in \G}
		{\subs{\G}{\bs{\set{(c\new{\mid\cls{\KIND{}}})}}}{\bs{C\new{\PROJ[\cls{\KIND{}}]}}}}

		\infrule[\textsc{sc-elem}]
		{{\bs{C_1}\new{\sqsubseteq} \bs{C_2}}}
		{\subs{\G}{\bs{C_1}}{\bs{C_2}}}

		\infrule[\ruledef{sc-set}]
		{\subs{\G}{\bs{C_1}}{\bs{C}}\andalso \subs{\G}{\bs{C_2}}{\bs{C}}}
		{\subs{\G}{\bs{C_1\cup C_2}}{\bs{C}}}

		\infrule[{\new{\textsc{sc-proj}}}]
		{\G \vdash \bs{C}: \cls{\KIND{}}}
		{\subs{\G}{\bs{C}}{\bs{C\PROJ[\cls{\KIND{}}]}}}

	\end{multicols}

	\flushleft{\textbf{Bound Subtyping} \quad \jbox{\G \vdash B_1 <:_{\mathsf{B}} B_2}}
	\begin{multicols}{3}
		\infrule[\textsc{b-set}]
		{\subs{\G}{\bs{C_1}}{\bs{C_2}}}
		{\G \vdash \bs{C_1} <:_{\mathsf{B}} \bs{C_2}}
		\infrule[\new{\textsc{b-kind}}]
		{\cls{\subk{\KIND_1}{\KIND_2}}}
		{\G \vdash \cls{\KIND_1} <:_{\mathsf{B}} \cls{\KIND_2}}
		\infrule[\new{\textsc{b-set-kind}}]
		{\G \vdash \bs{C}: \cls{\KIND}}
		{\G \vdash \bs{C} <:_{\mathsf{B}} \cls{\KIND}}

	\end{multicols}

	\flushleft{\textbf{Typing} \quad \jbox{\bs{C}; \G \vdash t : E}}

	\begin{multicols}{3}
		\infrule[\textsc{var}]
		{x: S\capt \bs{C} \in \G}
		{\bs{\set x}; \G \vdash x : S\capt\bs{\set x}}

		\infrule[\textsc{pack}]
		{\bs{C}; \G \vdash x : \bs{[D/c]}T \\
			\new{\G \vdash \bs D <:_{\mathsf{B}} {B}}}
		{\bs{\set{}}; \G \vdash \PACK\langle \bs D, x\rangle : \exists \bs c\new{: B}.\,{T}}

		\infrule[\ruledef{sub}]
		{\bs{C}; \G \vdash t : E \andalso \G \vdash E <: F \\
			\subs{\G}{\bs{C}}{\bs{C'}} \andalso \G \vdash \bs{C'}, F\ \textbf{\textsc{wf}}}
		{\bs{C'}; \G \vdash t : F}

		\infrule[\textsc{abs}]
		{\bs{C\uplus \set{x}}; (\G, x: T) \vdash t : E\\ \G \vdash T\ \textbf{\textsc{wf}}}
		{\bs{\set{}}; \G \vdash \lambda^{\bs{C}}(x: T)t : (\forall(x: T)E)\capt \bs{C}}

		\infrule[\textsc{app}]
		{\bs{C'}; \G \vdash x : (\forall (z: T)E)\capt \bs{C} \\
			\bs{C'}; \G \vdash y : T}
		{\bs{C'}; \G \vdash x\ y : \bs{[y/z]}E}

		\infrule[\ruledef{tabs}]
		{\bs{C}; (\G, X <: S) \vdash t : E \andalso \G \vdash S\ \textbf{\textsc{wf}}}
		{\bs{\set{}}; \G \vdash \lambda^{\bs{C}}[X <: S]t : (\forall[X <: S]E)\capt \bs{C}}

		\infrule[\ruledef{tapp}]
		{\bs{C'}; \G \vdash x : (\forall[X <: S]E)\capt \bs{C}}
		{\bs{C'}; \G \vdash x[S] : [S/X]E}

		\infrule[\ruledef{cabs}]
		{\bs{C}; (\G, c: B) \vdash t : E \andalso \G \vdash B\ \textbf{\textsc{wf}}}
		{\bs{\set{}}; \G \vdash \lambda^{\bs{C}}[c: B]t : (\forall[c: B]E)\capt \bs{C}}

		\infrule[\ruledef{capp}]
		{\bs{C}; \G \vdash x : (\forall[c: {B}]E)\capt \bs{C'}\\
			\G \vdash \bs D <:_{\mathsf{B}} {B}}
		{\bs{C}; \G \vdash x[\bs D] : \bs{[D/c]}E}

		\infrule[\textsc{let}]
		{\bs{C}; \G \vdash t : T\andalso \bs{C}; (\G, x: T) \vdash u : E \\
			\G \vdash \bs{C}, E\ \textbf{\textsc{wf}}}
		{\bs{C}; \G \vdash \LET x = t \IN u : E}

		\infrule[\textsc{let-e}]
		{\bs{C}; \G \vdash t : \exists c\new{: B}.\, T \andalso \G \vdash \bs{C}, F\ \textbf{\textsc{wf}} \\
			\bs{C}; (\G, \new{c: B}, x: T) \vdash u : F}
		{\bs{C}; \G \vdash \LET \langle c, x \rangle = t \IN u : E}

	\end{multicols}

	\flushleft{\textbf{Subtyping} \quad \jbox{\G \vdash E_1 <: E_2}}
	\begin{multicols}{4}
		\infax[\ruledef{top}]
		{\G \vdash S <: \top}

		\infax[\ruledef{refl}]
		{\G \vdash E <: E}

		\infrule[\ruledef{trans}]
		{\G \vdash E_1 <: E_2\\ \G \vdash E_2 <: E_3}
		{\G \vdash E_1 <: E_3}

		\infrule[\ruledef{tvar}]
		{X <: S \in \G}
		{\G \vdash X <: S}

		\infrule[\textsc{capt}]
		{\G \vdash S_1 <: S_2\\ \subs{\G}{\bs{C_1}}{\bs{C_2}}}
		{\G \vdash S_1\capt \bs{C_1} <: S_2\capt \bs{C_2}}

	\end{multicols}

	\begin{multicols}{2}
		\infrule[\textsc{exist}]
		{(\G, c <: \new{B_1}) \vdash T_1 <: T_2\andalso \new{\G \vdash B_1 <:_{\mathsf{B}} B_2}}
		{\G \vdash \exists c \new{: B_1}. T_1 <: \exists c \new{: B_2}.T_2}
		\infrule[\ruledef{fun}]
		{\G \vdash E_1 <: E_2\andalso \G \vdash T_2 <: T_1}
		{\G \vdash \forall (x: T_1) E_1 <: \forall (x: T_2) E_2}
		\infrule[\ruledef{tfun}]
		{(\G, X <: S_2) \vdash E_1 <: E_2\andalso \G \vdash S_2 <: S_1}
		{\G \vdash \forall (X <: S_1) E_1 <: \forall (X <: S_2) E_2}
		\infrule[\ruledef{cfun}]
		{(\G, c: {B_2}) \vdash E_1 <: E_2\andalso {\G \vdash B_2 <:_{\mathsf{B}} B_1}}
		{\G \vdash \forall (c: B_1) E_1 <: \forall (c: B_2) E_2}
	\end{multicols}

\end{minipage}}%
	\caption{Type system of System \calculus{} in full (copy of the condensed
	\Cref{fig:all-typing-paper}).
	Changes compared to System \capless{}~\cite{xu2025capless} are \tnew{marked}.}\label{fig:all-typing-full}

\end{figure*}
\begin{figure*}[htbp]
	\scriptsize

	\flushleft{\textbf{Intersection} \quad \jbox{\cls{\KIND{}_1 \land \KIND{}_2 = \KIND}}}

	\begin{multicols}{2}

		\infrule[\textsc{i-subtree-l}]
		{\cls{\CLS_1 \sqsubseteq \CLS_2}}
		{\cls{(\CLS_1 - \overline{\CLSS_1}) \land (\CLS_2 - \overline{\CLSS_2}) = \CLS_1 - (\overline{\CLSS_1}, \overline{\CLSS_2})}}

		\infrule[\ruledef{i-subtree-r}]
		{\cls{\CLS_2 \sqsubseteq \CLS_1}}
		{\cls{(\CLS_1 - \overline{\CLSS_1}) \land (\CLS_2 - \overline{\CLSS_2}) = \CLS_2 - (\overline{\CLSS_1}, \overline{\CLSS_2})}}

		\infrule[\textsc{i-disjoint}]
		{\cls{\disj{\CLS_1}{\CLS_2}}}
		{\cls{(\CLS_1 - \overline{\CLSS_1}) \land (\CLS_2 - \overline{\CLSS_2}) = \emptyset}}

		\infax[\ruledef{i-empty-r}]
		{\cls{\KIND{} \land \emptyset = \emptyset}}

		\infrule[\textsc{i-union-r}]
		{\cls{\KIND{} \land \KIND_1 = R_1}\andalso\cls{\KIND{}\land \KIND_2 = R_2}}
		{\cls{\KIND{} \land (\KIND_1 \lor \KIND_2) = R_1\lor R_2}}

		\infax[\ruledef{i-empty-l}]
		{\cls{\emptyset \land (\CLS - \overline{\CLS_i}) = \emptyset}}

		\infrule[\ruledef{i-union-l}]
		{\cls{\KIND_1 \land (\CLS - \overline{\CLS_i}) = R_1}\andalso\cls{\KIND_2\land (\CLS - \overline{\CLS_i}) = R_2}}
		{\cls{(\KIND_1 \lor \KIND_2)\land (\CLS - \overline{\CLS_i}) = R_1\lor R_2}}
	\end{multicols}

	\flushleft{\textbf{Subtraction} \quad \jbox{\cls{\ksub{\KIND_1}{\KIND_2} = \KIND}}}

	\begin{multicols}{2}

		\infax[\textsc{st-tree}]
		{\cls{\ksub{(\CLS_1 - \overline{\CLSS_1})}{(\CLS_2 - \epsilon)} = \CLS_1 - (\CLS_2, \overline{\CLSS_1})}}

		\infrule[\ruledef{st-absurd-r}]
		{\cls{\CLS_2 \sqsubset \CLSS}}
		{\cls{\ksub{(\CLS_1 - \overline{\CLSS_1})}{(\CLS_2 - (\CLSS, \overline{\CLSS_2}))} = \CLS_1 - \overline{\CLSS_1}}}

		\infrule[\ruledef{st-irrel-r}]
		{\cls{\disj{\CLS_2}{\CLSS}}\andalso \cls{\ksub{(\CLS_1 - \overline{\CLSS_1})}{(\CLS_2 - \overline{\CLSS_2})} = \KIND{}}}
		{\cls{\ksub{(\CLS_1 - \overline{\CLSS_1})}{(\CLS_2 - (\CLSS, \overline{\CLSS_2}))} = \KIND{}}}

		\infrule[\textsc{st-sub-r}]
		{\cls{\CLSS\sqsubseteq \CLS_2}\andalso \cls{\CLSS\sqsubseteq \CLS_1}\andalso \cls{\ksub{(\CLS_1 - \overline{\CLSS_1})}{(\CLS_2 - \overline{\CLSS_2})} = \KIND{}}}
		{\cls{\ksub{(\CLS_1 - \overline{\CLSS_1})}{(\CLS_2 - (\CLSS, \overline{\CLSS_2}))} = (\CLSS - \overline{\CLSS_1}) \lor \KIND{}}}

		\infrule[\ruledef{st-sub-l}]
		{\cls{\CLSS\sqsubseteq \CLS_2}\andalso \cls{\CLS_1\sqsubset \CLSS}}
		{\cls{\ksub{(\CLS_1 - \overline{\CLSS_1})}{(\CLS_2 - (\CLSS, \overline{\CLSS_2}))} = (\CLS_1 - \overline{\CLSS_1})}}

		\infrule[\ruledef{st-irrel-l}]
		{\cls{\CLSS\sqsubseteq \CLS_2}\andalso \cls{\disj{\CLS_1}{\CLSS}}\andalso \cls{\ksub{(\CLS_1 - \overline{\CLSS_1})}{(\CLS_2 - (\overline{\CLSS_2}))} = \KIND{}}}
		{\cls{\ksub{(\CLS_1 - \overline{\CLSS_1})}{(\CLS_2 - (\CLSS, \overline{\CLSS_2}))} = \KIND{}}}

		\infax[\ruledef{st-empty-r}]
		{\cls{\ksub{\KIND{}}\emptyset = \KIND{}}}

		\infrule[\textsc{st-union-r}]
		{\cls{\ksub\KIND{}\KIND_1 = R_1}\andalso \cls{\ksub{R_1}{\KIND_2} = R_2}}
		{\cls{\ksub\KIND{}{(\KIND_1 \lor \KIND_2)} = R_2}}

		\infax[\ruledef{st-empty-l}]
		{\cls{\ksub\emptyset{(\CLS - \overline{\CLS_i})} = \emptyset}}

		\infrule[\textsc{st-union-l}]
		{\cls{\ksub{\KIND_1}{(\CLS - \overline{\CLS_i})} = R_1}\andalso \cls{\ksub{\KIND_2}{(\CLS - \overline{\CLS_i})} = R_2}}
		{\cls{\ksub{(\KIND_1 \lor \KIND_2)}{(\CLS - \overline{\CLS_i})} = R_1 \lor R_2}}

	\end{multicols}

	\flushleft{\textbf{Emptiness} \quad \jbox{\cls{\KIND\EMPTY}}}

	\begin{multicols}{2}

		\infax[\textsc{e-empty}]
		{\cls{\emptyset\EMPTY}}

		\infrule[\textsc{e-absurd}]
		{\cls{k \sqsubseteq \CLSS_i}}
		{\cls{k - (\CLSS_1, \dots, \CLSS_i, \dots, \CLSS_n)\EMPTY}}

		\infrule[\textsc{e-union}]
		{\cls{\KIND_1\EMPTY}\andalso \cls{\KIND_2\EMPTY}}
		{\cls{(\KIND_1\lor\KIND_2)\EMPTY}}

	\end{multicols}

	\flushleft{\textbf{Disjoint} \quad \jbox{\cls{\disj{\KIND_1}{\KIND_2}}}}
	\quad Defined as \(\cls{(\KIND_1\land \KIND_2)\EMPTY}\).

	\flushleft{\textbf{Subkinding} \quad \jbox{\cls{\subk{\KIND_1}{\KIND_2}}}}
	\quad Defined as \(\cls{(\KIND_1\backslash \KIND_2)\EMPTY}\).

	\flushleft{\textbf{Inclusion} \quad \jbox{\cls{\CLS \in \KIND}}}

	\begin{multicols}{2}
		\infrule[\ruledef{in-base}]
		{\cls{\CLS \sqsubseteq \CLS_0}}
		{\cls{\CLS \in (\CLS_0 - \epsilon)}}

		\infrule[\ruledef{in-nonexcl}]
		{\cls{\CLS \not\sqsubseteq \CLS_1}\andalso \cls{\CLS \in (\CLS_0 - (\CLS_2, \ldots, \CLS_n))}}
		{\cls{\CLS \in (\CLS_0 - (\CLS_1, \ldots, \CLS_n))}}

		\infrule[\ruledef{in-union-l}]
		{\cls{\CLS \in \KIND_1}}
		{\cls{\CLS \in (\KIND_1\lor \KIND_2)}}

		\infrule[\ruledef{in-union-r}]
		{\cls{\CLS \in \KIND_2}}
		{\cls{\CLS \in (\KIND_1\lor \KIND_2)}}
	\end{multicols}

	\flushleft{\textbf{Capture Set Projection}\quad $\bs{C}\PROJ[\cls{\KIND}] \coloneq \bs{\left\{\overline{\sproj{\theta_i}{\cls{\KIND_i\land \KIND}}}\right\}}$ where $\bs{C} = \bs{\left\{\overline{\sproj{\theta_i}{\cls{\KIND_i}}}\right\}}$}

	\flushleft{\textbf{Exclusive Union} \quad
		\(\bs{C \uplus \set{x_1, \ldots, x_n}} = \bs{C \cup \set{\sproj{x_1}{\top}, \ldots, \sproj{x_n}{\top}}}\)
		where \(\forall i, \phi. \bs{\sproj{x_i}\phi} \not\in \bs{C}\)}

	\caption{Kind algebra of System \calculus{} in full (copy of the condensed
		\Cref{fig:kinds4}).}\label{fig:kinds-full}

\end{figure*}
 
\subsection{Relaxed Terms and the Relaxed Type System}\label{app:relaxed-typing}

\begin{figure*}[tp]
	\scriptsize

	\noindent
	\begin{minipage}[t]{0.48\textwidth}
		\flushleft{\textbf{Relaxed Syntax}}
		\begin{flalign*}
			s,\,t,\,u\coloneqq\  & \tag*{\textbf{Term}}                                  \\
			                     & x \BAR v                          \tag*{variable, value} \\
			                     & t\ s                              \tag*{application} \\
			                     & t[S]                              \tag*{type application} \\
			                     & t[\bs{C}]                         \tag*{capture application} \\
			                     & \LET x = t \IN u                  \tag*{let} \\
			                     & \LET \langle c, x\rangle = t \IN u \tag*{existential let} \\
			v\coloneqq\          & \tag*{\textbf{Value}}                                 \\
			                     & \lambda^{\bs{C}}(x{:}T)t \BAR \lambda^{\bs{C}}[X{<:}S]t \BAR \lambda^{\bs{C}}[c{:}B]t \tag*{functions} \\
			                     & \PACK\langle \bs{C}, t\rangle     \tag*{pack}        \\
			                     & \ls\ell                           \tag*{label}
		\end{flalign*}
	\end{minipage}%
	\hfill
	\begin{minipage}[t]{0.48\textwidth}
		\flushleft{\textbf{Typing (values and variables)} \quad \jbox{\bs{C}; \sta; \G \vdash_\star t : E}}

		\infrule[\ruledef{rt-var}]
		{x: S\capt \bs{C} \in \G}
		{\bs{\set x}; \sta; \G \vdash_\star x : S\capt\bs{\set x}}

		\infrule[\ruledef{rt-sub}]
		{\bs{C}; \sta; \G \vdash_\star t : E \andalso \sta; \G \vdash E <: F \\[3pt]
			\sta; \G \vdash \bs{C} <: \bs{C'}}
		{\bs{C'}; \sta; \G \vdash_\star t : F}

		\infrule[\textsc{rt-label}]
		{(\ls\ell : S :: \cls{\CLS}) \in \sta \andalso \cls{\CLS \in \KIND}}
		{\bs\emptyset; \sta; \G \vdash_\star \ls\ell : \BREAK[S]\capt \bs{\set{\sproj{\ls\ell}{\cls{\KIND}}}}}
	\end{minipage}

	\medskip

	\begin{multicols}{2}
		\infrule[\ruledef{rt-abs}]
		{\bs{C\uplus \set{x}}; \sta; (\G, x: T) \vdash_\star t : E}
		{\bs{\set{}}; \sta; \G \vdash_\star \lambda^{\bs{C}}(x: T)t : (\forall(x: T)E)\capt \bs{C}}

		\infrule[\ruledef{rt-tabs}]
		{\bs{C}; \sta; (\G, X <: S) \vdash_\star t : E}
		{\bs{\set{}}; \sta; \G \vdash_\star \lambda^{\bs{C}}[X <: S]t : (\forall[X <: S]E)\capt \bs{C}}

		\infrule[\ruledef{rt-cabs}]
		{\bs{C}; \sta; (\G, c: B) \vdash_\star t : E}
		{\bs{\set{}}; \sta; \G \vdash_\star \lambda^{\bs{C}}[c: B]t : (\forall[c: B]E)\capt \bs{C}}

		\infrule[\ruledefN{a-rt-app}{rt-app}]
		{\bs{C_1}; \sta; \G \vdash_\star t_1 : (\forall(z: S\capt \bs{C_0}) E)\capt \bs{C_1} \\[3pt]
			\bs{C_2}; \sta; \G \vdash_\star t_2 : S\capt \bs{C_2} \andalso \sta; \G \vdash \bs{C_2} <: \bs{C_0}}
		{\bs{C_1 \cup C_2}; \sta; \G \vdash_\star t_1\ t_2 : \bs{[C_2/z]}E}

		\infrule[\ruledef{rt-tapp}]
		{\bs{C}; \sta; \G \vdash_\star t : (\forall[X <: S]E)\capt \bs{C} \andalso \sta; \G \vdash R <: S}
		{\bs{C}; \sta; \G \vdash_\star t[R] : [R/X]E}

		\infrule[\ruledef{rt-capp}]
		{\bs{C}; \sta; \G \vdash_\star t : (\forall[c: B]E)\capt \bs{C} \\[3pt]
			\sta; \G \vdash \bs{D} <:_{\mathsf{B}} B}
		{\bs{C}; \sta; \G \vdash_\star t[\bs{D}] : \bs{[D/c]}E}

		\infrule[\ruledef{rt-let}]
		{\bs{C}; \sta; \G \vdash_\star t : T \andalso \bs{C}; \sta; (\G, x: T) \vdash_\star u : E}
		{\bs{C}; \sta; \G \vdash_\star \LET x = t \IN u : E}

		\infrule[\ruledef{rt-ex}]
		{\bs{C}; \sta; \G \vdash_\star t : \exists c: B.\, T \\[3pt]
			\bs{C}; \sta; (\G, c: B, x: T) \vdash_\star u : F}
		{\bs{C}; \sta; \G \vdash_\star \LET \langle c, x \rangle = t \IN u : F}

		\infrule[\ruledefN{a-rt-pack-var}{rt-pack-var}]
		{\bs{C'}; \sta; \G \vdash_\star x : \bs{[D/c]}T \andalso \sta; \G \vdash \bs{D} <:_{\mathsf{B}} B}
		{\bs{\emptyset}; \sta; \G \vdash_\star \PACK\langle \bs D, x\rangle : \exists c: B.\, T}

		\infrule[\ruledefN{a-rt-pack-val}{rt-pack-val}]
		{\bs{\emptyset}; \sta; \G \vdash_\star v : \bs{[D/c]}T \andalso \sta; \G \vdash \bs{D} <:_{\mathsf{B}} B}
		{\bs{\emptyset}; \sta; \G \vdash_\star \PACK\langle \bs D, v\rangle : \exists c: B.\, T}

		\infrule[\ruledef{rt-break}]
		{\bs{C}; \sta; \G \vdash_\star t : \BREAK[S]\capt \bs{C} \\[3pt]
			\bs{C}; \sta; \G \vdash_\star s : S}
		{\bs{C}; \sta; \G \vdash_\star t\ s : R}

		\infrule[\ruledef{rt-bnd}]
		{\bs{C \uplus \set{c, x}}; \sta; (\G, c : \cls{\CLS}, x : \BREAK[S]\capt \bs{\set{c}}) \vdash_\star t : S}
		{\bs{C}; \sta; \G \vdash_\star \BOUNDARY[S, \cls{\CLS}] \AS \langle c, x\rangle \IN t : S}

		\infrule[\ruledef{rt-icp-pass}]
		{\bs{C_t}; \sta; \G \vdash_\star t : E \\[3pt]
			\bs{C_h}; \sta; \G \vdash_\star h : \HANDLER_\mathsf{pass}[E, \cls{\KIND}, \bs{C_h}, \bs{C_t}]}
		{\bs{C_t\PROJ[\cls{\ksub{\top}{\KIND}}] \cup C_h}; \sta; \G \vdash_\star \INTERCEPT[E, \bs{C_t}, \cls{\KIND}]\WITH h \IN t : E}

		\infrule[\ruledef{rt-icp-gen}]
		{\bs{C_t}; \sta; \G \vdash_\star t : E \\[3pt]
			\bs{C_h}; \sta; \G \vdash_\star h : \HANDLER_\mathsf{gen}[E, \cls{\KIND}, \bs{C_h}, \bs{C_t}]}
		{\bs{C_t \cup C_h}; \sta; \G \vdash_\star \INTERCEPT[E, \bs{C_t}, \cls{\KIND}]\WITH h \IN t : E}
	\end{multicols}
	\vspace{-10pt}
	\caption{Relaxed terms and the full relaxed type system.
		Subtyping $\G \vdash E_1 <: E_2$, subcapturing
		$\subs{\G}{\bs{C_1}}{\bs{C_2}}$, capture kinding
		$\typs{\G}{\bs{C}}{\cls{\KIND}}$ and bound subtyping
		$\subb{\G}{B_1}{B_2}$ are as in Figures~\ref{fig:all-typing-full}
		and \ref{fig:kinds-full}, read with the label context $\sta$ and the
		label rules \ruleref{k-label}, \ruleref{k-label-absurd} of
		Figure~\ref{fig:sem-defs4} where labels occur.
		The rules \ruleref{a-rt-app}, \ruleref{a-rt-pack-var} and
		\ruleref{a-rt-pack-val} repeat Figure~\ref{fig:sem-defs4} for
		completeness.}\label{fig:relaxed-full}
\end{figure*}

Figure~\ref{fig:relaxed-full} defines the relaxed language and the complete
relaxed type system.
Relaxed terms extend the source syntax of Figure~\ref{fig:syntax-paper} in two
ways.
First, runtime labels $\ls\ell$ appear as terms, values and captures, exactly
as in the semantic domains of Figure~\ref{fig:bigstep}.
Second, the operand positions of applications, type applications and capture
applications, as well as the bodies of packs, may hold arbitrary terms rather
than variables.
The typing judgment $\bs{C}; \sta; \G \vdash_\star t : E$ carries the label
context $\sta$ in addition to $\G$, since labels are created only at
runtime and their types and classifiers are recorded in $\sta$.
The subtyping, subcapturing and kinding judgments are those of the main text,
used with the extended capture syntax and, where labels occur, with the label
context.
We omit the wellformedness premises of the corresponding MNF rules, they
carry over unchanged.

Most rules are the evident generalizations of their MNF counterparts,
replacing variable operands by typed sub-terms.
Three points deserve attention.

\paragraph{Use sets and capture sets align.}
In \ruleref{a-rt-app}, \ruleref{rt-tapp} and \ruleref{rt-capp} the function
sub-term is typed with a use set that coincides with the capture set of its
function type.
For a variable operand this is derivable by \ruleref{rt-var} and
\ruleref{rt-sub}, so the MNF rules are special cases.
For an arbitrary operand the alignment matters: the same set both types the
evaluation of the operand and bounds the labels charged when the resulting
closure is entered, which is exactly the invariant that checked evaluation
maintains (\Cref{app:full-bigstep}).

\paragraph{Two rules for pack.}
The relaxed system types $\PACK\langle \bs{D}, t\rangle$ by two rules.
\ruleref{a-rt-pack-var} admits a variable body, matching the source-level rule
\ruleref{pack} of Figure~\ref{fig:all-typing-paper}.
\ruleref{a-rt-pack-val} admits an arbitrary value body.
The distinction exists because evaluation never reduces under a pack.
Rule \ruleref{e-ex-v} substitutes the packed term directly into the
continuation of the consuming existential \textsf{let}.
Evaluating the body first could widen its use set and thereby break the
precise existential witness recorded by the pack, so instead typing insists
that the body is already a variable or a value.
Substitution maps variables to values, so this invariant is stable under
evaluation.

\paragraph{Breaks and labels.}
\ruleref{rt-label} types a runtime label at any kind $\cls{\KIND}$ that
contains the classifier recorded in $\sta$.
Labels arise only during evaluation, so the rule has no counterpart in the
main-text figures.
The kind annotation is chosen by the derivation and need not be the singleton
of the recorded classifier.
This flexibility is what lets a label allocated inside an intercepted body
satisfy the handler's capture requirements.
\ruleref{rt-break} generalizes \ruleref{t-break}: a break invocation never
returns, so it can be assigned any wellformed result type.

\subsection{Relating the MNF and Relaxed Systems}\label{app:mnf-relaxed}

We write $\bs{C}; \sta; \G \vdash t : E$ for the MNF typing judgment of the
main text (Figures~\ref{fig:all-typing-paper} and \ref{fig:extension-paper}),
extended pointwise with the label context in the same way as the relaxed
system.
A relaxed term is \emph{in MNF} if every operand of an application, type
application or capture application, and every pack body, is a variable.
Source terms are in MNF by construction.

\begin{lemma}[MNF typing embeds into relaxed typing]\label{lem:mnf-to-relaxed}
	If $\bs{C}; \sta; \G \vdash t : E$ then
	$\bs{C}; \sta; \G \vdash_\star t : E$.
\end{lemma}

\begin{lemma}[Relaxed typing restricts to MNF typing]\label{lem:relaxed-to-mnf}
	If $\bs{C}; \sta; \G \vdash_\star t : E$ and $t$ is in MNF, then
	$\bs{C}; \sta; \G \vdash t : E$.
\end{lemma}

For MNF terms the two systems therefore coincide.
To transport arbitrary relaxed terms back into MNF we use a standard
normalization function.

\begin{definition}[MNF normalization]\label{def:mnf-normalization}
	The function $\widehat{t}$ let-binds every operand that is not already a
	variable and is a homomorphism on all remaining term forms.
	The characteristic equations are
	\begin{align*}
		\widehat{t_1\ t_2}                        & = \LET x = \widehat{t_1} \IN \LET y = \widehat{t_2} \IN x\ y \\
		\widehat{t[S]}                            & = \LET x = \widehat{t} \IN x[S]                              \\
		\widehat{t[\bs{C}]}                       & = \LET x = \widehat{t} \IN x[\bs{C}]                          \\
		\widehat{\PACK\langle \bs{D}, t\rangle}   & = \LET x = \widehat{t} \IN \PACK\langle \bs{D}, x\rangle
	\end{align*}
	with $x$, $y$ fresh, where the extra binding is omitted whenever the
	operand already is a variable.
\end{definition}

\begin{lemma}[Normalization preserves relaxed typing]\label{lem:mnf-normalization}
	$\widehat{t}$ is in MNF, and if
	$\bs{C}; \sta; \G \vdash_\star t : E$ then
	$\bs{C}; \sta; \G \vdash_\star \widehat{t} : E$.
\end{lemma}

Combining Lemma~\ref{lem:mnf-normalization} with
Lemma~\ref{lem:relaxed-to-mnf} shows that every relaxed-typed term normalizes
to an MNF-typed term of the same type and use set.

\subsection{Checked Big-Step Evaluation in Full}\label{app:full-bigstep}

\begin{figure*}[tp]
	\scriptsize

	\flushleft{\textbf{Big-step Evaluation (complete core rules)} \quad \jbox{\sta \vdash t \Downarrow_{\ls{F}} a}}

	\medskip
	\noindent
	\begin{minipage}[t]{0.44\textwidth}
		\infrule[\ruledefN{a-e-val}{e-val}]
		{}
		{\sta \vdash v \Downarrow_{\ls{F}} \langle \sta, v \rangle}
		\medskip
		\bigskip
		\infrule[\ruledefN{a-e-app}{e-app}]
		{\sta \vdash e_1 \Downarrow_{\ls{F}} \langle \sta', \lambda^{\bs D}(x: T)t \rangle \andalso \sta' \vdash e_2 \Downarrow_{\ls{F}} \langle \sta'', v \rangle\\[3pt]
			\ls{D^*_{\sta''}} \subseteq \ls F	\andalso \ls{v^*_{\sta''}} \subseteq \ls F \\[3pt]
			\sta'' \vdash \bs{[\erase{v}/x]}[v/x]t \Downarrow_{\ls{D^*_{\sta''} \cup\ v^*_{\sta''}}} a
		}
		{\sta \vdash e_1 e_2 \Downarrow_{\ls{F}} a}
		\medskip
		\bigskip
		\infrule[\ruledefN{a-e-tapp}{e-tapp}]
		{\sta \vdash e \Downarrow_{\ls{F}} \langle \sta', \lambda^{\bs D}[X <: S_0]t \rangle \andalso \ls{D^*_{\sta'}} \subseteq \ls F\\[3pt]
			\sta' \vdash [S/X]t \Downarrow_{\ls{D^*_{\sta'}}} a
		}
		{\sta \vdash e[S] \Downarrow_{\ls{F}} a}
		\medskip
		\bigskip
		\infrule[\ruledefN{a-e-capp}{e-capp}]
		{\sta \vdash e \Downarrow_{\ls{F}} \langle \sta', \lambda^{\bs D}[c : B]t \rangle \andalso \ls{D^*_{\sta'}} \subseteq \ls F\\[3pt]
			\sta' \vdash \bs{[C/c]}t \Downarrow_{\ls{D^*_{\sta'}}} a
		}
		{\sta \vdash e\bs{[C]} \Downarrow_{\ls{F}} a}
	\end{minipage}%
	\hfill
	\begin{minipage}[t]{0.54\textwidth}
		\infrule[\ruledefN{a-e-let-v}{e-let-v}]
		{\sta \vdash t_1 \Downarrow_{\ls F} \langle \sta', v\rangle \\[3pt]
			\sta' \vdash \bs{[\erase{v}/x]}[v/x]t_2 \Downarrow_{\ls F} a}
		{\sta \vdash \LET x = t_1 \IN t_2 \Downarrow_{\ls F} a}
		\bigskip
		\medskip
		\infrule[\ruledefN{a-e-ex-v}{e-ex-v}]
		{\sta \vdash t_1 \Downarrow_{\ls F} \langle \sta', \PACK\langle \bs D, v\rangle\rangle \\[3pt]
			\sta' \vdash \bs{[D/c]}\bs{[\erase{v}/x]}[v/x]t_2 \Downarrow_{\ls F} a}
		{\sta \vdash \LET \langle c, x \rangle = t_1 \IN t_2 \Downarrow_{\ls F} a}
		\bigskip
		\medskip
		\infrule[\ruledefN{a-e-break}{e-break}]
		{\sta \vdash e_1 \Downarrow_{\ls{F}} \langle \sta', \ls\ell \rangle \andalso \sta' \vdash e_2 \Downarrow_{\ls{F}} \langle \sta'', v \rangle\\[3pt]
			\ls{l} \in \ls F}
		{\sta \vdash e_1 e_2 \Downarrow_{\ls F} \langle \sta'', \BRK(\ls\ell, v) \rangle}
	\end{minipage}

	\bigskip

	\flushleft{\textbf{Break Propagation}}

	\begin{multicols}{2}
		\infrule[\ruledefN{a-e-app-b1}{e-app-b1}]
		{\sta \vdash e_1 \Downarrow_{\ls{F}} \langle \sta', \BRK(\ls\ell, v) \rangle}
		{\sta \vdash e_1 e_2 \Downarrow_{\ls F} \langle \sta', \BRK(\ls\ell, v) \rangle}

		\infrule[\ruledefN{a-e-app-b2}{e-app-b2}]
		{\sta \vdash e_1 \Downarrow_{\ls{F}} \langle \sta', v_1 \rangle \andalso \sta' \vdash e_2 \Downarrow_{\ls{F}} \langle \sta'', \BRK(\ls\ell, v) \rangle}
		{\sta \vdash e_1 e_2 \Downarrow_{\ls F} \langle \sta'', \BRK(\ls\ell, v) \rangle}

		\infrule[\ruledef{e-tapp-b}]
		{\sta \vdash e \Downarrow_{\ls{F}} \langle \sta', \BRK(\ls\ell, v) \rangle}
		{\sta \vdash e[S] \Downarrow_{\ls F} \langle \sta', \BRK(\ls\ell, v) \rangle}

		\infrule[\ruledef{e-capp-b}]
		{\sta \vdash e \Downarrow_{\ls{F}} \langle \sta', \BRK(\ls\ell, v) \rangle}
		{\sta \vdash e\bs{[C]} \Downarrow_{\ls F} \langle \sta', \BRK(\ls\ell, v) \rangle}

		\infrule[\ruledef{e-let-b}]
		{\sta \vdash t_1 \Downarrow_{\ls{F}} \langle \sta', \BRK(\ls\ell, v) \rangle}
		{\sta \vdash \LET x = t_1 \IN t_2 \Downarrow_{\ls F} \langle \sta', \BRK(\ls\ell, v) \rangle}

		\infrule[\ruledef{e-ex-b}]
		{\sta \vdash t_1 \Downarrow_{\ls{F}} \langle \sta', \BRK(\ls\ell, v) \rangle}
		{\sta \vdash \LET \langle c, x \rangle = t_1 \IN t_2 \Downarrow_{\ls F} \langle \sta', \BRK(\ls\ell, v) \rangle}
	\end{multicols}
	\vspace{-10pt}
	\caption{Checked big-step evaluation, complete core rules.
		The first block repeats Figure~\ref{fig:bigstep} for completeness,
		the second block lists the break propagation rules omitted there.
		The boundary and intercept rules of Figure~\ref{fig:bigstep-ext} are
		already complete.}\label{fig:bigstep-full}
\end{figure*}

Figure~\ref{fig:bigstep-full} lists the complete core rules of the checked
big-step semantics, including the break propagation rules that
Figure~\ref{fig:bigstep} elides.
A propagating break simply short-circuits the surrounding evaluation context.
Together with the boundary and intercept rules of
Figure~\ref{fig:bigstep-ext}, these are all evaluation rules of the system.

\paragraph{Where the checks fire.}
The \ls{green} side conditions perform all access checking.
\ruleref{a-e-app} checks that the runtime labels of the closure and of the
argument are within the current allowance and then evaluates the body under
the tightened allowance
$\ls{D^*_{\sta''} \cup v^*_{\sta''}}$.
\ruleref{a-e-tapp} and \ruleref{a-e-capp} check the closure's runtime labels
only, type and capture arguments carry no labels of their own.
\ruleref{a-e-break} checks that the invoked label is permitted, and
\ruleref{e-icp-p} rechecks a passing break against the outer allowance.
Boundary rules extend the allowance with the fresh label for the extent of
the body, and \ruleref{e-icp-v}, \ruleref{e-icp-m} and \ruleref{e-icp-p} run
the intercepted body under the allowance
$\ls{(C_t)^*_\sta}$ induced by its static use set.

\paragraph{Usage is checked, capture is not.}
No rule inspects the labels captured inside a closure that is merely passed
around.
A closure may travel through contexts whose allowance excludes its captured
labels.
The check fires when the closure is entered, at which point its annotation
must be within the allowance.
This is the operational counterpart of the distinction between capture sets
and use sets in the type system.

\paragraph{Unchecked evaluation.}
Deleting the allowance subscripts and every \ls{green} side condition of the
forms $\ls{U} \subseteq \ls{F}$ and $\ls{l} \in \ls{F}$ yields an ordinary
call-by-value big-step semantics for the relaxed language, written
$\sta \vdash t \Downarrow a$.
The corollaries of \Cref{app:corollaries} relate the two judgments: for
well-typed terms the checks never fail, so checked and unchecked evaluation
agree.

\subsection{Safety}\label{app:safety}

Answers are typed by the judgment
$\ls{F}; \sta; \G \vDash_\mathsf{a} a : E$ of Figure~\ref{fig:sem-defs4}.
A returned value must be relaxed-typed at $E$ with an empty use set
(\ruleref{sa-ret}).
A propagating break $\BRK(\ls\ell, v)$ must invoke a label that is bound in
$\sta$ and permitted by the allowance $\ls{F}$, carrying a payload of the
label's answer type (\ruleref{sa-brk}).

Throughout, we use the convention that the label context only grows during
evaluation, and we identify labels across this growth.
The mechanization instead threads explicit injective renamings between label
contexts, see \Cref{app:lean-diff}.

\begin{theorem}[Safety of Checked Evaluation]\label{thm:safety-relaxed}
	If\;
	$\bs{C}; \sta; \emptyset \vdash_\star t : E$\;
	and\; $\ls{C^*_\sta} \subseteq \ls{F}$\;
	and\; $\sta \vdash t \Downarrow_{\ls{F}}^? a$,\;
	then\; evaluation does not get stuck\;
	and\; $\exists \mathit{ans}, \sta' \supseteq \sta$\;
	such that\; $a = \langle \sta', \mathit{ans} \rangle$\;
	and\; $\ls{F}; \sta'; \emptyset \vDash_\mathsf{a} \mathit{ans} : E$.
\end{theorem}

The proof is by induction on the evaluation derivation, using inversion of
the relaxed typing judgment in each case and the substitution and renaming
lemmas of the mechanization to push typing through the value substitutions
performed by the rules.
The full development is mechanized (theorem \texttt{Eval.safety},
\Cref{app:mechanization}).

Transporting the theorem along the round trip of
\Cref{app:mnf-relaxed} yields safety for the MNF system of the main text.

\begin{theorem}[MNF Safety]\label{thm:safety-mnf}
	If\;
	$\bs{C}; \sta; \emptyset \vdash t : E$\;
	and\; $\ls{C^*_\sta} \subseteq \ls{F}$\;
	and\; $\sta \vdash t \Downarrow_{\ls{F}} \langle \sta', \mathit{ans} \rangle$,\;
	then $\sta' \supseteq \sta$ and
	\begin{enumerate}
		\item if $\mathit{ans} = v$ then
		      $\bs{\emptyset}; \sta'; \emptyset \vdash \widehat{v} : E$,
		\item if $\mathit{ans} = \BRK(\ls\ell, v)$ then $\ls{\ell} \in \ls{F}$,
		      $(\ls\ell : S :: \cls{\CLS}) \in \sta'$, and
		      $\bs{\emptyset}; \sta'; \emptyset \vdash \widehat{v} : S$.
	\end{enumerate}
\end{theorem}

Starting from an MNF-typed term, evaluation produces an answer whose payload
normalizes to an MNF-typed value of the expected type.
The main text states the relaxed version as Theorem~\ref{thm:safety} in
Section~\ref{sec:metatheory}.

\subsection{Corollaries}\label{app:corollaries}

The safety theorem yields the runtime guarantees stated in
Section~\ref{sec:metatheory}.
We restate them here in full.

\begin{corollary}[Effect Safety]\label{cor:effect-safety-full}
	If $\bs{\emptyset}; \emptyset; \emptyset \vdash_\star t : T \capt \bs{\emptyset}$
	and $\emptyset \vdash t \Downarrow_{\ls{\emptyset}} \langle \sta', \mathit{ans} \rangle$,
	then $\mathit{ans} = v$ for some value $v$.
\end{corollary}

A well-typed closed program with empty capture always returns a value, no
break goes unhandled.

\begin{corollary}[Used Label Prediction]\label{cor:use-prediction}
	If\;
	$\bs{C}; \sta; \emptyset \vdash t : E$\;
	and\; $\sta \vdash t \Downarrow_{\ls{C^*_\sta}}^?\ a$,
	then evaluation does not get stuck.
\end{corollary}
That is, evaluation of $t$ relies only on access to runtime labels of $\bs C$.
In other words, the typing judgment's use set correctly predicts the set of labels used during evaluation of $t$.

\begin{corollary}[Capture Prediction]\label{cor:capture-prediction}
	If $\bs{D}; \sta; \emptyset \vdash t : S\capt\bs C$
	and $\sta \vdash t \Downarrow_{\ls{D^*_\sta}} \langle \sta', v \rangle$, then
	\[
		\sta'; \emptyset \vdash \bs{\lfloor v \rfloor} <: \bs{C}
		\qquad\text{and}\qquad
		\ls{\lfloor v \rfloor^*_{\sta'}} \subseteq \ls{C^*_{\sta'}}
	\]
\end{corollary}

Recall that a potential use set \(\bs{\lfloor v \rfloor}\) bounds the captures that a value reaches upon
usage. We can predict the potential use set of the answer returned by
the evaluation of a term from its static typing judgment.
This is similar to the Capture Prediction for Terms lemma in $\text{CC}_{<:\square}$ \cite{boruchgruszecki2023capturing}.

\begin{corollary}[Closure Tightness]\label{cor:closure-tightness}
	If
	$\bs{\emptyset}; \sta; \emptyset \vdash_\star v_f : (\forall(z: T\capt \bs{C_1})E)\capt \bs{C_0}$,\;
	$\bs{\emptyset}; \sta; \emptyset \vdash_\star v_x : T\capt \bs{C_2}$,\;
	$\sta; \emptyset \vdash \bs{C_2} <: \bs{C_1}$,
	and $\sta \vdash v_f\ v_x \Downarrow a$,
	then $\sta \vdash v_f\ v_x \Downarrow_{\ls{(C_0)^*_\sta \cup\, (C_2)^*_\sta}} a$.
\end{corollary}

Applying a closure to a fitting argument needs access to exactly the labels
predicted by the two capture sets and nothing more.
A closure's annotation is thus a tight interface to its runtime behavior.

\begin{corollary}[Handler Coverage]\label{cor:handler-coverage-full}
	If $\bs{D}; \sta; \G \vdash t : E$ and $\sta \vdash t \Downarrow_{\ls{D^*_\sta}} \langle \sta', \BRK(\ls\ell, v) \rangle$, then:
	\begin{enumerate}
		\item \emph{Breaks.} \(\ls\ell \in \ls{D^*_{\sta'}}\).
		\item \emph{Intercepts.}
		      If \(t = \INTERCEPT[R, \bs{C_t}, \cls\KIND] \WITH h \IN t'\), and \((\ls\ell : S :: \cls\CLS) \in \sta\),
		      then either \(h\) was evaluated, or \(\cls\CLS \in \cls{\ksub{\top}\KIND}\).
	\end{enumerate}
\end{corollary}

No label can be used at runtime that was not statically accounted for.
An escaping break was visible in the use set, and an unhandled label passing through an $\INTERCEPT$
fits the required shape. Note that the second point is trivial from the evaluation rules.

\begin{corollary}[Boundary Safety]\label{cor:boundary-safety}
	If $\bs{C}; \sta; \emptyset \vdash_\star \BOUNDARY[S, \cls{\CLS}] \AS \langle c, x\rangle \IN t : E$
	and $\sta \vdash \BOUNDARY[S, \cls{\CLS}] \AS \langle c, x\rangle \IN t \Downarrow_{\ls{C^*_\sta}} \langle \sta', v \rangle$,
	then the fresh label $\ls\ell$ allocated by the boundary satisfies
	$\ls{\ell} \notin \ls{\erase{v}^*_{\sta'}}$.
\end{corollary}

A boundary label is local to its scope.
It cannot occur in the potential use set of any value that leaves the
boundary, even nested inside closures.

\begin{corollary}[Pass-Handler Safety]\label{cor:pass-handler-safety}
	Let $t = \INTERCEPT[R, \bs{C_t}, \cls{\KIND}] \WITH h \IN t_0$ be typed by
	\ruleref{rt-icp-pass} with use set
	$\bs{C} = \bs{C_t\PROJ[\cls{\ksub{\top}{\KIND}}] \cup C_h}$,
	and suppose the handler's captures exclude the intercepted kind,
	$\sta; \emptyset \vdash \bs{C_h} <: \bs{C_h\PROJ[\cls{\ksub{\top}{\KIND}}]}$.
	If $\sta \vdash t \Downarrow_{\ls{C^*_\sta}} \langle \sta', \BRK(\ls\ell, v) \rangle$
	with $(\ls\ell : S :: \cls{\CLS}) \in \sta'$,
	then $\cls{\CLS \notin \KIND}$.
\end{corollary}

When a pass handler does not itself retain labels of the intercepted kind,
the intercept is exhaustive for that kind.
No break classified under $\cls{\KIND}$ can escape it.

\section{The Lean 4 Mechanization}\label{app:mechanization}

All definitions and results of \Cref{app:formal-system} are mechanized in
Lean~4.
The development is called \textsc{CaplessK}, comprises about 10\,000 lines of
Lean in 59 files, and is free of \texttt{sorry} placeholders and of axioms
beyond Lean's kernel.
This section is a guide for readers who want to relate the pencil and paper
formulation to the mechanized one.

\subsection{Proof Structure}\label{app:lean-structure}

The central mechanized result is type safety for MNF terms, obtained by the
round trip described in \Cref{app:formal-system}.

\begin{enumerate}
	\item \textbf{MNF typing to relaxed typing.}
	      The inductive \texttt{MNF.HasTyp} (\texttt{MNF/Typing.lean})
	      mirrors the typing figures of the main text.
	      Lemma~\ref{lem:mnf-to-relaxed} is
	      \texttt{HasTyp.from\_mnf\_typing}: every \texttt{MNF.HasTyp}
	      derivation is a \texttt{HasTyp} derivation of the relaxed system
	      (\texttt{Typing.lean}).
	\item \textbf{Type-safe evaluation.}
	      The fuel-indexed evaluator \texttt{Term.eval} (\texttt{Eval.lean})
	      implements checked big-step evaluation of relaxed terms.
	      Theorem~\ref{thm:safety-relaxed} is \texttt{Eval.safety}
	      (\texttt{Safety.lean}), proved by well-founded induction on fuel
	      with the four invariants of the theorem statement maintained
	      through every recursive call.
	      Each case applies an inversion lemma
	      (\texttt{Inversion/Typing.lean}) to decompose the typing
	      derivation, the induction hypothesis to each sub-evaluation, and
	      the structural lemmas \texttt{HasTyp.subst} and
	      \texttt{HasTyp.rename} to reassemble the result.
	\item \textbf{Back to MNF.}
	      The normalization function \texttt{Term.as\_mnf}
	      (\texttt{MNF/Core.lean}) implements
	      Definition~\ref{def:mnf-normalization}.
	      Lemma~\ref{lem:mnf-normalization} combines
	      \texttt{Term.as\_mnf.is\_mnf} and \texttt{HasTyp.as\_mnf}, and
	      Lemma~\ref{lem:relaxed-to-mnf} is
	      \texttt{MNF.HasTyp.from\_regular\_typing}.
	      Chaining all three steps gives Theorem~\ref{thm:safety-mnf} as
	      \texttt{MNF.safety} (\texttt{MNF/Basic.lean}).
\end{enumerate}

The corollaries of \Cref{app:corollaries} are derived from
\texttt{Eval.safety} in \texttt{Safety/Corollaries.lean}.

\subsection{Substitutions}\label{app:lean-subst}

Every substitution performed by the evaluation rules is an instance of a
single structure \texttt{Subst} (\texttt{Subst.lean}) that bundles four
simultaneous maps.
\begin{description}\itemsep1pt
	\item[\normalfont$\mathtt{rebind}$] sends a term variable to a term.
	\item[\normalfont$\mathtt{rebind\_capt}$] sends a term variable to a
	      capture set, its capture annotation.
	\item[\normalfont$\mathtt{trebind}$] sends a type variable to a shape type.
	\item[\normalfont$\mathtt{crebind}$] sends a capture variable to a capture
	      set.
\end{description}
The two maps on term variables are separate because a term variable $x$
occurs in two syntactic roles, as a term and as the capture annotation
$\bs{\set{x}}$ inside types and capture sets.
Both roles must be rewritten together to preserve typing, which is why the
evaluation rules of \Cref{app:full-bigstep} always pair a term substitution
$[v/x]$ with a capture substitution $\bs{[\erase{v}/x]}$.
The named substitutions used by the rules are the following instances.

\begin{description}\itemsep2pt
	\item[\normalfont\texttt{Subst.open} $v\ \bs{C}$]
	      Term variable $0$ becomes $v$ as a term and $\bs{C}$ as an
	      annotation, $\mathtt{rebind}(0) = v$ and
	      $\mathtt{rebind\_capt}(0) = \bs{C}$.
	      This is the substitution of \ruleref{e-app}, \ruleref{e-let-v} and
	      the invoke rules, with $\bs{C}$ the effective use set of $v$.
	\item[\normalfont\texttt{Subst.exlet\_open} $\bs{D}\ v\ \bs{C}$]
	      Additionally sends capture variable $0$ to the witnessed set,
	      $\mathtt{crebind}(0) = \bs{D}$, and is used by \ruleref{e-ex-v}.
	\item[\normalfont\texttt{Subst.boundary\_open}]
	      At boundary entry a fresh label $\ls\ell$ replaces the break-return
	      variable $x$ both as a term and as its annotation, and the
	      classifier variable $c$, so
	      $\mathtt{rebind}(0) = \ls\ell$,
	      $\mathtt{rebind\_capt}(0) = \bs{\set{\ls\ell}}$ and
	      $\mathtt{crebind}(0) = \bs{\set{\ls\ell}}$.
	      This is the $\bs{[\ls\ell/c]}\bs{[\ls\ell/x]}[\ls\ell/x]$ of the
	      boundary rules.
	\item[\normalfont\texttt{Subst.handler\_open} $\ls\ell\ \cls{\KIND}\ S\ v$]
	      At an intercept match the four handler binders are filled in,
	      the break value $y \mapsto v$, the break label $b \mapsto \ls\ell$,
	      the shape variable $X \mapsto S$ and the capture set 
	      $c \mapsto \bs{\set{\sproj{\ls\ell}{\cls{\KIND}}}}$, matching the
	      substitution in \ruleref{e-icp-m}.
\end{description}

\noindent
Narrowing is not a separate lemma.
The three instances \texttt{CtxSubst.narrow}, \texttt{.tnarrow} and
\texttt{.cnarrow} express it as an identity substitution from a context with
a weaker bound into the same context with a stronger bound, so replacing a
bound by a subtype follows from the substitution lemma below.

\subsection{Structural and Inversion Lemmas}\label{app:lean-structural}

The structural lemmas are stated through two predicates that abstract the
notion of a well-typed context morphism.
$\mathsf{CtxRename}(\G, f, \Delta)$ (\texttt{Renaming/Morphism.lean}) holds
when the renaming $f$ carries every term, type, capture and label binding of
$\G$ to the corresponding binding of $\Delta$.
$\mathsf{CtxSubst}(\G, s, \Delta)$ (\texttt{Subst/Morphism.lean}) holds when
the substitution $s$ is admissible from $\G$ to $\Delta$, meaning every
substituted term is well-typed at the substituted binding type, capture
annotations remain consistent under subcapturing, and type and capture
variable images satisfy the respective subtyping and bound constraints.
Admissibility carries one extra requirement, \texttt{var\_or\_val}, that each
substituted term is a variable or a value.
The safety proof relies on it because it substitutes evaluation results,
which are values, without a term store.

\begin{lemma}[Renaming, \texttt{HasTyp.rename}]\label{lem:rename}
	If $\bs{C}; \sta; \G \vdash_\star t : E$ and
	$\mathsf{CtxRename}(\G, f, \Delta)$, then
	$\bs{C[f]}; \sta; \Delta \vdash_\star t[f] : E[f]$.
	The analogue holds for subtyping, subcapturing and kinding.
\end{lemma}

\begin{lemma}[Substitution, \texttt{HasTyp.subst}]\label{lem:subst}
	If $\bs{C}; \sta; \G \vdash_\star t : E$ and
	$\mathsf{CtxSubst}(\G, s, \Delta)$, then
	$\bs{C[s]}; \sta; \Delta \vdash_\star t[s] : E[s]$.
	The analogue holds for subtyping, subcapturing and kinding.
\end{lemma}

Inversion lemmas (\texttt{Inversion/Typing.lean},
\texttt{Inversion/Subtyping.lean}) decompose a typing or subtyping derivation
for a fixed syntactic form into its premises up to subsumption.
Table~\ref{tab:inversion} lists them.
Two further lemmas single out values.
\texttt{HasTyp.value\_use\_set\_is\_empty} states that a well-typed value can
always be retyped with an empty use set, since a value uses no labels at the
time it is formed.
\texttt{HasTyp.tighten\_value} (\texttt{Eval/Props.lean}) states that a value
typed at $S \capt \bs{C}$ can be retyped at $S \capt \bs{\erase{v}}$ with
$\sta; \G \vdash \bs{\erase{v}} <: \bs{C}$, so its potential use set
$\bs{\erase{v}}$ is the tightest capture annotation it admits.
These are the lemmas behind Closure Tightness
(Corollary~\ref{cor:closure-tightness}) and Boundary Safety
(Corollary~\ref{cor:boundary-safety}).

\begin{table}[htbp]
	\caption{Inversion lemmas of the mechanization.}\label{tab:inversion}
	\footnotesize
	\begin{tabular}{@{}ll@{}}
		\toprule
		\textbf{Lemma} & \textbf{Form inverted} \\
		\midrule
		\texttt{HasTyp.var\_inv}       & $\bs{C}; \sta; \G \vdash_\star x : E$ \\
		\texttt{HasTyp.abs\_inv}       & $\bs{C}; \sta; \G \vdash_\star \lambda^{\bs{C_0}}(x{:}T)\,t : E$ \\
		\texttt{HasTyp.tabs\_inv}      & $\bs{C}; \sta; \G \vdash_\star \lambda^{\bs{C_0}}[X{<:}S]\,t : E$ \\
		\texttt{HasTyp.cabs\_inv}      & $\bs{C}; \sta; \G \vdash_\star \lambda^{\bs{C_0}}[c{:}B]\,t : E$ \\
		\texttt{HasTyp.app\_inv}       & $\bs{C}; \sta; \G \vdash_\star t\ s : E$ (application or invoke) \\
		\texttt{HasTyp.tapp\_inv}      & $\bs{C}; \sta; \G \vdash_\star t[S] : E$ \\
		\texttt{HasTyp.capp\_inv}      & $\bs{C}; \sta; \G \vdash_\star t[\bs{D}] : E$ \\
		\texttt{HasTyp.lett\_inv}      & $\bs{C}; \sta; \G \vdash_\star \LET x = t \IN u : E$ \\
		\texttt{HasTyp.exlet\_inv}     & $\bs{C}; \sta; \G \vdash_\star \LET \langle c, x\rangle = t \IN u : E$ \\
		\texttt{HasTyp.pack\_inv}      & $\bs{C}; \sta; \G \vdash_\star \PACK\langle \bs{D}, t\rangle : E$ \\
		\texttt{HasTyp.boundary\_inv}  & $\bs{C}; \sta; \G \vdash_\star \BOUNDARY[S, \cls{\CLS}] \dots : E$ \\
		\texttt{HasTyp.intercept\_inv} & $\bs{C}; \sta; \G \vdash_\star \INTERCEPT[\dots] \WITH h \IN t : E$ \\
		\texttt{HasTyp.label\_inv}     & $\bs{C}; \sta; \G \vdash_\star \ls\ell : E$ \\
		\midrule
		\texttt{SSubtyp.top\_inv}      & $\sta; \G \vdash \top <:_s S$ \\
		\texttt{SSubtyp.break\_inv}    & $\sta; \G \vdash \BREAK[S_1] <:_s S$ \\
		\texttt{SSubtyp.ffun\_inv}     & $\sta; \G \vdash (\forall(x{:}T_1)E_1) <:_s S$ \\
		\texttt{SSubtyp.tfun\_inv}     & $\sta; \G \vdash (\forall[X{<:}S_1]E_1) <:_s S$ \\
		\texttt{SSubtyp.cfun\_inv}     & $\sta; \G \vdash (\forall[c{:}B_1]E_1) <:_s S$ \\
		\bottomrule
	\end{tabular}
\end{table}

\subsection{Semantic Predicates}\label{app:lean-sem}

Three predicates carry the invariants of the safety proof.
The mechanization separates the runtime store $\varphi$, which records the
shape type and classifier of each allocated label, from the static context
$\G$.
In the notation of \Cref{app:formal-system} the store $\varphi$ and the
label bindings of $\G$ are the single label context $\sta$.

\begin{description}\itemsep2pt
	\item[\normalfont$\mathsf{SemStore}(\G, \varphi)$
	      (\texttt{Store/Props.lean})]
	      The store is consistent with the context, every label $\ls\ell$
	      satisfies $(\ls\ell : \varphi.\mathsf{type}(\ls\ell) ::
	      \varphi.\mathsf{class}(\ls\ell)) \in \G$.
	\item[\normalfont$\mathsf{TypedAnswer}(\G, \varphi, \ls{F}, a, E)$
	      (\texttt{Sem.lean})]
	      The answer is well-typed.
	      A value answer $v$ satisfies
	      $\bs{\emptyset}; \G \vdash_\star v : E$.
	      A break answer $\BRK(\ls\ell, v)$ satisfies $\ls\ell \in \ls{F}$ and
	      $\bs{\emptyset}; \G \vdash_\star v : \varphi.\mathsf{type}(\ls\ell)$,
	      typing the payload at the label's stored shape type with empty
	      capture.
	      This is the answer typing $\ls{F}; \sta; \G \vDash_\mathsf{a} a : E$
	      of Figure~\ref{fig:sem-defs4}.
	\item[\normalfont$\mathsf{TypedRes}(\G, \varphi, \ls{F}, \mathit{res}, E)$
	      (\texttt{Sem.lean})]
	      The evaluation result is well-typed up to a label renaming $f$,
	      there are $\G'$ and $\varphi'$ with
	      $\mathit{res} = \mathsf{ok}(f, a, \varphi')$,
	      $\mathsf{CtxRename}(\G, f, \G')$,
	      $\mathsf{SemStore}(\G', \varphi')$, and
	      $\mathsf{TypedAnswer}(\G', \varphi', f(\ls{F}), a, E[f])$.
	      In particular $\mathit{res} \neq \mathsf{stuck}$.
\end{description}

The renaming $f$ accommodates the fresh labels allocated by boundaries,
which grow the label index space, and is the mechanized counterpart of the
paper convention that the label context only grows.

\subsection{Type Safety}\label{app:lean-safety}

In Theorem~\ref{thm:safety-relaxed} the single label context $\sta$ serves
at once as the static context of the typing judgment and the runtime store of
the evaluator, so their agreement holds by construction.
The mechanization separates the static context $\G$ from the runtime store
$\varphi$, so \texttt{Eval.safety} (\texttt{Safety.lean}) carries the extra
store-consistency premise $\mathsf{SemStore}(\G, \varphi)$, premise~(4) below,
that Theorem~\ref{thm:safety-relaxed} leaves implicit.
Its conclusion $\mathsf{TypedRes}$ (\Cref{app:lean-sem}) likewise replaces the
growing-context convention $\sta' \supseteq \sta$ with an explicit label
renaming and records that evaluation does not get stuck.
Otherwise the statement is Theorem~\ref{thm:safety-relaxed}.
\[
	\begin{array}{c}
		\underbrace{\varphi \vdash t \Downarrow_{\ls{F}} \mathit{res}}_{\text{(1) evaluation terminates}}
		\;\wedge\;
		\underbrace{\bs{C}; \G \vdash_\star t : E}_{\text{(2) $t$ is well typed}}
		\\[2.5ex]
		{}\wedge\;
		\underbrace{\ls{C^*_\varphi} \subseteq \ls{F}}_{\text{(3) the allowance covers the use set}}
		\;\wedge\;
		\underbrace{\mathsf{SemStore}(\G, \varphi)}_{\text{(4) store consistent with $\G$}}
		\\[2.5ex]
		\Longrightarrow\;
		\mathsf{TypedRes}(\G, \varphi, \ls{F}, \mathit{res}, E)
	\end{array}
\]
Premise (1) is expressed as
$\mathtt{eval}(t, \varphi, \ls{F}, \mathit{fuel}) = \mathtt{some}(\mathit{res})$
for some finite fuel, and the proof is by well-founded induction on that
fuel, maintaining all four premises across every recursive call.
Each case runs in three steps.
\begin{enumerate}\itemsep2pt
	\item \textbf{Inversion.}
	      The inversion lemma for the term's form
	      (Table~\ref{tab:inversion}) decomposes premise (2) into sub-term
	      typings, and premise (3) is propagated to the sub-terms using
	      subcapturing and the structure of the evaluation rule.
	\item \textbf{Induction.}
	      The induction hypothesis applies to each sub-evaluation at strictly
	      smaller fuel, yielding a $\mathsf{TypedRes}$ for each sub-result.
	\item \textbf{Reassembly.}
	      The sub-results recombine into a $\mathsf{TypedRes}$ for the whole
	      term.
	      \texttt{HasTyp.subst} (Lemma~\ref{lem:subst}) discharges the value
	      substitutions of \ruleref{e-app}, \ruleref{e-let-v},
	      \ruleref{e-ex-v} and the boundary and intercept rules, and
	      \texttt{HasTyp.rename} (Lemma~\ref{lem:rename}) transports typing
	      across the label growth at \ruleref{e-bnd-v}, \ruleref{e-bnd-c} and
	      \ruleref{e-bnd-p}.
\end{enumerate}
Chaining the round trip of \Cref{app:mnf-relaxed} around this theorem gives
\texttt{MNF.safety}, the mechanized form of Theorem~\ref{thm:safety-mnf}.

\subsection{Module Overview}\label{app:lean-modules}

\begin{table}[htbp]
	\caption{Modules of the \textsc{CaplessK} development.}\label{tab:lean-modules}
	\small
	\begin{tabular}{@{}ll@{}}
		\toprule
		\textbf{Module}                       & \textbf{Contents}                                                    \\
		\midrule
		\texttt{Classifier/*}                 & classifier tree, kinds, intersection, subtraction,                   \\
		                                      & subkinding, disjointness, inclusion, set semantics                   \\
		\texttt{CaptureSet}, \texttt{CaptureSet/*} & capture sets, projection, renaming, substitution                \\
		\texttt{Term}, \texttt{Term/*}        & de Bruijn terms, values, potential use sets                          \\
		\texttt{Type}, \texttt{Type/*}        & shape, capturing and existential types                               \\
		\texttt{Context}                      & typing contexts with four binding forms                              \\
		\texttt{Subcapt}                      & subcapturing and capture kinding                                     \\
		\texttt{Subtyping}, \texttt{Subtyping/*} & subtyping and its transitivity                                    \\
		\texttt{Typing}                       & the relaxed type system (\texttt{HasTyp})                            \\
		\texttt{MNF/*}                        & MNF predicate, normalization, MNF typing, MNF safety                 \\
		\texttt{Renaming/*}, \texttt{Subst/*} & context morphisms and the structural lemmas                          \\
		\texttt{Inversion/*}                  & inversion lemmas for typing and subtyping                            \\
		\texttt{Store}, \texttt{Store/*}      & label stores and their consistency with contexts                     \\
		\texttt{Filter}, \texttt{Filter/*}    & runtime label allowances                                             \\
		\texttt{Eval}, \texttt{Eval/*}        & the checked big-step evaluator and its properties                    \\
		\texttt{Sem}                          & answer and result typing                                             \\
		\texttt{Safety}, \texttt{Safety/*}    & the safety theorem and its corollaries                               \\
		\texttt{Try}                          & the mechanized \textsf{Try} encoding of Section~\ref{sec:calculus}   \\
		\texttt{FinFun}, \texttt{Set}, \texttt{Tactics} & infrastructure                                             \\
		\bottomrule
	\end{tabular}
\end{table}

Table~\ref{tab:lean-modules} maps the modules of the development to their
contents.
All paths are relative to the \texttt{CaplessK/} source directory.

\subsection{Differences from the Paper Presentation}\label{app:lean-diff}

The mechanization makes nine representation choices that differ from the
pencil and paper formulation.
None of them affects the meaning of the theorems.

\begin{itemize}
	\item \textbf{de Bruijn indices.}
	      Terms are intrinsically scoped over a quadruple index counting
	      term variables, type variables, capture variables and labels.
	      Wellformedness premises of the paper rules are subsumed by this
	      indexing and do not appear in the mechanization.
	\item \textbf{Labels live in the context.}
	      The paper separates the label context $\sta$ from $\G$.
	      The mechanization uses a single context with label bindings, and a
	      predicate \texttt{SemStore} states that these bindings agree with
	      the runtime label store.
	      In the paper formulation this agreement holds by construction
	      because the same $\sta$ appears in typing and evaluation.
	\item \textbf{Explicit label renamings.}
	      Where the paper lets the label context grow monotonically, the
	      evaluator returns an injective renaming between the initial and
	      final label index spaces, and the safety proof transports typing
	      along it (\texttt{CtxRename}, \texttt{HasTyp.rename}).
	\item \textbf{Functional, fuel-indexed evaluation.}
	      \texttt{Term.eval} is a total function on a fuel argument
	      returning either an answer or an explicit \texttt{stuck} marker.
	      The relational judgment
	      $\sta \vdash t \Downarrow_{\ls{F}} a$ of the paper corresponds
	      to \texttt{Term.eval} returning an answer at some fuel, and
	      Corollary~\ref{cor:capture-prediction} appears as the statement
	      that evaluation of a well-typed term never returns
	      \texttt{stuck}.
	\item \textbf{Annotated applications and packing.}
	      Application nodes in the mechanization carry the use set of the
	      argument, and intercept nodes carry the static capture sets of
	      handler and body.
	      The paper instead reads the potential use set $\bs{\erase{v}}$
	      off the evaluated operand.
	      The typing rules constrain the stored annotations to coincide
	      with the sets the paper computes, so the two formulations check
	      the same labels.

        Similarly, packing stores the capture set of the value during typing
        as part of its term, which gets substituted during unpacking instead
        of \(\bs{\erase{v}}\).
	\item \textbf{Handlers as open terms.}
	      The mechanization types the handler body in a context extended
	      with its four binders, whereas the paper packages the handler as
	      a value of the $\HANDLER$ type abbreviation.
	      The two presentations are interderivable by abstracting and
	      applying the handler.
	\item \textbf{Substitutions in bulk.}
	      A single structure \texttt{Subst} bundles the term, capture
	      annotation, type and capture variable substitutions, with
	      admissibility captured by the \texttt{CtxSubst} predicate.
	      The simultaneous substitutions of the evaluation rules, such as
	      $\bs{[\erase{v}/x]}[v/x]$, are single instances of this
	      structure, and one lemma \texttt{HasTyp.subst} covers all of
	      them.
	      Narrowing is the special case of an identity substitution into a
	      context with a stronger bound.
	\item \textbf{Split subtyping.}
	      The single subtyping judgment of
	      Figure~\ref{fig:all-typing-full} is split into three mutually
	      defined judgments for shape, capturing and existential types
	      (\texttt{SSubtyp}, \texttt{CSubtyp}, \texttt{ESubtyp}).
	\item \textbf{Algorithmic kinding.}
	      The mechanized capture kinding has no counterpart of
	      \ruleref{k-sub}, subkind reasoning is inlined into the rule for
	      kind-bounded capture variables, and the general rule is derived
	      as a lemma (\texttt{CaptureKind.sub}).
	      Similarly, \ruleref{sc-proj} is mechanized for singleton sets
	      only, with the general form derived by induction.
\end{itemize}

\subsection{Correspondence of Rules and Theorems}\label{app:lean-tables}

\begin{table}[htbp]
	\caption{Correspondence between paper judgments and Lean types.}\label{tab:lean-judgments}
	\small
	\begin{tabular}{@{}lll@{}}
		\toprule
		\textbf{Paper}                                                & \textbf{Lean type}      & \textbf{Source file}     \\
		\midrule
		$\cls{\CLS_1 \sqsubseteq \CLS_2}$                             & \texttt{Classifier.Subclass} & \texttt{Classifier/Core.lean} \\
		$\cls{\KIND_1 \land \KIND_2}$, $\cls{\ksub{\KIND_1}{\KIND_2}}$ & \texttt{Kind.Intersect}, \texttt{Kind.Subtract} & \texttt{Classifier/*.lean} \\
		$\cls{\KIND\EMPTY}$, $\cls{\subk{\KIND_1}{\KIND_2}}$, $\cls{\CLS \in \KIND}$ & \texttt{Kind.IsEmpty}, \texttt{Kind.Subkind}, \texttt{Kind.Contains} & \texttt{Classifier/*.lean} \\
		$\bs{C_1} \sqsubseteq \bs{C_2}$                               & \texttt{CaptureSet.Subset} & \texttt{CaptureSet.lean} \\
		$\sta; \G \vdash \bs{C} : \cls{\KIND}$                      & \texttt{CaptureKind}    & \texttt{Subcapt.lean}    \\
		$\sta; \G \vdash \bs{C_1} <: \bs{C_2}$                      & \texttt{Subcapt}        & \texttt{Subcapt.lean}    \\
		$\G \vdash B_1 <:_{\mathsf{B}} B_2$                                      & \texttt{Subbound}       & \texttt{Subtyping.lean}  \\
		$\sta; \G \vdash E_1 <: E_2$                                & \texttt{SSubtyp}, \texttt{CSubtyp}, \texttt{ESubtyp} & \texttt{Subtyping.lean} \\
		$\bs{C}; \sta; \G \vdash t : E$ (MNF)                       & \texttt{MNF.HasTyp}     & \texttt{MNF/Typing.lean} \\
		$\bs{C}; \sta; \G \vdash_\star t : E$                       & \texttt{HasTyp}         & \texttt{Typing.lean}     \\
		$t$ in MNF, $\widehat{t}$                                     & \texttt{Term.IsMNF}, \texttt{Term.as\_mnf} & \texttt{MNF/Core.lean} \\
		$\sta \vdash t \Downarrow_{\ls{F}} a$                       & \texttt{Term.eval}      & \texttt{Eval.lean}       \\
		$\ls{C^*_\sta}$, $\bs{\erase{v}}$                           & \texttt{CaptureSet.runtime\_labels}, \texttt{Value.use\_set} & \texttt{Eval.lean}, \texttt{Term/Ops.lean} \\
		$\ls{F}; \sta; \G \vDash_\mathsf{a} a : E$                  & \texttt{TypedAnswer}    & \texttt{Sem.lean}        \\
		\bottomrule
	\end{tabular}
\end{table}

\begin{table}[htbp]
	\caption{Correspondence between paper results and Lean theorems.}\label{tab:lean-theorems}
	\small
	\begin{tabular}{@{}lll@{}}
		\toprule
		\textbf{Result}                                & \textbf{Lean theorem}                & \textbf{Source file}          \\
		\midrule
		Lemma~\ref{lem:mnf-to-relaxed}                 & \texttt{HasTyp.from\_mnf\_typing}    & \texttt{MNF/Typing.lean}      \\
		Lemma~\ref{lem:relaxed-to-mnf}                 & \texttt{MNF.HasTyp.from\_regular\_typing} & \texttt{MNF/Typing.lean} \\
		Lemma~\ref{lem:mnf-normalization}              & \texttt{Term.as\_mnf.is\_mnf}, \texttt{HasTyp.as\_mnf} & \texttt{MNF/Core.lean} \\
		Theorem~\ref{thm:safety-relaxed}               & \texttt{Eval.safety}                 & \texttt{Safety.lean}          \\
		Theorem~\ref{thm:safety-mnf}                   & \texttt{MNF.safety}                  & \texttt{MNF/Basic.lean}       \\
		Corollary~\ref{cor:effect-safety-full}         & \texttt{Eval.effect\_safety}         & \texttt{Safety/Corollaries.lean} \\
		Corollary~\ref{cor:capture-prediction}         & \texttt{Eval.capture\_prediction}    & \texttt{Safety/Corollaries.lean} \\
		Corollary~\ref{cor:closure-tightness}          & \texttt{Eval.closure\_tightness}     & \texttt{Safety/Corollaries.lean} \\
		Corollary~\ref{cor:handler-coverage-full}      & \texttt{Eval.handler\_coverage}      & \texttt{Safety/Corollaries.lean} \\
		Corollary~\ref{cor:boundary-safety}            & \texttt{Eval.boundary\_safety}       & \texttt{Safety/Corollaries.lean} \\
		Corollary~\ref{cor:pass-handler-safety}        & \texttt{Eval.pass\_handler\_safety}  & \texttt{Safety/Corollaries.lean} \\
		Set semantics of kinds (Section~\ref{sub:set-semantics}) & equivalence lemmas         & \texttt{Classifier/Semantics.lean} \\
		\bottomrule
	\end{tabular}
\end{table}

Table~\ref{tab:lean-judgments} relates the judgments of the paper to the
corresponding Lean types, and Table~\ref{tab:lean-theorems} locates each
stated result in the sources.
The constructors of \texttt{HasTyp} correspond one to one to the rules of
Figure~\ref{fig:relaxed-full}:
\texttt{var}, \texttt{sub}, \texttt{abs}, \texttt{tabs}, \texttt{cabs},
\texttt{app}, \texttt{tapp}, \texttt{capp}, \texttt{lett}, \texttt{exlet},
\texttt{pack\_var}, \texttt{pack\_val}, \texttt{label}, \texttt{invoke},
\texttt{boundary}, \texttt{intercept\_pass} and \texttt{intercept\_gen}
implement
\ruleref{rt-var}, \ruleref{rt-sub}, \ruleref{rt-abs}, \ruleref{rt-tabs},
\ruleref{rt-cabs}, \ruleref{a-rt-app}, \ruleref{rt-tapp}, \ruleref{rt-capp},
\ruleref{rt-let}, \ruleref{rt-ex}, \ruleref{a-rt-pack-var},
\ruleref{a-rt-pack-val}, \ruleref{rt-label}, \ruleref{rt-break},
\ruleref{rt-bnd}, \ruleref{rt-icp-pass} and \ruleref{rt-icp-gen}
respectively.
The cases of \texttt{Term.eval} implement the evaluation rules of
Figures~\ref{fig:bigstep-full} and \ref{fig:bigstep-ext} in the same way,
and the constructors of the kind algebra carry the names of the rules in
Figure~\ref{fig:kinds-full}.

\subsection{The Try Example}\label{app:lean-try}

The encoding of \lstinline|Try[T]| from Section~\ref{sec:calculus} is
mechanized in full in \texttt{Try.lean}, from the Church encoded type down to
the constructor and its typing derivation.
We give the development here in the notation of the paper.

\paragraph{Comparison to Scala's \lstinline|Try|.}
Scala's \lstinline|Try[T]| constructor has the capture-typed signature
\begin{scalacode}
def apply[T](body: => T): Try[T]^{body.only[Control]}
\end{scalacode}
so the returned \lstinline|Try[T]| captures only the control-kinded effects
of \lstinline|body|.
The encoding matches this exactly.
\texttt{Try.mk} produces a value of type
$\mathsf{Try}(X, \bs{\set{\sproj{c}{\cls{\text{Control}}}}})$, where only the
control-kinded part of the thunk's capture set $\bs{\set{c}}$ is retained.
Although a $\mathsf{Try}$ value is itself pure, its own use set is empty,
applying it, that is pattern matching on the result, charges the retained
control captures to the caller through the capture annotation on the outer
binder of its type.
The control effects therefore resurface at the use site, which is the
intended behavior of a value that stands for a suspended control effect.

\paragraph{The Try type.}
Given a shape type $T$ and a capture set $\bs{C}$, the type
$\mathsf{Try}(T, \bs{C})$ is the Church encoded sum of a success and a failure
case (\texttt{Try}).
\begin{align*}
	\mathsf{Try}(T, \bs{C}) \;=\;
	 & \forall[X{<:}\top].\;\forall[c{:}\top].                   \\
	 & \quad\Bigl(
	\forall\!\bigl(s{:}\underbrace{(\forall(x{:}T)\,X)\capt\bs{\{c\}}}_{\text{success continuation}}\bigr). \\
	 & \qquad
	\forall\!\bigl(f{:}\underbrace{\bigl(\forall[Y{<:}\top].\;
		\forall(b{:}(\BREAK[Y])\capt\bs{C})\,(\forall(r{:}Y)\,X)\capt\bs{\{c\}}
		\bigr)\capt\bs{\{c\}}}_{\text{failure continuation}}\bigr).\;X
	\Bigr)\capt\bs{\{c\mid\top\}\cup\bs{C}}
\end{align*}
A value of type $\mathsf{Try}(T, \bs{C})$ is a function that, given a success
continuation and a failure continuation, invokes the appropriate one.
The failure continuation receives the break payload $b$ together with a
resumption $r$.
The type is proved covariant in both $T$ and $\bs{C}$ (\texttt{Try.sub}), so a
$\mathsf{Try}$ over a smaller result type and a smaller capture set is a
subtype.

\paragraph{Constructors.}
\texttt{Try.success} takes a value $v : T$ and produces an MNF term of type
$\mathsf{Try}(T, \bs{C})$ that ignores the failure continuation and applies
the success continuation to $v$ (\texttt{Try.success.typeable}).
\texttt{Try.failure} takes a label $\ls\ell : (\BREAK[U]) \capt \bs{C}$ and a
value $v : U$ and produces an MNF term of type $\mathsf{Try}(T, \bs{C})$ that
ignores the success continuation and passes $\ls\ell$ and $v$ to the failure
continuation (\texttt{Try.failure.typeable}).

\paragraph{The constructor \texttt{Try.mk}.}
The main entry point has type
\[
	\mathsf{Try.mk} :
	\forall[X{<:}\top].\;
	\forall[c{:}\cls{\top}]^{\bs{\set{\sproj{c}{\cls{\ksub{\top}{\text{Control}}}}}}}\!.\;
	\forall\!\bigl(t{:}(\forall(x{:}\top)\,X)\capt\bs{\{c\}}\bigr).\;
	\mathsf{Try}(X, \bs{\set{\sproj{c}{\cls{\text{Control}}}}}).
\]
It takes a thunk $t$ and proceeds in three steps
(\texttt{Try.mk}, \texttt{Try.mk'}).
\begin{enumerate}\itemsep1pt
	\item It runs the thunk inside an $\INTERCEPT$ that intercepts
	      control-kinded breaks, that is $\cls{\KIND} = \cls{\text{Control}}$.
	\item If the thunk returns a value $w$, it wraps $w$ with
	      \texttt{Try.success} into a success result.
	\item If a break $\BRK(\ls\ell, w)$ escapes the thunk, the intercept
	      handler invokes \texttt{Try.failure} with $\ls\ell$ and $w$.
\end{enumerate}
The capture set of the resulting $\mathsf{Try}$ value is
$\bs{C}\PROJ[\cls{\text{Control}}]$, the control-kinded part of the thunk's
capture set $\bs{C}$, reflecting that only control-kinded captures can surface
as failures.
Because the handler is a pass handler, the use set of the whole term is the
non-control part $\bs{C}\PROJ[\cls{\ksub{\top}{\text{Control}}}]$
(\texttt{Try.mk'.typeable}).
The entire term is in MNF and is proved well-typed in the MNF system
(\texttt{Try.mk.typeable}), so it is a genuine program of System \calculus{}
and not only of the relaxed system.

\subsection{Checking the Development}\label{app:lean-build}

The development builds with Lean 4.30.0 using
\texttt{lake build}.
The main safety theorems are in \texttt{Safety.lean} and
\texttt{MNF/Basic.lean}, and the corollaries in
\texttt{Safety/Corollaries.lean}.
The sources contain no \texttt{sorry} and declare no axioms, so the kernel
checks every result down to Lean's foundations.
 
\end{document}